\documentclass[12pt]{article}
\usepackage{amsmath}
\usepackage{graphicx}
\usepackage{enumerate}
\usepackage{natbib}
\usepackage{url} 
\usepackage{bm}
\usepackage{amsfonts}
\usepackage[flushleft]{threeparttable}
\usepackage{booktabs}
\usepackage{xcolor}
\usepackage[medium,compact]{titlesec}
\usepackage{tikz}
\usepackage{multirow}
\usepackage{subcaption}
\usepackage{bigints}
\usepackage{amsthm}
\usepackage{hyperref}

\allowdisplaybreaks

\newcommand{\blind}{1}

\newcommand{\bbI}{\mathbb{I}}
\newcommand{\Pn}{\mathbb{P}_n}
\newtheorem*{myexample}{Example 1}
\newtheorem{remark}{Remark}
\newtheorem{lemma}{Lemma}
\newtheorem{example}{Example}
\newtheorem{assumption}{Assumption}
\newtheorem{theorem}{Theorem}
\newtheorem{proposition}{Proposition}
\newtheorem{definition}{Definition}
\theoremstyle{plain}

\begin{document}

\def\spacingset#1{\renewcommand{\baselinestretch}%
{#1}\small\normalsize} \spacingset{1}


\if1\blind
{
  \title{\bf \Large Inverting estimating equations for causal inference on quantiles}
  \author{ Chao Cheng$^*$ and Fan Li$^\dagger$\hspace{.2cm}\\
    Department of Biostatistics, \\
    Center for Methods in Implementation and Prevention Science, \\ Yale School of Public Health, CT, USA\\
    $^*$\href{mailto:c.cheng@yale.edu}{c.cheng@yale.edu}  \quad $^\dagger$\href{mailto:fan.f.li@yale.edu}{fan.f.li@yale.edu}}
  \maketitle

  \vspace{-1cm} 
  
} \fi

\if0\blind
{
  \bigskip
  \bigskip
  \bigskip
  \begin{center}
    {\LARGE\bf Inverting estimating equations for causal inference on quantiles}
\end{center}
  \medskip
} \fi

\begin{abstract}
\noindent The causal inference literature frequently focuses on estimating the mean of the potential outcome, whereas quantiles of the potential outcome may carry important additional information. We propose a unified approach, based on the inverse estimating equations, to generalize a class of causal inference solutions from estimating the mean of the potential outcome to its quantiles. 
We assume that a moment function is available to identify the mean of the threshold-transformed potential outcome, based on which a convenient construction of the estimating equation of quantiles of potential outcome is proposed. In addition, we give a general construction of the efficient influence functions of the mean and quantiles of potential outcomes, and explicate their connection. We motivate estimators for the quantile estimands with the efficient influence function, and develop their asymptotic properties when either parametric models or data-adaptive machine learners are used to estimate the nuisance functions. A broad implication of our results is that one can rework the existing result for mean causal estimands to facilitate causal inference on quantiles. Our general results are illustrated by several analytical and numerical examples.
\end{abstract}

\noindent%

\spacingset{1.75} 

\section{Introduction}

The potential outcomes framework is typically used to describe the population-level causal effect as a contrast between a pair of mean potential outcomes. Examples include the average treatment effect, the natural indirect and direct effects in mediation analysis \citep{imai2010general}, and the principal causal effect \citep{frangakis2002principal}. Beyond a simple average, the quantiles of the potential outcome distribution can often serve as more complete summary measures. 
Assessing the quantile causal effect offers important insights into the treatment benefit over different regions of the potential outcome distribution. Recently, there has been a flourishing line of research dedicated to investigating quantile causal effects. For example, \cite{firpo2007efficient} and \cite{cattaneo2010efficient} considered the estimation of the quantile causal effect with a point treatment under the ignorability assumption; \citet{cheng2024doubly} studied identification results for the marginal structural quantile models with a time-varying treatment. In cases where endogeneity is a concern, \cite{frolich2013unconditional} used instrumental variables to identify complier quantile treatment effect. In the mediation context, \cite{hsu2023doubly} proposed a doubly robust estimator for the quantile indirect and direct effects. 

Despite the increasing interests in causal inference for quantiles, there is generally a paucity of work providing a unified procedure to streamline causal inference for quantiles when their counterparts for the average causal effect estimands are well-understood. To fill in this gap, we propose a generic framework to generalize causal inference from the average to quantiles of a potential outcome $Y_d$, where $Y_d$ is the outcome $Y$ that would have been observed under certain condition $d$.  
We start from a moment function that can identify the mean of $Y_d$ (subject to a threshold transformation). 
Based on this identifying moment function, we elicit the corresponding estimating equation targeting the quantiles of $Y_d$. Our development is built on a simple but important connection between quantiles and mean---the quantile of $Y_d$ is the inverse of the cumulative distribution function (CDF) of $Y_d$, and the latter can be characterized by an expectation of the threshold-transformed potential outcome $\bbI(Y_{d}\leq \theta)$, where $\bbI(\cdot)$ is the indicator function and $\theta$ is a threshold within the support of outcome. Based on this connection, we demonstrate that an appropriate estimating equation for quantiles can be directly obtained from switching the roles of the parameters of interest in the original identifying moment function.  
We refer to this approach as an \textit{inverse estimating equation} (IEE) approach. 

{A notable feature of the IEE is that it typically includes certain nuisance functions, $\gamma$, of the observed data distribution, such as the conditional mean and CDF of the observed variables.} As a  preliminary step, we may first obtain an estimator for the nuisance $\widehat\gamma$ and then solve a plug-in IEE based on this first-step estimator. However, this plug-in estimator may be not preferred for several reasons. First, the IEE based solely on the identifying moment function can be sensitive to misspecifications of $\gamma$, such that the plug-in IEE estimator is biased if $\widehat\gamma$ is not consistent. Second, even if data-adaptive machine learners are used to minimize the bias for estimating $\gamma$, the plug-in estimator may still fail to be $\sqrt{n}$-consistent, leading to inferential challenges at the usual parametric rate. Building upon advances in statistical learning of nonparametric functionals \citep{van2006targeted,chernozhukov2018double}, we further propose a debiased IEE approach for estimating the causal quantiles; this approach combines the IEE with the efficient influence function (EIF) of the quantile of the potential outcome distribution to ensure valid inference at the parametric rate, even if the machine learners used to estimate the nuisance functions converge at a slower rate. We further explicate the relationship between the EIFs for the mean and the quantile of potential outcome distribution, and show that one does not have to re-derive the EIF for a quantile causal estimand whenever the counterpart for a mean causal estimand already exists. We establish simple regularity conditions for $\sqrt{n}$-consistency and asymptotic normality of the debiased IEE estimator for quantile estimands, when either parametric models or nonparametric machine learners with cross-fitting are used to estimate the nuisance functions. {We involve both analytical and numerical examples to illustrate our general theoretical results.} 

\section{Inverting the mean estimating equation to target quantile estimands}\label{sec:method}

\subsection{A general class of problems concerning quantile estimands}
Suppose that we observe $\{W_i\}_{i=1}^n$ as $n$ independent copies of a random vector $W$ with an unknown CDF, $F$. Oftentimes, we can write $W = \{Y,A,X\}$, where $Y$ is a univariate outcome, $A$ is the treatment assignment, and $X$ includes all other variables. For example, $X$ usually involves a set of baseline covariates $L$ but can also include post-treatment variables $M$. 
While $Y$ is assumed to be a scalar, we place no restriction on $A$ and $X$ so they can be vectors. For each unit, we let $Y_d$ be the potential outcome, which is the value of $Y$ that would have taken had this unit received certain path $d \in \mathcal D$ underlying $\{A,M\}$.
For example, with a binary treatment $A\in\{0,1\}$, we can define $Y_d :=Y_a$ as the value of $Y$ would have taken if $A$ were set to level $a\in\{0,1\}$ (see Example \ref{example:example1}). With a binary treatment and a post-treatment variable $M$ (mediator), one can set $Y_d := Y_{am}$ as the value of $Y$ would have taken had $A$ been set to level $a$ and $M$ been set to level $m$ (see Example 2). Notice that the definition of potential outcome implicitly imposes a consistency assumption requiring that (i) all variables $W$ are defined unambiguously and (ii) there is no between-unit interference (the potential outcome of one unit does not depend on other units' treatment or covariates). 
Throughout, for two variables $U$ and $Z$, let $F_{U\mid Z}(u\mid z)$ be the true CDF of $U=u$ given $Z=z$, $f_{U\mid Z}(u\mid z)$ be its corresponding probability density or mass function, $E[U\mid Z=z]=\int u dF_{U\mid Z}(u\mid z)$ be the expectation of $U$ given $Z=z$, and $Q_{U\mid Z}(q\mid z) = \text{inf} \{u : F_{U\mid Z}(u\mid z) \geq q\}$ be its $q$-quantile. 

Frequently, we are interested in identification and estimation of the mean potential outcome $Y_d$, possibly conditional on $V$. Usually, $V$ is a subset of $\{A,X\}$ (including $V=\emptyset$); for example, if we set $V=A$, the expectation of $Y_d$ conditional on $A=a$ represents the mean potential outcome on the sub-population who received the treatment $A=a$. Under the principal stratification framework \citep{frangakis2002principal}, $V$ can depend on potential values of post-treatment variables $M$ to represent principal strata (Example 3).  
Under certain identifying assumptions required in each setup, the mean of the threshold-transformed potential outcome, denoted by $\tau_0(\theta)$, can be identified based on $W$: 
\begin{equation}\label{eq:counterfactual_mean}
\tau_0(\theta) := E\left[\bbI(Y_d\leq \theta)\mid V=v\right],
\end{equation}
where $\theta\in\mathcal Y$ is a threshold, and $\mathcal Y$ is the sample space of $Y_d$. Notice that $\tau_0(\theta)$ equals to the probability of $Y_d$ at or below $\theta$ conditional on $V=v$, and lies in the unit interval $[0,1]$.  By definition, $\tau_0(\theta)$ connects the mean to the quantiles of the potential outcome distribution and has a dual role. On the one hand, investigating $\tau_0(\theta)$ over $\theta$ provides a complete assessment of the CDF of $Y_d|\{V=v\}$, facilitating assessment of the quantile. On the other hand, $\tau_0(\theta)$ can be regarded as a expectation of a threshold-transformed potential outcome, $\bbI(Y_d\leq\theta)$. 

\begin{remark}[Identification of the mean threshold-transformed outcome]
If we can identify the mean of the potential outcome, identifying the mean of the threshold-transformed potential outcome is trivial. To elaborate, suppose we have a set of assumptions to point identify $E[Y_d|V=v]$; then, if we view $Y^*:=\bbI(Y\leq\theta)$ as the new outcome, it is straightforward to identify $E[Y_d^*|V=v]$ based on the original identifying assumptions (subject to at most slight modifications), where $Y_d^*:=\{\bbI(Y\leq\theta)\}_d$ is the value of $\bbI(Y\leq\theta)$ that would have been observed under path $d$. Because $Y_d^*\equiv\bbI(Y_d\leq\theta)$ (the interchange of $d$ and indicator function is guaranteed by the consistency assumption), we can ensure the identification of $\tau_0(\theta)$.  
\end{remark}

To be more concrete, suppose that $\tau_0(\theta)$ can be identified based on a moment function $g(W,\tau,\theta,\gamma)$, where $W$ is the observed variables, $\tau \in [0,1]$ is a scalar parameter, $\theta$ is the threshold in \eqref{eq:counterfactual_mean}, and $\gamma \equiv \gamma(W,F,\theta) := \{\gamma_{1}(W,F,\theta),\dots,\gamma_{R}(W,F,\theta)\}$ is a set of nuisance functions of the observed data CDF, 
which may themselves depend on $\theta$. Importantly, our framework accommodates the scenario where $\gamma$ depends on $\tau$; we omit this dependency in our notation for ease of exposition. 
Below are two examples of nuisance functions: $\gamma_{1}(L,F):=f_{A|L}(1|L)$, and  $\gamma_{2}(L,F,\theta):=F_{Y|A,L}(\theta|1,L)$, where the first is the propensity score and the second is the probability of the outcome at or below $\theta$ among treated units given $L$. 
We suppress the arguments of $\gamma$ in $g(W,\tau,\theta,\gamma) \equiv g(W,\tau,\theta,\gamma(W,F, \theta))$ for brevity.  
To proceed, we assume $\tau_0(\theta)$ is identified based on $g(W,\tau,\theta,\gamma)$ in the sense of the following condition. 
\begin{assumption}\label{assump:identification_mean}
(Point identification of $\tau_0(\theta)$) For any $\theta\in\mathcal Y$, $\tau_0(\theta)$ defined in \eqref{eq:counterfactual_mean} is the unique solution of $E[g(W,\tau,\theta,\gamma)] = 0$ with respect to $\tau\in[0,1]$. 
\end{assumption}
{ Assumption \ref{assump:identification_mean} essentially requires that the original causal assumptions for identifying the mean of the threshold-transformed potential outcome hold.} To elaborate, 
Assumption \ref{assump:identification_mean} suggests that $g(W,\tau,\theta,\gamma)$ is an unbiased estimating score for $\tau_0(\theta)$ such that
\begin{equation}\label{eq:mean_identification}
E[g(W,\tau_0(\theta),\theta,\gamma)] = 0.
\end{equation}
If $\gamma$ is known, an unbiased estimating equation for $\tau_0(\theta)$ can be obtained by setting $\Pn[g(W,\tau,\theta,\gamma)] = 0$, where $\Pn[Z] = \frac{1}{n}\sum_{i=1}^n Z_i$ defines the empirical average. 
To contextualize our discussion, we start with the following classic causal inference problem as a special case of our general setup. 

\begin{example}\label{example:example1}
\textit{(Causal effect under ignorability)} Suppose we observe $W=\{Y,A,L\}$ with $A\in \{0,1\}$ a binary treatment and $L$ baseline covariates. The estimand of interest is $\tau_0(\theta):=E[\bbI(Y_a\leq \theta)]$ based on which the average treatment effect on the threshold outcome $E[\bbI(Y_1\leq \theta)-\bbI(Y_0\leq\theta)]$ can be defined. Under (I) consistency ($Y_a=Y$ if $A=a$), (II) ignorability ($Y_a\perp A\mid L$), and (III) positivity ($f_{A\mid L}(a\mid l)>0$ for any $a$ and $l$), $\tau_0(\theta)$ can be identified by an inverse probability weighted average, $\tau_0(\theta) = E\left[\frac{\bbI(A=a)\bbI(Y\leq\theta)}{f_{A\mid L}(a\mid L)}\right]$. Therefore, the identifying moment function of $\tau_0(\theta)$ is  $g(W,\tau,\theta,\gamma) = \frac{\bbI(A=a)\bbI(Y\leq\theta)}{f_{A\mid L}(a\mid l)}-\tau$ with nuisance $\gamma = f_{A\mid L}(a\mid L)$. 
\end{example}

The goal of this paper is one step further, that is, to identify the quantile of the potential outcome distribution. Specifically, for a fixed real $q\in(0,1)$, we develop a unified strategy to estimate the $q$-quantile of the potential outcome distribution $F_{Y_d\mid V}$, denoted by $\theta_0$:
\begin{equation}\label{eq:counterfactual_quantile}
\theta_0 := Q_{Y_d\mid V}(q\mid v). 
\end{equation}
Based on \eqref{eq:counterfactual_quantile}, a quantile causal effect can be defined as the difference between $Q_{Y_d\mid V}(q|v)$ and $Q_{Y_{d'}\mid V}(q|v)$ with $d'\neq d\in \mathcal D$. This estimand precisely measures the effect of $d$ versus $d'$ on the $q$-quantile of the potential outcome conditional on the sub-population with $V=v$. In the context of Example \ref{example:example1}, the central task is to obtain an identification formula for $\theta_0$, given well-established identification results for estimating the mean potential outcome. We characterize a unified strategy below to point identify $\theta_0$ in our general problem setup.

\subsection{Constructing the inverse estimating equation}\label{sec:IEE_2}

As a prerequisite for studying causal inference on quantiles, we make the following assumption to ensure that the $q$-quantile of $Y_d\mid \{V=v\}$ is unique:
\begin{assumption}\label{assump:unique_quantile}
(Uniqueness of the $q$-quantile) 
There exists $c_1>0$ and an interval $\Theta:=(c_2,c_3) \subseteq \mathcal Y$ such that $F_{Y_d|V}(\theta|v)$ is continuous and strictly increasing in $\Theta$ with $F_{Y_d|V}(c_2|v)<q-c_1$ and $F_{Y_d|V}(c_3|v)>q+c_1$.
\end{assumption}
{Assumption \ref{assump:unique_quantile}  assumes that the CDF of $Y_d\mid \{V=v\}$ is continuously increasing and not flat everywhere in a neighborhood $\Theta$ around its $q$-quantile, which often holds when the outcome is a continuous  variable.} 
Importantly, Assumption \ref{assump:unique_quantile} ensures that the quantile function $Q_{Y_d|V}$ is unambiguously defined by the inverse function of  $F_{Y_d|V}(\theta|v)$ for $\theta \in \Theta$ so that 
\begin{equation}\label{eq:inverse_function}
\theta_0:=Q_{Y_d\mid V}(q\mid v) = F_{Y_d\mid V}^{-1}(q\mid v). 
\end{equation}
If one plugs $\theta=\theta_0$ into \eqref{eq:counterfactual_mean}, we shall have
$$
\tau_0(\theta_0)  = E[\bbI(Y_d\leq \theta_0)\mid V=v] = F_{Y_d|V}(\theta_0|v) = F_{Y_d|V} \circ F_{Y_d|V}^{-1}(q|v) = q,
$$
where the first to the last equalities hold by \eqref{eq:counterfactual_mean}, definition of CDF, \eqref{eq:inverse_function}, and the property of inverse functions, respectively. Finally, substituting $\theta=\theta_0$ into equation \eqref{eq:mean_identification} and noting $\tau_0(\theta_0)=q$, we can verify that $E[g(W,q,\theta_0,\gamma)] = 0$; that is, $\theta_0$ is a solution of $E[g(W,q,\theta,\gamma)] = 0$ with respect to $\theta \in \Theta$. The following Theorem further shows that $\theta_0$ is also the unique solution so that $\theta_0$ is identified based on $g$, although this moment function is originally motivated to identify the mean of the threshold-transformed potential outcome.

\begin{theorem}\label{thm:identification}
(Point identification of $\theta_0$) Under Assumptions \ref{assump:identification_mean} and \ref{assump:unique_quantile}, $\theta_0$  is the unique solution of $E[g(W,q,\theta,\gamma)] = 0$ in term of $\theta \in \Theta$.
\end{theorem}

Theorem \ref{thm:identification} suggests that we can repurpose $g(W,\tau,\theta,\gamma)$ as an identifying moment function for  $\theta_0$, by now treating $\theta$ as the new unknown parameter but fixing the original unknown parameter $\tau=q$. Therefore, an estimating equation for $\theta_0$ can be constructed by setting the empirical average of $g(W,q,\theta,\gamma)$ to zero such that
\begin{equation}\label{eq:iee}
\Pn\left[g(W,q,\theta,\gamma)\right] = 0.
\end{equation}
We refer to \eqref{eq:iee} as the inverse estimating equation (IEE) for $\theta_0$, as we have intentionally switched the roles of parameters in the original estimating equation for \eqref{eq:counterfactual_mean}. 

\begin{myexample} (Continued) Now suppose our interest lies in assessing the quantile treatment effect (QTE), defined as $Q_{Y_1}(q)-Q_{Y_0}(q)$, where the essential task is estimating $\theta_0 := Q_{Y_a}(q)$ with $a\in \{0,1\}$. Based on the identifying moment function for $\tau_0(\theta)$ and Assumption \ref{assump:unique_quantile}, it is immediate that an identifying moment function of $\theta_0$ is $g(W,q,\theta,\gamma)=\frac{\bbI(A=a)\bbI(Y\leq\theta)}{f_{A\mid L}(a\mid L)}-q$. It follows that the IEE is $\Pn\left[\frac{\bbI(A=a)\bbI(Y\leq\theta)}{f_{A\mid L}(a\mid L)}-q\right]=0$, where we now treat $\theta$ as the unknown parameter to be solved for. This is precisely the estimating equation in  \cite{firpo2007efficient} for estimating  $\theta_0$. 
\end{myexample}

In an ideal setting when $\gamma$ is known, an estimator $\widehat\theta^{o}$ of $\theta$ can be obtained by solving \eqref{eq:iee} with respect to $\theta$. Oftentimes, the IEE is a nonsmooth function with respect to $\theta$ (see Example \ref{example:example1}) and one can use a derivative-free algorithm to find an approximate solution. 
In Proposition S1 of the Supplementary Material, we show that $\widehat\theta^{o}$ is consistent and asymptotically normal (CAN) under mild regularity conditions. 
In practice, however, $\gamma$ is often unknown. When $\gamma$ is unknown, one should first obtain a consistent estimator $\widehat\gamma\equiv\widehat\gamma(W,\theta)$ and then solve $\Pn\left[g(W,q,\theta,\widehat\gamma)\right] = 0$ with respect to $\theta$ to obtain a plug-in IEE estimator, $\widehat\theta^{pi}$. 
As previously mentioned, $\widehat\theta^{pi}$ is not preferred because it may be subject to bias when $\gamma$ 
is estimated by machine learners or misspecified parametric models, and is not fully efficient. 
These limitations can be alleviated by using the debiased inverse estimating equation (debiased IEE)  in Section \ref{sec:debiased}. 

\begin{remark}[Clarification on the nuisance functions]
\label{remark:nuisance_function}
Estimation of $\gamma$ in the IEE is often more complicated than in original estimating equations for $\tau_0(\theta)$. This is because $\theta$, which may also be nested in $\gamma$, is a parameter to be solved for in the IEE but was treated as a fixed point in the original estimating equation. In the IEE, we are required to estimate $\gamma$ for all $\theta\in\Theta$. 
For instance, across all examples, we need to estimate a nuisance function of the form $F_{Y|U}(\theta|u)$ with $U$ being a subset of $\{A,X\}$. In this case, one can use parametric or machine learning methods to estimate the outcome CDF $F_{Y|U}(y|u)$ for $y\in \mathcal{Y}$ and then set $\widehat F_{Y|U}(\theta|u)$ as  $\widehat F_{Y|U}(y|u)$ with $y=\theta$. 
\end{remark}

\begin{remark}[Comparison to the inverse CDF quantile estimator]\label{remark:inverset_CDF_approach}
{An alternative approach for estimating $\theta_0$ exists by inverting the entire CDF of $Y_{d}|\{V=v\}$ \citep{hsu2023doubly}. In this approach, one selects a set of grid points $\underline\theta_1<\underline\theta_2<\cdots < \underline\theta_R$ in $\mathcal Y$,  obtains $\widehat\tau_0(\underline\theta_r) = \widehat F_{Y_d|V}(\underline\theta_r|v)$ for each $\underline\theta_r$ based on original estimating equation $g(W,\tau,\underline\theta_r,\widehat\gamma)$, and then empirically inverts the discrete CDF estimate $\{\underline\theta_r,\widehat F_{Y_d|V}(\underline\theta_r|v)\}$ to obtain the desired quantile estimate of $\theta_0$. Theoretically, this estimator should bear similar performance to the plug-in IEE approach when the grid points are sufficiently dense, as it solves a discrete analogue of the IEE. In practice, however, this inverse CDF approach requires additional work on specifying grid points, and could be subject to bias when $R$ is small or the location of $\underline \theta_r$ is sparse around $\theta_0$. To illustrate the feature of each estimator, we numerically compare the inverse CDF estimator of \cite{hsu2023doubly} with the proposed debiased IEE estimator in the context of quantile mediation analysis; see Section \ref{sec:sim_qme} for details.} 
\end{remark}

\section{Improving efficiency and robustness via the debiased IEE}\label{sec:debiased}

\subsection{Connections between the efficient influence functions for $\tau_0(\theta)$ and $\theta_0$}\label{sec:debiased_IEE2}

The debiased IEE relies on the concept of efficient influence function (EIF) employed to estimate nonparametric functionals. To describe the EIF, one could define a parametric submodel $F_t:=(1-t) F + t \widetilde F$ indexed by a one-dimensional parameter $t\in [0,1]$, where $\widetilde F$ is another CDF of $W$. 
Let $z(F)$ be a nonparametric functional of $F$ and $z(F_t)$ be its value evaluated under the parametric submodel $F_t$. Following \citet{bickel1993efficient}, the EIF of $z(F)$ (under the nonparametric model), whenever exists, is the unique function $\psi_z^{\text{eff}}$ satisfying $\frac{d}{dt}z(F_t)|_{t=0} = \int \psi_z^{\text{eff}}(w)d\widetilde F(w)$ with $E[\psi_z^{\text{eff}}(W)]=0$ and $E[\psi_z^{\text{eff}}(W)^2]<\infty$, where $\frac{dz(F_t)}{d t}$ is the pathwise derivative with respect to $t$. 
The variance of the EIF is the semiparametric efficiency lower bound for estimating $z(F)$. 
To construct the debiased IEE, we first derive the EIF for $\theta_0$, 
which is a zero-mean finite-variance function $\psi_{\theta_0}^{\text{eff}}$ satisfying $\frac{d \theta_0(F_t)}{dt}|_{t=0}  = \int \psi_{\theta_0}^{\text{eff}}(w) d \widetilde F(w)$, where $\theta_0(F_t)$ is the value of $\theta_0$ under $F_t$.  
Our central point is to establish a connection between the EIF of $\tau_0(\theta)$ (denoted by $\psi_{\tau_0(\theta)}^{\text{eff}}$) and $\psi_{\theta_0}^{\text{eff}}$ such that knowing $\psi_{\tau_0(\theta)}^{\text{eff}}$ (as developed extensively in existing literature) provides a direct path for obtaining $\psi_{\theta_0}^{\text{eff}}$.

Define $\bar g (F_t;\tau,\theta) := E[g(W,\tau,\theta,\gamma_t)]$, where   $\gamma_t\equiv\gamma(w,F_t,\theta)$ is the value of the nuisance functions under $F_t$.
In order to offer deeper insights into the relationships between $\psi_{\tau_0(\theta)}^{\text{eff}}$ and $\psi_{\theta_0}^{\text{eff}}$, we assume that there exists a function $\phi(W,\tau,\theta,h)$ with $E[\phi(W,\tau,\theta,h)] = 0$ and $E[\phi(W,\tau,\theta,h)^2] < \infty$ such that 
\begin{equation}\label{eq:fsif}
\frac{d}{d t} \bar g (F_t;\tau,\theta) \Big|_{t=0} = \int \phi(w,\tau,\theta,h) d \widetilde F(w), 
\end{equation}
where the arguments $\{W,\tau,\theta\}$ are defined as in $g(W,\tau,\theta,\gamma)$ and $h\equiv \{\gamma(W,F,\theta),\alpha(W,F,\theta)\}$ is a set of  nuisance functions. Notice that $h$ may involve additional nuisance functions $\alpha$ besides $\gamma$. The function $\phi(W,\tau,\theta,h)$ is the {adjustment term} \citep{newey1994asymptotic} for investigating the general constructions of the EIF of nonparametric functionals---also referred to as the {first-step influence function} in \citet{chernozhukov2022locally}. By definition, $\phi(W,\tau,\theta,h)$ is the EIF for $\bar g (F;\tau,\theta)= E[g(W,\tau,\theta,\gamma)]$, which measures how sensitive the expectation of the identifying moment function $g(W,\tau,\theta,\gamma)$ would change in the presence of a small perturbation of $\gamma$ toward $\gamma_t$. In other words, the assumption in \eqref{eq:fsif} states that the EIF for $\bar g (F;\tau,\theta)$ exists. 
Given this assumption, the following theorem reveals the relationship between $\psi_{\tau_0(\theta)}^{\text{eff}}$ and $\psi_{\theta_0}^{\text{eff}}$. 
\begin{theorem}\label{thm:EIF_relationship}
Assume that (i) for any $\theta\in \mathcal Y$ and $\tau\in [0,1]$, the EIF for $\bar g(F;\tau,\theta)$ exists  (i.e., there exists a zero-mean finite-variance function $\phi(W,\tau,\theta,h)$ satisfying \eqref{eq:fsif}); (ii) for any $t\in [0,1]$, $\bar g(F_t,\tau,\theta)$ is continuous and differentiable of $\{\tau,\theta\}\in[0,1]\times \mathcal Y$ with $E[g(W,\tau,\theta,\gamma)^2]<\infty$. 
Then, for any $\theta\in\mathcal Y$, if $C(\theta):=\frac{d}{ d \tau}\bar g(F,\tau,\theta)|_{\tau=\tau_0(\theta)}$ is non-singular, the EIF for $\tau_0(\theta)$ exists and equals to
\begin{align*}
\psi_{\tau_0(\theta)}^{\text{eff}}(W,\tau_0(\theta),\theta,h) & = -\frac{1}{C(\theta)} \Big\{g(W,\tau_0(\theta),\theta,\gamma) + \phi(W,\tau_0(\theta),\theta,h)\Big\}
\end{align*}
Similarly, if $B:=\frac{d}{ d \theta}\bar g(F,q,\theta)|_{\theta=\theta_0}$ is non-singular, the EIF for $\theta_0$ exists and equals to
\begin{equation}\label{eq:theta_eif}
\psi_{\theta_0}^{\text{eff}}(W,q,\theta_0,h) = - \frac{1}{B} \Big\{g(W,q,\theta_0,\gamma) + \phi(W,q,\theta_0,h)\Big\}.
\end{equation}
\end{theorem}
In Theorem \ref{thm:EIF_relationship}, condition (i) is discussed previously; condition (ii) requires that $\bar g(F_t;\tau,\theta)$ is a smooth function of $\tau$ and $\theta$. Moreover, we also require the partial derivatives of $\bar g(F;\tau,\theta)$ with respect to $\tau$ and $\theta$ evaluated at their true values $\tau_0(\theta)$ and $\theta_0$ (i.e., $C(\theta)$ and $B$) are non-singular in order to ensure the EIF of $\tau_0(\theta)$ and $\theta_0$ exist because they appear in the denominator. 
This result reveals that the EIFs for $\tau_0(\theta)$ and $\theta_0$ carry very similar patterns, both consisting of three components---the identifying moment function $g$, the adjustment term $\phi$, and a normalizing denominator ($C(\theta)$ or $B$). 
Although $\psi_{\tau_0(\theta)}^{\text{eff}}$ and $\psi_{\theta_0}^{\text{eff}}$ share the same identifying moment function $g$ and adjustment term $\phi$, the values of their second and third arguments in $g$ and $\phi$ are different---in $\psi_{\tau_0(\theta)}^{\text{eff}}$ they were set to $\tau_0(\theta)$ and $\theta$ but in $\psi_{\theta_0}^{\text{eff}}$ they were set to $q$ and $\theta_0$.

Theorem \ref{thm:EIF_relationship} further indicates that if the expression for any one of (i) the adjustment term $\phi$, (ii) the EIF of $\tau_0(\theta)$, or (iii) the EIF of $\theta_0$ is known, we can immediately obtain expressions of the other two. To elaborate, if the adjustment term is known, then $\psi_{\tau_0(\theta)}^{\text{eff}}$ and $\psi_{\theta_0}^{\text{eff}}$ can be obtained directly as shown Theorem \ref{thm:EIF_relationship}. Alternatively, if $\psi_{\tau_0(\theta)}^{\text{eff}}(W,\tau_0(\theta),\theta,h)$ is known, the EIF for $\theta_0$ is
\begin{equation}\label{eq:eif_relationship}
\psi_{\theta_0}^{\text{eff}}(W,q,\theta_0,h) = \frac{C(\theta_0)}{B} \psi_{\tau_0(\theta)}^{\text{eff}}(W,q,\theta_0,h),
\end{equation}
by changing the value of second and third arguments in $\psi_{\tau_0(\theta)}^{\text{eff}}$ from $\tau_0(\theta)$ and $\theta$ to $q$ and $\theta_0$. Moreover, the adjustment term is $\phi(W,\tau,\theta,h) = -C(\theta)\psi_{\tau_0(\theta)}^{\text{eff}}(W,\tau,\theta,h) - g(W,\tau,\theta,h)$. Similarly, if $\psi_{\theta_0}^{\text{eff}}(W,q,\theta_0,h)$ is known, then the EIF of $\tau_0(\theta)$ is $\psi_{\tau_0(\theta)}^{\text{eff}}(W,\tau_0(\theta),\theta,h) = \frac{B}{C(\theta)} \psi_{\theta_0}^{\text{eff}}(W,\tau_0(\theta),\theta,h)$, where one should  change the value of second and third arguments in $\psi_{\theta_0}^{\text{eff}}$ from $q$ and $\theta_0$ to $\tau_0(\theta)$ and $\theta$; the adjustment term is $\phi(W,\tau,\theta,h) = -B\psi_{\theta_0}^{\text{eff}}(W,\tau,\theta,h) - g(W,\tau,\theta,h)$. Therefore, besides direct calculation of $\psi_{\theta_0}^{\text{eff}}$ based on the pathwise derivative $\theta_0(F_t)$, we have two additional strategies to calculate the EIF for $\theta_0$. As a first strategy, if the EIF for $\tau_0(\theta)$ already exists, then we can obtain $\psi_{\theta_0}^{\text{eff}}$ based on \eqref{eq:eif_relationship} (Strategy I). Alternatively, if we know the identifying moment function such that the EIF for $\bar g(F;\tau,\theta)$ (i.e., the adjustment term) can be easily constructed, then we can obtain $\psi_{\theta_0}^{\text{eff}}$ based on \eqref{eq:theta_eif} (Strategy II). This is further illustrated using our running example. 

\begin{myexample} (Continued)
\cite{hahn1998role} studies the EIF of average treatment effect, and have shown that the EIF for $\tau_0(\theta)$, $\psi_{\tau_0(\theta)}^{\text{eff}}(W,\tau_0(\theta),\theta,h)$, has the following form
$$
\frac{\bbI(A=a)}{f_{A|X}(a|L)}\left\{\bbI(Y\leq\theta)-E[\bbI(Y\leq\theta)|A=a,L]\right\} + E[\bbI(Y\leq\theta)|A=a,L] - \tau_0(\theta),
$$ 
where $h = \{f_{A|L}(a|L),F_{Y|A,X}(\theta|a,L)\}$ because $E[\bbI(Y\leq\theta)|A=a,L] = F_{Y|A,X}(\theta|a,L)$. The EIF for the quantile estimand $\theta_0:=Q_{Y_a}(q)$ has been derived in \cite{firpo2007efficient}, and here we re-derive it using Strategy I.  Observing that $\bar g(F;\tau,\theta)=E\left[\frac{\bbI(A=a)\bbI(Y\leq\theta)}{f_{A\mid L}(a\mid L)}-\tau \right]= E\left[E\left\{\bbI(Y\leq\theta)|A=a,L\right\}\right]-\tau  = E[F_{Y|A,L}(\theta|a,L)]-\tau$ by law of iterated expectation, we obtain that $C(\theta):=\frac{d \bar g(F;\tau,\theta)}{d \tau}  |_{\tau=\tau_0(\theta)} = -1$ and $B:=\frac{d \bar g(F;q,\theta)}{d \tau}  |_{\theta=\theta_0} = E[f_{Y|A,X}(\theta_0|a,L)]$. Following \eqref{eq:eif_relationship}, we immediately know that  $\psi_{\theta_0}^{\text{eff}}(W,q,\theta_0,h)$ is given by 
$$
-\frac{1}{B}\times \left[\frac{\bbI(A=a)}{f_{A|L}(a|L)}\left\{\bbI(Y\leq\theta_0)-F_{Y|A,L}(\theta_0|a,L)\right\} + F_{Y|A,L}(\theta_0|a,L) - q\right].
$$ 
The derivation of the EIF of $\theta_0$ based on Strategy II is presented in the Supplementary Material. 
\end{myexample}

\subsection{The debiased IEE and its asymptotic properties}\label{sec:DIEE3}

An improved estimator for $\theta_0$ can be obtained by solving the following debiased IEE:
\begin{equation}\label{eq:debiased_IEE}
\Pn[\psi_{\theta_0}^{\text{eff}}(W,q,\theta,\widehat h)] = 0,
\end{equation}
where we replace $h$ with its estimator $\widehat h \equiv \widehat h(W,\theta)$. We denote the solution to \eqref{eq:debiased_IEE} as $\widehat\theta^{de}$. When \eqref{eq:debiased_IEE} is nonsmooth of $\theta$, derivative-free algorithms can be used to obtain an approximate solution. Because the normalizing denominator $B$ in the EIF does not affect $\widehat\theta^{de}$, the debiased IEE can be re-expressed as adding the adjustment term to the plug-in IEE, $\Pn[g(W,q,\theta,\widehat \gamma)+ \phi(W,q,\theta,\widehat h)] = 0$. As shown below, the adjustment term can help achieve additional robustness as compare to the plug-in IEE alone, which is borne out by the following mixed bias property. 
\begin{definition}\label{def:mixed_bias}
(Mixed bias property) Assume that  $h(W,F,\theta)$ has $J\geq 2$ components with $\{h_{1}(W,F,\theta),\dots,h_{J}(W,F,\theta)\}$. 
We say $\psi_{\theta_0}^{\text{eff}}(W,q,\theta, h)$ is mixed biased, if there exists $K\geq 1$ distinct pairs $\{a_k,b_k\}$ with $a_k\neq b_k\in \mathcal J$ and $\cup_{k=1}^K\{a_k,b_k\}=\mathcal J$, such that, for any  $\theta\in\Theta$,
\begin{equation}\label{eq:mixed_bias}
E[\psi_{\theta_0}^{\text{eff}}(W,q,\theta, \widehat h)-\psi_{\theta_0}^{\text{eff}}(W,q,\theta, h)] = \sum_{k=1 }^K E\left[ S_{k}\{\widehat h_{a_k}-h_{a_k}\}\{\widehat h_{b_k}-h_{b_k}\}\right],
\end{equation}
where $\mathcal J:=\{1,\dots,J\}$, $S_k \equiv S_{k}(W, h, \widehat h)$ is a known function of $W$, $h$, and $\widehat h$ with $E[S_k^2]<\infty$.
\end{definition}
The mixed bias property was first discussed in \cite{rotnitzky2021characterization} with $J=2$ nuisance functions, and Definition \ref{def:mixed_bias} offers a generalization to quantile estimands and $J>2$. The mixed bias property offers important advantages when either data-adaptive machine learners or parametric models are used to obtain $\widehat h$. We first consider machine learning estimators for nuisance functions. With data-adaptive machine learning estimators, it is necessary to adopt the cross-fitting scheme to control for the empirical process term \citep{zheng2010asymptotic,chernozhukov2018double}. Under this scheme, one evenly splits the data into $R$ groups (e.g., $R=5$). For each unit $i$ from group $r\in\{1,\dots,R\}$, the EIF for that unit used in \eqref{eq:debiased_IEE} is evaluated based on an estimator $\widehat h$ trained on data for all units not included in the $r$-th group. Following the convention, 
we assume that $\widehat h_{j}$ converges to the true $h_{j}$ at a $\xi_{j,n}$-order such that $\sup_{\theta\in\Theta}\|\widehat h_{j}-h_{j}\|=O_p(\xi_{j,n})$, where $\|h\|=\{\int h^2dF\}^{1/2}$ defines the $L_2(F)$-norm. The following theorem summarizes the asymptotic property of $\widehat \theta^{de}$. A key condition in Theorem \ref{thm:machine_learner} is that we require the product rate of convergence of $\widehat h$ to satisfy $\sum_{k=1}^K\xi_{a_k,n}\xi_{b_k,n}=o(n^{-{1}/{2}})$, where $a_k$ and $b_k$ are defined in \eqref{eq:mixed_bias}. Therefore, a standard $o_p(n^{-{1}/{4}})$-type rate condition for each component of $\widehat h$ are sufficient for the $\widehat \theta^{de}$ to be consistent, asymptotically normal, and efficient.
  
\begin{theorem}\label{thm:machine_learner}
Suppose that $\widehat h$ is consistently estimated by data-adaptive machine learners based on cross-fitting and (i) conditions in Theorem \ref{thm:EIF_relationship} hold with $B\neq 0$; (ii) the function $\theta \to \psi_{\theta_0}^{\text{eff}}(W,q,\theta,h)$ is P-Donsker for $\theta\in\Theta$; (iii) the EIF for $\theta_0$ satisfies the mixed bias property \eqref{eq:mixed_bias} with all $E[S_k^2]<\infty$; (iv) $\widehat \theta^{de}$ is an approximate solution of the debiased IEE in the sense that $\Pn[\psi_{\theta_0}^{\text{eff}}(W,q,\widehat\theta^{de},\widehat h)] = o_p(n^{-{1}/{2}})$. If the rate of convergence of $\widehat h$ satisfies $\sum_{k=1}^K\xi_{a_k,n}\xi_{b_k,n}=o(n^{-{1}/{2}})$, then $\sqrt{n}(\widehat \theta^{de}-\theta_0)$ is asymptotically normal with zero-mean and its variance achieves the semiparametric efficiency lower bound characterized by $E[\psi_{\theta_0}^{\text{eff}}(W,q,\theta_0, h)^2]$. 
\end{theorem}


\begin{remark}
{Condition (ii) in Theorem \ref{thm:machine_learner} ensures that the estimating function of the debiased IEE with known nuisance, $\psi_{\theta_0}^{\text{eff}}(W,q,\theta,h)$, satisfies the stochastic equicontinuity property with respect to $\theta\in\Theta$, and places no restriction on the complexity of the nuisance function estimators. Let $G_n(\theta) = \mathbb{P}_n[\psi_{\theta_0}^{\text{eff}}(W,q,\theta,h)]$ and $G(\theta) = E[\psi_{\theta_0}^{\text{eff}}(W,q,\theta,h)]$, then condition (ii) implies  (ii$'$) $\sup_{\theta\in\Theta}\frac{\|G_n(\theta)-G(\theta)\|}{1+\|G_n(\theta)\|+\|G(\theta)\|}=o_p(1)$ and  (ii$''$)  $\sup_{|\theta-\theta_0|<\delta_n} \frac{\|G_n(\theta)-G(\theta)+G_n(\theta_0)\|}{n^{-1/2}+\|G_n(\theta)\|+\|G(\theta)\|}=o_p(1)$ for all  $\delta_n=o_p(1)$, which are two Huber-type regularity conditions that are commonly used in semparametric analysis with non-smooth estimating equations \citep{pakes1989simulation}. Without using the empirical process language, Theorem \ref{thm:machine_learner} continues to hold even if we replace condition (ii) by (ii$'$)---required for proving consistency---and (ii$''$)---required for proving asymptotic normality. }
\end{remark}

Next, we consider the case of using finite-dimensional parametric working models to estimate $h$,  where certain components of $\widehat h$ may be inconsistent and converge to an incorrect probability limit. For $j\in \mathcal J$, let $\mathcal M_j$ be a parametric working model for $h_{j}$, where $\widehat h_j$ is consistent to $h_{j}$ if $\mathcal M_j$ is correctly specified but asymptotically biased otherwise. We use ``$\cup$" and ``$\cap$" to denote union and intersection of models so that $\mathcal M_j \cup \mathcal M_l$ means that either $\mathcal M_j$ or $\mathcal M_l$ is correctly specified and $\mathcal M_j \cap \mathcal M_l$ means that both $\mathcal M_j$ and $\mathcal M_l$ are correctly specified. 

\begin{theorem}\label{thm:parametric}
Suppose that $\widehat h$ is obtained by parametric working models and regularity conditions in the Supplementary Material hold. If the union model $\mathcal M_{\text{union}} := \cap_{k=1}^K\{\mathcal M_{a_k}\cup \mathcal M_{b_k}\}$ is correctly specified, then $\sqrt{n}(\widehat \theta^{de}-\theta_0)$ converges to a zero-mean normal distribution with its variance $V_{\theta_0}$ given in the Supplementary Material. Moreover, if $\cap_{j=1}^J \mathcal M_j$ is correct, $V_{\theta_0}$ achieves the semiparametric efficiency lower bound characterized by $E[\psi_{\theta_0}^{\text{eff}}(W,q,\theta_0, h)^2]$. 
\end{theorem}
Theorem \ref{thm:parametric} 
shows that $\widehat \theta^{de}$, constructed from the debiased IEE, is a multiply robust estimator. Specifically, $\mathcal M_{\text{union}}$ holds if, for all $k=1,\dots,K$, either $\mathcal M_{a_k}$ or $\mathcal M_{b_k}$ is correctly specified. This provides at most $2^K$ opportunities to achieve consistency if the $2^K$ intersection union models $\{\cap_{k=1}^K \mathcal M_{c_k}:c_k\in\{a_k,b_k\} \text{ for } k=1,\dots,K\}$ are mutually distinct. If we only have $J=2$ nuisance functions, $\mathcal M_{\text{union}}$ degenerates to $\mathcal M_1 \cup \mathcal M_2$ so $\widehat \theta^{de}$ becomes a doubly robust estimator. {In Theorem \ref{thm:machine_learner}, we assume that $h$ is always consistently estimated when machine learners are used, which is plausible when the specified machine learners explore a sufficiently large parameter space. If $\mathcal M_j$ is a machine learning model and $\widehat h_j$ becomes inconsistent under misspecification, we show in Proposition S2 of the Supplementary Material that $\widehat \theta^{de}$ is still a consistent estimator under $\mathcal M_{union}$, but the $\sqrt{n}$-consistency and asymptotic normality are no longer guaranteed.}


\begin{myexample} (Continued)
Through some algebra, we can show that $E[\psi_{\theta_0}^{\text{eff}}(W,q,\theta, \widehat h)-\psi_{\theta_0}^{\text{eff}}(W,q,\theta, h)]=E\left[ S_{1}\{\widehat f_{A|L}(a|L)-f_{A|X}(a|L)\}\{\widehat F_{Y|A,L}(\theta|a,L)-F_{Y|A,L}(\theta|a,L)\}\right]$, 
with $S_1^{-1} = - B\times \widehat f_{A|L}(a|L)$. If  $\widehat f_{A|L}(a|L)$ is bounded away from 0, then $E[S_1^2]<\infty$ and we confirm that $\psi_{\theta_0}^{\text{eff}}$ satisfies the mixed bias property. Thus, if the machine learners are used for $\widehat h$, then a $o_p(n^{-{1}/{4}})$-rate consistency of both $\widehat F_{Y|A,L}(\theta|a,L)$ and $\widehat f_{A|L}(a|L)$ are sufficient for $\widehat \theta^{de}$ to be consistent, asymptotically normal and efficient---a special case included in Theorem \ref{thm:machine_learner}.  Otherwise, if parametric working models are used, $\widehat \theta^{de}$ is consistent and asymptotically normal if either $\widehat F_{Y|A,L}(\theta|a,L)$ or $\widehat f_{A|L}(a|L)$ is correct, and is a doubly robust estimator (Theorem \ref{thm:parametric}). {We conduct simulations in the Supplementary Material to investigate finite-sample behavior of $\widehat \theta^{de}$ and show that it improves the existing estimator of \cite{firpo2007efficient}.}
\end{myexample}

For inference, one can use nonparametric bootstrap to estimate the asymptotic variance of $\widehat \theta^{de}$ if parametric models are used to estimate each nuisance. However, bootstrap is not justified when machine learners are used for the nuisance. In this case, it is recommended to use the empirical variance of the EIF \citep{chernozhukov2018double} to estimate the asymptotic variance of $\widehat \theta^{de}$ such that $\text{Var}(\widehat \theta^{de})\approx \frac{1}{n}\mathbb{P}_n[\widehat B^{-2} \{g(W,q,\widehat \theta^{de},\widehat \gamma) + \phi(W,q,\widehat \theta^{de},\widehat h)\}^2]$, where $\widehat B$ is the estimator of $B$; notice that $\widehat B$ is not required for point estimation. 
Estimation of $B$ should be analyzed in each specific example. In Example 1, $B=E[f_{Y|A,L}(\theta_0|a,L)]$ can be estimated by $\widehat B=\mathbb{P}_n  [\widehat f_{Y|A,L}(\widehat{\theta}^{de}|a,L) ]$, where $\widehat f_{Y|A,L}(\widehat{\theta}^{de}|a,L)$ is an estimate of $f_{Y|A,L}(\theta|a,L)$ at $\theta =\widehat{\theta}^{de}$.

Below, we illustrate our general results in the context of quantile mediation analysis (Example 2). Additional examples (Example 3: survivor quantile causal effect and Example 4: quantile causal effect with a time-varying treatment), numerical studies, as well as a real data application are provided in the Supplementary Material.

\vspace{-0.5cm}

\section{Example 2: Quantile causal mediation analysis}

\subsection{Theoretical development}\label{sec:qme_theory}


Suppose that we observe $W=\{Y,M,A,L\}$ with $A\in\{0,1\}$ and $L$ as a set of baseline covariates. Following \cite{imai2010general}, we denote $Y_a$ and $M_a$ as the potential outcome and mediator had $A=a$ and $Y_{am}$ as the potential outcome had $A=a$ and $M=m$.  In causal mediation analysis, we can decompose the treatment effect on the $q$-quantile of the outcome, $Q_{Y_{1M_1}}(q)-Q_{Y_{0M_0}}(q)$, into a natural quantile indirect effect, $\text{NQIE}(q)= Q_{Y_{1M_1}}(q)-Q_{Y_{1M_0}}(q)$, and a natural quantile direct effect, $\text{NQDE}(q) = Q_{Y_{1M_0}}(q)-Q_{Y_{0M_0}}(q)$. Noting that $Q_{Y_{1M_1}}(q) \equiv Q_{Y_{1}}(q)$ and $Q_{Y_{0M_0}}(q) \equiv Q_{Y_{0}}(q)$ due to composition of potential outcomes, and both can be analyzed following Example 1. Below, we focus on estimation of $\theta_0:=Q_{Y_{1M_0}}(q)$.

As is shown in \cite{imai2010general}, under (I) consistency ($Y_{am}=Y$ if $A=a$ and $M=m$ and $Y_a=Y$ and $M_a=M$ for all $a$ and $m$), (II) sequantial ignorability ($\{Y_{am},M_{a^*}\}\perp A|L$ for all $a$, $a^*$ and $m$),  (III) positivity ($f_{A|L}(a|l)>0$ and $f_{A|M,L}(a|m,l)>0$ for any $a$, $m$, and $l$), the  mean of the threshold-transformed potential outcome  $\tau_0(\theta):=E[\bbI(Y_{1M_0}\leq\theta)]$ can be identified based on the identifying moment function
$$
g(W,\tau,\theta,\gamma) =  \mu(L,\theta) - \tau
$$
with $\gamma \equiv \mu(L,\theta) = E\left[F_{Y|A,M,L}(\theta|1,M,L)|A=0,L\right]$, where the expectation is taken with respect to $M$ given $A=0$ and $L$. If the $q$-quantile of $Y_{1M_0}$ is unique, Theorem 1 suggests that $\theta_0$ is identified and can be estimated by solving 
$
\Pn[\widehat \mu(L,\theta) - q] = 0,
$
where $\widehat \mu (L,\theta)$ is the estimate of $\mu (L,\theta)$. Estimation strategies for all nuisance functions involved in this section are presented in the Supplementary Material. 


We leverage the debiased IEE for improved estimation of $\theta_0$. The EIF for the mean estimand $\tau_0(\theta)$ was developed previously in \cite{tchetgen2012semiparametric} and \cite{zhou2022semiparametric} with $\psi_{\tau_0(\theta)}^{\text{eff}}(W,\tau_0(\theta),\theta,h)=-\frac{1}{C(\theta)}\{g(W,\tau_0(\theta),\theta,\gamma)+\phi(W,\tau_0(\theta),\theta,h)\}$, where $C(\theta)=-1$, 
\begin{align*}
\phi(W,\tau,\theta,h) = & \frac{A}{1-f_{A|L}(1|L)}\frac{1-f_{A|M,L}(1|M,L)}{f_{A|M,L}(1|M,L)}\{\bbI(Y\leq\theta)-F_{Y|A,M,L}(\theta|1,M,L)\} \\
& +\frac{1-A}{1-f_{A|L}(1|L)}\{F_{Y|A,M,L}(\theta|1,M,L)-\mu(L,\theta)\},
\end{align*}
is the adjustment term, and $h=\{\mu(L,\theta),f_{A|L}(1|L),f_{A|M,L}(1|M,L),F_{Y|A,M,L}(\theta|1,M,L)\}$. We abbreviate the four nuisance functions as $h_1$ to $h_4$ for simplicity. 
Applying Theorem 2, the EIF of $\theta_0$ is  $\psi_{\theta_0}^{\text{eff}}(W,q,\theta_0,h) = -\frac{1}{B}\{g(W,q,\theta_0,\gamma)+\phi(W,q,\theta_0,h)\}$ with $B:=\frac{dE[g(W,q,\theta,\gamma)]}{d\theta}|_{\theta=\theta_0}  = E\left\{E\left[f_{Y|A,M,L}(\theta|1,M,L)|A=0,L\right]\right\}$. Therefore, one can solve the following debiased IEE with respect to $\theta$ to obtain $\widehat\theta^{de}$:
$$
-B^{-1}\Pn\left[\widehat h_1 - q + \frac{(1-\widehat h_3)A}{\widehat h_3 (1-\widehat h_2)}\{\bbI(Y\leq\theta)-\widehat h_4\}+\frac{1-A}{1-\widehat h_2}\{\widehat h_4-\widehat h_1\}\right] = 0,
$$ 
where $\widehat h$ is the estimate of $h$. 
Alternatively, \cite{hsu2023doubly} proposed a double machine learning estimator of $\theta_0$ based on the inverse CDF approach (Remark \ref{remark:inverset_CDF_approach}); details of their approach are presented in the Supplementary Material and we provide a numerical comparison in Section \ref{sec:sim_qme}. 


We show in the Supplementary Material that $\psi_{\theta_0}^{\text{eff}}$ satisfies the mixed bias property with $K=2$ and $[\{a_1,b_1\},\{a_2,b_2\}]=[\{1,2\},\{3,4\}]$. It follows  from Theorem 3 that $\widehat\theta^{de}$ is consistent and asymptotically normal if $\widehat h$ is consistently estimated based machine learners with rate of convergence satisfying  $\xi_{1,n}\xi_{2,n}+\xi_{3,n}\xi_{4,n}=o(n^{-1/2})$. Otherwise, if we estimate $h$ based on parametric working models, Theorem 4 indicates that $\widehat \theta^{de}$ is consistent and asymptotically normal under  
$\{\mathcal M_{1}\cup \mathcal M_{2}\}\cap  \{\mathcal M_{3}\cup \mathcal M_{4}\}$; i.e., either  (i) $\mathcal M_1 \cap \mathcal M_{3}$, (ii) $\mathcal M_1 \cap \mathcal M_{4}$, (iii) $\mathcal M_2 \cap \mathcal M_{3}$, (iv) $\mathcal M_2 \cap \mathcal M_{4}$ is correctly specified. Notice that the two parametric models, $\mathcal M_1$ and  $\mathcal M_4$, are not variationally independent because $F_{Y|A,M,L}(\theta|1,M,L)$ (i.e., $\mathcal M_4$) is nested in $\mu(L,\theta)$ (i.e., $\mathcal M_1$), thus $\mathcal M_1$ is likely misspecified if $\mathcal M_4$ is misspecified. This suggests that (i) $\mathcal M_1 \cap \mathcal M_{3}$ is more rigid than (ii) $\mathcal M_1 \cap \mathcal M_{4}$ since (i) implicitly assumes correct specification of $\mathcal M_4$. Therefore, $\widehat \theta^{de}$ is triply robust under the union model of $\mathcal M_1 \cap \mathcal M_4$, $\mathcal M_2 \cap \mathcal M_3$, and $\mathcal M_2 \cap \mathcal M_4$. 

\subsection{Numerical studies}\label{sec:sim_qme}

We conduct numerical studies to (i) evaluate finte-sample performance of the proposed method for estimating $\theta_0:=Q_{1M_0}(q)$ and (ii) compare our method to the inverse CDF method in \cite{hsu2023doubly}. We consider $n=1,000$ with a binary treatment $A\in\{0,1\}$, four baseline covariates $L=\{L_1,L_2,L_3,L_4\}$, and a bivariate mediator $M=\{M_1,M_2\}$. The observed data $W=\{L,A,M,Y\}$ are simulated based on the specification in the Supplementary Material. 

We compare the performance across three estimators, with the same set of nuisance functions $h=\{\mu(L,\theta),f_{A|L}(1|L),f_{A|M,L}(1|M,L),F_{Y|A,M,L}(\theta|1,M,L)\}$ (details in Supplementary Material). 
First, we consider the inverse CDF estimator by \cite{hsu2023doubly}; this approach first estimates $\tau_0(\underline\theta_r) = F_{Y_{1M_0}}(\underline\theta_r)$ on a grid points $\{\underline\theta_r\}_{r=1}^R \subset\mathcal Y$, and the nuisance function estimates $\widehat h(W,\underline \theta_r)$ are obtained via Post-Lasso (as in the original approach of \citet{hsu2023doubly}). Next, we approximate the entire CDF $F_{Y_{1M_0}}$ by linearly interpolating the discrete points $\{\theta_r,\widehat \tau_0(\underline\theta_r)\}_{r=1}^R$, and then invert the CDF to obtain the quantile estimate of $\theta_0$. 
We denote this estimator by \texttt{hsu-pl($R$)}, with $R$ representing the number of grid points. Second, we consider the proposed debiased IEE estimator, \texttt{de-pl($R$)}, for which all nuisance function estimates are recycled from constructing \texttt{hsu-pl($R$)}. To elaborate, the nuisance functions in \texttt{de-pl($R$)}, i.e., $h(W,F,\theta)$ for $\theta\in\mathcal Y$, are approximated by linearly interpolating the set of Post-Lasso estimators $\{\underline\theta_r,\widehat h(W,\underline \theta_r)\}_{r=1}^R$ in $\mathcal Y$. Note that \texttt{de-pl($R$)} and \texttt{hsu-pl($R$)} leverage the same nuisance function estimates, and the only difference is that \texttt{hsu-pl($R$)} estimates $\theta_0$ by inverting a discrete CDF whereas \texttt{de-pl($R$)} directly estimates $\theta_0$ by solving the debiased IEE. To demonstrate potential improvement of using alternative machine learners, we consider an additional debiased IEE estimator, \texttt{de-ml}, with nuisance functions estimated from the Super Learner \citep{phillips2023practical} that fuses random forest, extreme gradient boosting, and generalized linear model to minimize the prediction error. One advantage of \texttt{de-ml} is that \texttt{de-ml} obtains the nuisance functions $h(W,F,\theta)$ directly without any discretization.

We use $1,000$ Monte Carlo simulations to evaluate the performance of \texttt{hsu-pl($R$)}, \texttt{de-pl($R$)}, and \texttt{de-ml}, and apply a five-fold cross-fitting procedure to obtain each estimator. We select $R\in\{4,10,40,100\}$ to evaluate performance of \texttt{hsu-pl($R$)} and \texttt{de-pl($R$)}. We consider 6 scenarios regarding correct and incorrect specification of the nuisance models: (a) all models  correctly specified; (b) $\widehat f_{A|L}(1|L)$ incorrect; (c) $\widehat f_{A|M,L}(1|M,L)$ incorrect; (d) $\widehat F_{Y|A,M,L}(\theta|1,M,L)$ incorrect; (e) $\widehat \mu(\theta,L)$ incorrect; (f) all models incorrect. For correctly specified models, we input the true baseline covariates $L$ into the the Post-Lasso regression or Super Learner. Otherwise, we input a set of transformed covariates $\widetilde L=[\widetilde L_1,\widetilde L_2,\widetilde L_3,\widetilde L_4]^T$ if the nuisance function is  misspecified, where $\widetilde L_1 = \exp(0.5L_1)$, $\widetilde L_2 = {L_2}/{(1+L_1)}$, $\widetilde L_3 = (L_2L_3/25+0.6)^3$, and $\widetilde L_4=(L_2+L_4+20)^2$. 

\begin{figure}[ht]
\begin{center}
\includegraphics[width=0.99\textwidth]{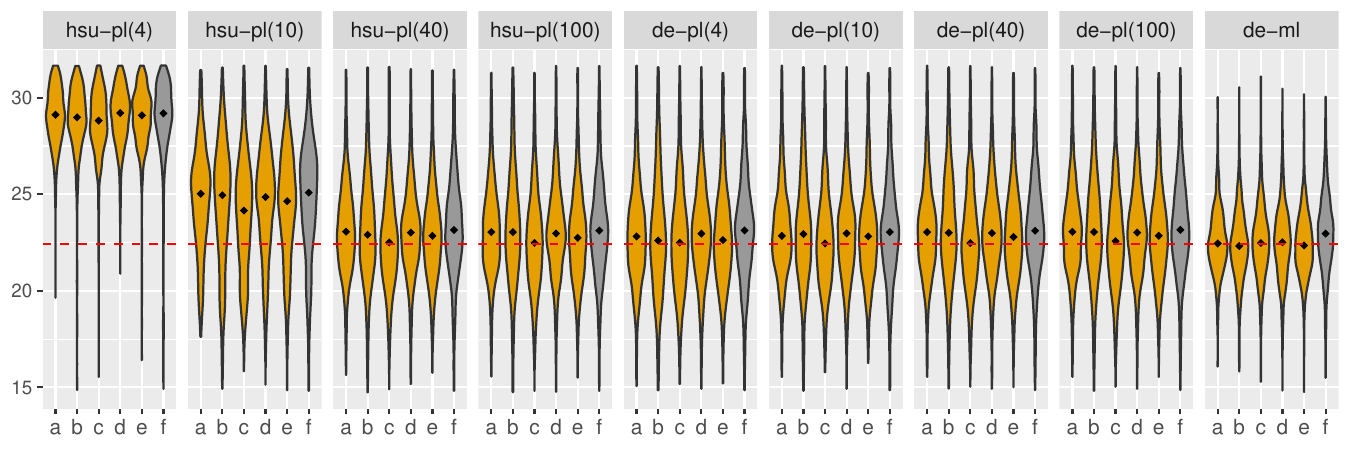}
\end{center}
\caption{Violin plots of the point estimates of $Q_{Y_1M_0}(0.9)$ under 6 scenarios (Scenarios a--f) regarding correct and incorrect specification of nuisance models. The red dotted line is the true value of $Q_{Y_1M_0}(0.9)$. For each scenario, the violin plot filled with orange color indicates the estimator is consistent based on theory.
}
\label{fig:QME-90A}
\end{figure}

Figure \ref{fig:QME-90A} presents the sampling distribution of each estimator on assessing $Q_{Y_{1M_0}}(0.9)$. Except for \texttt{hsu-pl(4)} and \texttt{hsu-pl(10)}, all other estimators are perform as expected with their sampling distribution centered around the truth under Scenarios (a)--(e). In contrast, \texttt{hsu-pl(4)} and \texttt{hsu-pl(10)} carry non-negligible bias among all scenarios, likely because few grid points are used to approximate the CDF $F_{Y_{1M_0}}$. However, \texttt{de-pl($R$)} provides much lower bias even with $R=4$. This is because in \texttt{de-pl($R$)}, the number of grid points only affects the quality of nuisance function estimation, without fluctuating the final inverse estimating equation for the quantile estimand. Meanwhile, we observe that the violin plots of \texttt{de-ml} are more concentrated around the truth, indicating higher efficiency than \texttt{de-pl($R$)} and \texttt{hsu-pl($R$)}. 

\begin{figure}[ht]
\begin{center}
\includegraphics[width=0.99\textwidth]{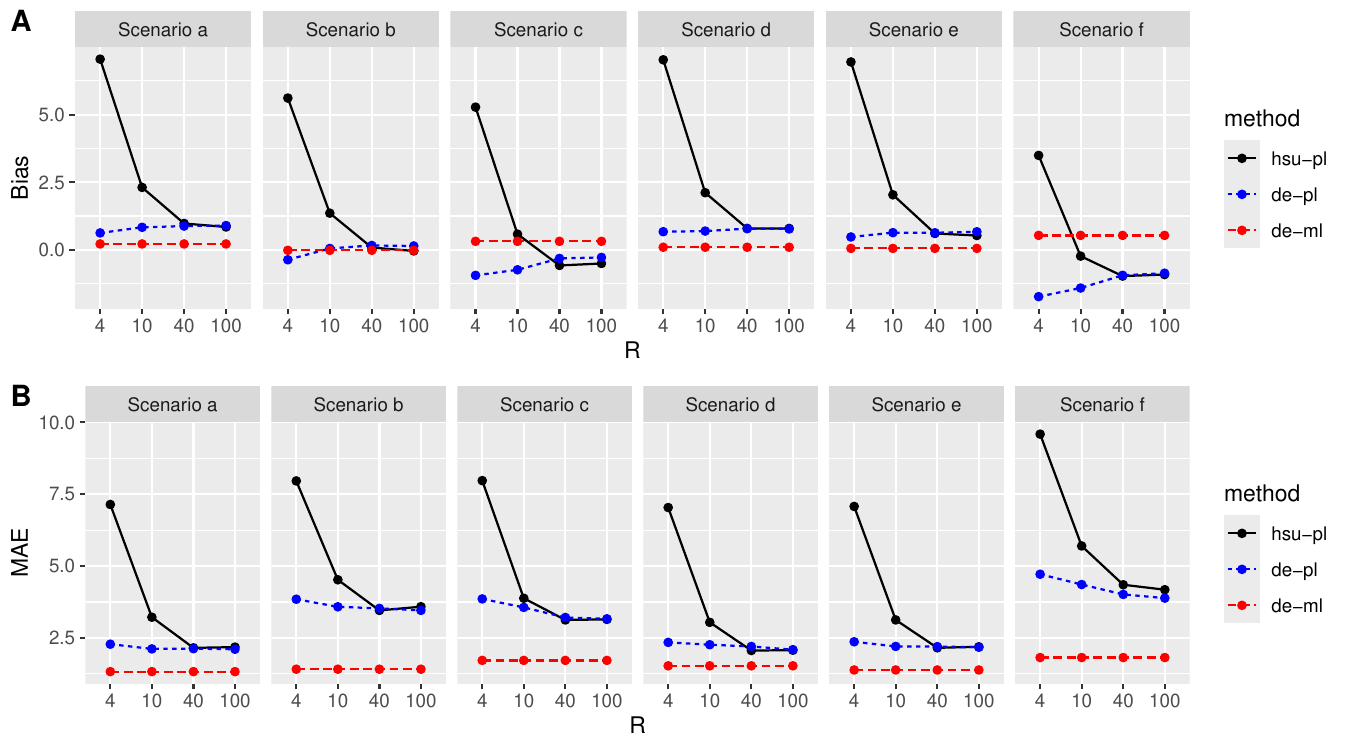}
\end{center}
\caption{Bias and MAE for point estimates of $Q_{1M_0}(0.9)$ with different number of thresholds (i.e., $R$) used to construct \texttt{hsu-pl($R$)} and \texttt{de-pl($R$)}. The bias and MAE of \texttt{de-ml} (red horizontal line) are added as reference.
}
\label{fig:QME-90BC}
\end{figure}

To further explore the bias and relative efficiency between \texttt{hsu-pl($R$)} and \texttt{de-pl($R$)} across different values of $R$, Figure \ref{fig:QME-90BC} presents the trend of the bias and mean absolute error (MAE) of both estimators with $R$ increasing from 4 to 100; the bias and MAE of \texttt{de-ml} are also added as a benchmark (red horizontal line). It is shown that \texttt{de-pl($R$)} is always less biased with a smaller MAE. As $R$ increases, the MAE of \texttt{de-pl($R$)} also shrinks much faster as compared to \texttt{hsu-pl($R$)}. When $R$ is large enough ($R \geq 40$), both \texttt{de-pl($R$)} and \texttt{hsu-pl($R$)} perform simiarly. Finally, \texttt{de-ml} always outperforms \texttt{de-pl($R$)} and \texttt{hsu-pl($R$)} in terms of the bias and MAE. In the Supplementary Material, we replicate the simulations on $Q_{Y_{1M_0}}(q)$ for $q\in\{0.1, 0.25, 0.5, 0.75\}$. The findings are qualitatively similar to that of $Q_{Y_{1M_0}}(0.9)$. 

\section{Discussion}

This paper provides a unified framework to generalize a class of causal inference problems from mean estimands to their quantile analogues. Our framework depends on two assumptions: (i) an identification formula for the mean of threshold-transformed potential outcome (i.e., $\tau_0(\theta)$) is available (Assumption \ref{assump:identification_mean}) and (ii) the $q$-th quantile of potential outcome is unique (Assumption \ref{assump:unique_quantile}). In principle, Assumption \ref{assump:identification_mean} depends on the causal assumptions used to identify the threshold-transformed mean, which are more extensively studied in the existing literature. When certain causal assumptions are violated, sensitivity analysis has been developed and could be considered. Assumption \ref{assump:unique_quantile} is plausible if the outcome is continuously distributed, but may not hold for discrete outcomes (e.g., count, ordinal data). However, as a possible path forward, we can leverage the jittering approach \citep{machado2005quantiles} to construct a ``jittered" outcome $Y_d^*$, by adding to the original potential outcome ($Y_d$) with a continuous and bounded noise term. Because the jittered outcome is continuous, we can estimate the quantile of $Y_d^*|\{V=v\}$ based on the proposed approach before transforming back to the original potential outcome $Y_d$. The theoretical property of this approach requires further investigation.

The estimated quantile curve is not guaranteed to be monotonic with respect to $q$ in finite samples, which is a common problem in quantile inference. An effective approach to ensure monotonicity is through quantile rearrangement \citep{chernozhukov2010quantile}. Consider a sequence of quantiles $q_1<q_2<\cdots<q_J$ within the interval $(0,1)$ and $\{\widehat Q_{Y_d|V}(q_1|v),\dots,\widehat Q_{Y_d|V}(q_J|v)\}$ be their initial estimates. The rearranged quantile estimates are then given by  $\widehat Q_{Y_d|V}^{\text{(R)}}(q_j|v) = \text{inf}\left\{y:\sum_{l=1}^J \mathbb{I}(\widehat Q_{Y_d|V}(q_l|v)\leq y) (q_l-q_{l-1})\geq q_j\right\}$, for each $j\in\{1,\dots,J\}$, where $q_0=0$. 
An empirical evaluation of this strategy will be pursued in future work. 

To facilitate implementation of the proposed methods, we have included reproducible R code for all  numerical analyses in the Supplementary Material.

\section*{Acknowledgement}
This work is supported by the Patient-Centered Outcomes Research Institute\textsuperscript{\textregistered} (PCORI\textsuperscript{\textregistered} Award ME-2023C1-31350). The statements are solely the responsibility of the authors and do not necessarily represent the views of PCORI\textsuperscript{\textregistered}, its Board of Governors or Methodology Committee.

\section*{Supplementary material}
Supplementary Material includes technical proofs, supporting information for Examples 1--2, two additional Examples 3--4, a real data application on the National Jobs Corps study, and  reproducible R code for all analyses.

\spacingset{1}

\bibliographystyle{jasa3}

\bibliography{Bibliography-MM-MC}

\clearpage

\section*{\centering Supplementary material to ``Inverting estimating equations for causal inference on quantiles"}

Section 1 provides Propositions \ref{thm:oracle}--\ref{thm:mr_machine_learning} and the proofs of all Theorems; Sections 2--3 provides additional supporting information for Examples 1--2 in the main manuscript; Sections 4--5 provides two additional examples (Examples 3 and 4) to further illustrate our general theoretical results; Section 4 additionally includes a simulation study and a real data application in the context of Example 3.

\renewcommand{\theproposition}{S\arabic{proposition}}
\renewcommand{\theequation}{S\arabic{equation}}

\setcounter{section}{0}
\setcounter{figure}{0}
\setcounter{assumption}{0}
\setcounter{proposition}{0}
\setcounter{equation}{0}

\section{Technical details}

\subsection{Proportions \ref{thm:oracle}--\ref{thm:mr_machine_learning}}

In Proposition \ref{thm:oracle}, we state the asymptotic properties of the IEE estimator with known nuisance, $\widehat\theta^o$. While we do not impose smoothness assumptions directly on $g(W,q,\theta,\gamma)$, a smoothness condition for the expectation of $g(W,q,\theta,\gamma)$ is required to enable inference.

\begin{proposition}\label{thm:oracle}
(Large-sample property of the IEE with known nuisance) 
Suppose that Assumptions 1--2 hold. We assume (i) the class of functions $\theta\to g(W,q,\theta,\gamma)$ is $P$-Donsker and $E[g(W,q,\theta,\gamma)^2]<\infty$ for $\theta\in\Theta$, (ii) $E[g(W,q,\theta,\gamma)]$ is differentiable of $\theta\in\Theta$ with non-singular derivative $B$ at $\theta_0$, (iii) $\widehat\theta^o$ is an approximate solution of the IEE in the sense that $\Pn[g(W,q,\widehat\theta^o,\gamma)]=o_p(n^{-{1}/{2}})$. Then, $\sqrt{n}(\widehat{\theta}^o-\theta_0)$ is asymptotically normal with mean 0 and  variance $B^{-2} E[g(W,q,\theta_0,\gamma)^2]$. 
\end{proposition}

In Proposition \ref{thm:mr_machine_learning}, we show the multiple robustness of $\widehat \theta^{de}$ when $\mathcal M_j$ is a data-adaptive machine learning model and the resulting $\widehat h_j$ can be inconsistent if $\mathcal M_j$ is misspecified. Specifically, we now assume that $\|\widehat h_j - h_j \| = o_p(1)$ if  $\mathcal M_j$ is correctly specified but $\|\widehat h_j - h_j \| = O_p(1)$ if $\mathcal M_j$ is misspecified. This is a generalization of Theorem 3 to the scenario that $h$ can be inconsistently estimated.

\begin{proposition}\label{thm:mr_machine_learning}
(Multiple robustness for the machine learning estimator) Suppose that $\widehat h$ is estimated through data-adaptive machine learners based on cross-fitting. Under regularity conditions (i)--(iv) in Theorem 3, $\widehat \theta^{de}$ is consistent if the union model $\mathcal M_{union}$ is correctly specified. 
\end{proposition}

\subsection{Proofs}

\begin{proof}[Proof of Theorem 1]
Since we have shown that $\theta_0$ is a solution of $E[g(W,q,\theta,\gamma)]=0$ in the main manuscript, here we only need to show the uniqueness of $\theta_0$. We prove this by contradiction. Suppose that there exists another solution $\theta^*\neq\theta_0 \in \Theta$ of $E[g(W,q,\theta,\gamma)]=0$ such that
\begin{equation}\label{eq:prop1_1}
E[g(W,q,\theta^*,\gamma)] = 0.
\end{equation}
Therefore, by equation (2), we have
\begin{equation}\label{eq:prop1_2}
E[g(W,\tau_0(\theta^*),\theta^*,\gamma)] = 0.
\end{equation}
Because Assumption 1 suggests that $\tau_0(\theta^*)$ is the unique solution of $E[g(W,\tau,\theta^*,\gamma)] = 0$ in terms of $\tau\in [0,1]$, we have $\tau_0(\theta^*)=q$ by comparing \eqref{eq:prop1_1} and \eqref{eq:prop1_2}. This further implies that
\begin{equation}\label{eq:prop1_3}
\tau_0(\theta^*)=\tau_0(\theta_0) \quad (\theta^*\neq\theta_0),
\end{equation}
because $\tau_0(\theta_0)=q$ as explained after equation (4). But Equation \eqref{eq:prop1_3} contradicts Assumption 2, as the latter requires that $\tau_0(\theta):=F_{Y_d|V}(\theta|v)$ is a monotonically increasing function of $\theta$. Therefore, $\theta_0$ must be the unique solution $E[g(W,q,\theta,\gamma)]=0$ in terms of $\theta$ over $\Theta$. 
\end{proof}

\begin{proof}[Proof of Proposition \ref{thm:oracle}]
The proof follows \cite{pakes1989simulation} by setting $G_n(\theta) = \Pn[g(W,q,\theta,\gamma)]$ and $G(\theta) = E[g(W,q,\theta,\gamma)]$. To prove consistency, we need to verify conditions (i)--(iii) in Corollary 3.2 of \cite{pakes1989simulation}. First, conditions (i) and (ii) can be easily verified based on our regularity conditions (ii) and (iii), respectively. To verify condition (iii) in Corollary 3.2 of \cite{pakes1989simulation}, it is sufficient to show that 
$\sup_{\theta \in \Theta}\|G_n(\theta)-G(\theta)\|=o_p(1)$, which is already implied from the P-Donsker condition in (i). Therefore, we have $\widehat\theta^o = \theta_0 + o_p(1)$. 

To establish the $\sqrt{n}$-consistency and asymptotic normality, one can use Theorem 3.3 of \cite{pakes1989simulation} by noting the consistency result and verifying their conditions (i)--(v), where conditions (i), (ii) and (v) are directly implied from the conditions in Proposition \ref{thm:oracle}. To verify their condition (iii), our P-Donsker condition suggests that $\sup_{|\theta-\theta_0|\leq \delta_n}\|G_n(\theta)-G(\theta)-G_n(\theta_0)\|=o_p(n^{-1/2})$ for all positive values $\delta_n=o_p(1)$, which implies their condition (iii). Moreover, by appealing to the Central Limit Theorem, one can show that $\sqrt{n}G_n(\theta_0)$ converges weakly to a normal distribution with mean $0$ with finite variance $E[g(W,q,\theta,\gamma)^2]$ (by our regularity condition (i)), therefore condition (iv) in Theorem 3.3 of \cite{pakes1989simulation} is verified. Finally, by applying Theorem 3.3 of \cite{pakes1989simulation}, we have that $\sqrt{n}(\widehat{\theta}^o-\theta_0)$ is asymptotically normal with mean 0 and  variance $B^{-2} E[g^2(W,q,\theta_0,\gamma)]$. 
\end{proof} 

\begin{proof}[Proof of Theorem 2]
First consider the EIF of $\tau_0(\theta)$. We shall derive $\psi_{\tau_0(\theta)}^{\text{eff}}$ by directly calculating the pathwise derivative $\frac{d}{dt} \tau_0(F_t;\theta)$, where $\tau_0(F_t;\theta)$ is the root of the implicit function $E_{F_t}[g(W,\tau,\theta,\gamma_t)]=0$ in terms of $\tau$ and $E_{F_t}$ is the expectation under $F_t$. Noting that $E_{F_t}[g(W,\tau,\theta,\gamma_t)]$ is differentiable with respect to $\tau$ (condition (ii)), we have
\begin{align*}
\frac{d}{dt} \tau_0(F_t,\theta) \Big|_{t=0} & = -\left[\frac{d}{d\tau}E_{F_0}[g(W,\tau,\theta,\gamma_0)]\Big|_{\tau=\tau_0(\theta)}\right]^{-1} \cdot \frac{d}{dt}E_{F_t}[g(W,\tau_0(\theta),\theta,\gamma_t)]\Big|_{t=0} \\
& = - \frac{1}{C(\theta)} \cdot \frac{d}{dt}E_{F_t}[g(W,\tau_0(\theta),\theta,\gamma_t)]\Big|_{t=0},
\end{align*}
by using the implicit function theorem. Furthermore, because $E_{F_t}[g(W,\tau_0(\theta),\theta,\gamma_t)] = \int g(w,\tau_0(\theta),\theta,\gamma_t) d F_t(w)$, we can show
\begin{align*}
\frac{d}{dt}E_{F_t}[g(W,\tau_0(\theta),\theta,\gamma_t)]\Big|_{t=0} & = \int g(w,\tau_0(\theta),\theta,\gamma) d \widetilde F(w) + \frac{d}{dt}\int g(w,\tau_0(\theta),\theta,\gamma_t) d F(w) \Big|_{t=0} \\
& = \int g(w,\tau_0(\theta),\theta,\gamma) d \widetilde F(w) + \int \phi(w,\tau_0(\theta),\theta,h) d \widetilde F(w) \\
& = \int g(w,\tau_0(\theta),\theta,\gamma) + \phi(w,\tau_0(\theta),\theta,h) d \widetilde F(w),
\end{align*}
where the first equality holds by the chain rule and the second equality holds by condition (i). Next, observing that $C(\theta)$  does not depend on $w$, we have
$$
\frac{d}{dt} \tau_0(F_t,\theta) \Big|_{t=0}= \int - \frac{1}{C(\theta)}\left\{g(w,\tau_0(\theta),\theta,\gamma) + \phi(w,\tau_0(\theta),\theta,h)\right\} d \widetilde F(w)
$$
and we confirm that $\psi_{\tau_0(\theta)}^{\text{eff}} (W,\tau_0(\theta),\theta,h)= - \frac{1}{C(\theta)}\left\{g(W,\tau_0(\theta),\theta,\gamma) + \phi(W,\tau_0(\theta),\theta,h)\right\}$ satisfies the constraint $\frac{d}{dt} \tau_0(F_t,\theta)|_{t=0} = \int \psi_{\tau_0(\theta)}^{\text{eff}} (w,\tau_0(\theta),\theta,h) d\widetilde F(w)$. Finally, because the expectations of both $g(W,\tau_0(\theta),\theta,\gamma)$ and $\phi(W,\tau_0(\theta),\theta,h)$ are 0 (condition (i)), we immediately have that 
$E[\psi_{\tau_0(\theta)}^{\text{eff}} (W,\tau_0(\theta),\theta,h)] = 0$. Also, conditions (i) and (ii) along with $C(\theta)\neq 0$ suggest that $E[\psi_{\tau_0(\theta)}^{\text{eff}} (W,\tau_0(\theta),\theta,h)^2] < \infty$. This then helps establish that $\psi_{\tau_0(\theta)}^{\text{eff}} (W,\tau_0(\theta),\theta,h)$ is the EIF of $\tau_0(\theta)$. 

The derivation of $\psi_{\theta_0}^{\text{eff}}$ follows a similar strategy. Notice that $\theta_0(F_t)$ is characterized by the root of the implicit function $E_{F_t}[g(W,q,\theta,\gamma_t)]=0$ in terms of $\theta$. By the implicit function theorem, we have 
\begin{align*}
\frac{d}{dt} \theta_0(F_t) \Big|_{t=0} & = -\left[\frac{d}{d\theta}E_{F_0}[g(W,q,\theta,\gamma_0)]\Big|_{\theta=\theta_0}\right]^{-1} \cdot \frac{d}{dt}E_{F_t}[g(W,q,\theta_0,\gamma_t)]\Big|_{t=0} \\
& = - \frac{1}{B} \cdot \frac{d}{dt}E_{F_t}[g(W,q,\theta_0,\gamma_t)]\Big|_{t=0},
\end{align*}
where 
\begin{align*}
\frac{d}{dt}E_{F_t}[g(W,q,\theta_0,\gamma_t)]\Big|_{t=0} & = \int g(w,q,\theta_0,\gamma) d \widetilde F(w) + \frac{d}{dt}\int g(w,q,\theta,\gamma_t) d F(w) \Big|_{t=0} \\
& = \int g(w,q,\theta_0,\gamma) + \phi(w,q,\theta_0,h) d \widetilde F(w),
\end{align*}
and thus
$$
\frac{d}{dt} \theta_0(F_t) \Big|_{t=0}= \int - \frac{1}{B}\left\{g(w,q,\theta_0,\gamma) + \phi(w,q,\theta_0,h)\right\} d \widetilde F(w).
$$
We have now verified that $\psi_{\theta_0}^{\text{eff}} (W,q,\theta_0,h)= -\frac{1}{B}\left\{g(W,q,\theta_0,\gamma) + \phi(W,q,\theta_0,h)\right\}$ satisfies the constraint $\frac{d}{dt} \theta_0(F_t)|_{t=0} = \int \psi_{\theta_0}^{\text{eff}} (w,q,\theta_0,h) d\widetilde F(w)$. Next, using a similar procedure, we can show $E[\psi_{\theta_0}^{\text{eff}} (W,q,\theta_0,h)]=0$ and $E[\psi_{\theta_0}^{\text{eff}} (W,q,\theta_0,h)^2]<\infty$, which conclude that $\psi_{\theta_0}^{\text{eff}} (W,q,\theta_0,h)$ is the EIF of $\theta_0$.
\end{proof}

Before we proceed to the proof of Theorem 3, we first introduce the following Lemma.

\begin{lemma}\label{lemma:s1}
if $\widehat h$ is obtained based on  cross-fitting and its rate condition satisfies what is listed in Theorem 3, then we have
\begin{equation}\label{eq:thm3_1}
\Pn[\psi_{\theta_0}^{\text{eff}}(W,q,\theta,\widehat h)] = \Pn[\psi_{\theta_0}^{\text{eff}}(W,q,\theta,h)]+o_p(n^{-1/2}),
\end{equation}
uniformly for $\theta\in\Theta$.
\end{lemma}

\begin{proof}[Proof of Lemma \ref{lemma:s1}]
Equation \eqref{eq:thm3_1} essentially shows that the \emph{plug-in} debiased IEE converges to the \emph{true} debiased IEE with a rate faster than $n^{-1/2}$. According to the cross-fitting procedure, we split the data into $R$ groups and let $\mathcal W_r$ be the the $r$-th fold and $\mathcal W_{-r}$ be remaining fold. For each $r\in\{1,\dots,R\}$, let $\widehat h^{r}$ be the nuisance function estimate on the $r$-th fold which is computed through machine learners trained based on $\mathcal W_{-r}$. 
Let $n_r$ be the sample size of the $r$-th fold and $\mathbb{P}_{n_r}$ be the corresponding empirical average operator on the $r$-th fold, then one can rewrite \eqref{eq:thm3_1} as
\begin{equation}\label{eq:thm3_2}
\frac{1}{n}\sum_{r=1}^R n_r\mathbb{P}_{n_r}[\psi_{\theta_0}^{\text{eff}}(W,q,\theta,\widehat h^r)] = \frac{1}{n}\sum_{r=1}^R n_r\mathbb{P}_{n_r}[\psi_{\theta_0}^{\text{eff}}(W,q,\theta,h)]+o_p(n^{-1/2}).
\end{equation}
Next, we can decompose $\mathbb{P}_{n_r}[\psi_{\theta_0}^{\text{eff}}(W,q,\theta,\widehat h^r)]$ into
\begin{align*}
\mathbb{P}_{n_r}[\psi_{\theta_0}^{\text{eff}}(W,q,\theta,\widehat h^r)] = & \mathbb{P}_{n_r}[\psi_{\theta_0}^{\text{eff}}(W,q,\theta,h)] +  (\mathbb{P}_{n_r}-E)[\psi_{\theta_0}^{\text{eff}}(W,q,\theta,\widehat h^r) - \psi_{\theta_0}^{\text{eff}}(W,q,\theta,h) ] \\
& + E[\psi_{\theta_0}^{\text{eff}}(W,q,\theta,\widehat h^r) - \psi_{\theta_0}^{\text{eff}}(W,q,\theta,h)],
\end{align*}
where the second and third terms on the right-hand side are called the empirical process term and the remainder term, denoted by $\text{Res}_{1,r}$ and $\text{Res}_{2,r}$, respectively. Next we show that $\text{Res}_{1,r}=o_p(n^{-1/2})$ due to cross-fitting and $\text{Res}_{2,r}=o_p(n^{-1/2})$ due to the mixed bias property, which will conclude equation \eqref{eq:thm3_2} and thus \eqref{eq:thm3_1} hold. To bound $\text{Res}_{1,r}$, first notice that the conditional variance is  
\begin{align*}
& \text{Var}\left\{(\mathbb{P}_{n_r} - E )\left[\psi_{\theta_0}^{\text{eff}}(W,q,\theta,\widehat h^r) - \psi_{\theta_0}^{\text{eff}}(W,q,\theta,h)\right]\Big|\mathcal W_{-r}\right\}\\ 
= &  \text{Var}\left\{\mathbb{P}_{n_r} \left[\psi_{\theta_0}^{\text{eff}}(W,q,\theta,\widehat h^r) - \psi_{\theta_0}^{\text{eff}}(W,q,\theta,h)\right]\Big|\mathcal W_{-r}\right\} \\
= &  \frac{1}{n_r}\text{Var}\left\{\psi_{\theta_0}^{\text{eff}}(W,q,\theta,\widehat h^r) - \psi_{\theta_0}^{\text{eff}}(W,q,\theta,h)\Big|\mathcal W_{-r}\right\} \\
= &  \frac{1}{n_r}\|\psi_{\theta_0}^{\text{eff}}(W,q,\theta,\widehat h^r) - \psi_{\theta_0}^{\text{eff}}(W,q,\theta,h)\|^2.
\end{align*}
This further suggests that, for any $\epsilon>0$,
\begin{align*}
& \text{Pr}\left\{\frac{\sqrt{n_r}\left|(\mathbb{P}_{n_r} - E )\left[\psi_{\theta_0}^{\text{eff}}(W,q,\theta,\widehat h^r) - \psi_{\theta_0}^{\text{eff}}(W,q,\theta,h)\right]\right|}{\|\psi_{\theta_0}^{\text{eff}}(W,q,\theta,\widehat h^r) - \psi_{\theta_0}^{\text{eff}}(W,q,\theta,h)\|} \geq \epsilon \right\} \\
= &  E\left[\text{Pr}\left\{\frac{\sqrt{n_r}\left|(\mathbb{P}_{n_r} - E )\left[\psi_{\theta_0}^{\text{eff}}(W,q,\theta,\widehat h^r) - \psi_{\theta_0}^{\text{eff}}(W,q,\theta,h)\right]\right|}{\|\psi_{\theta_0}^{\text{eff}}(W,q,\theta,\widehat h^r) - \psi_{\theta_0}^{\text{eff}}(W,q,\theta,h)\|} \geq \epsilon \Big| \mathcal W_{-r}\right\}\right] \\
\leq &\frac{1}{\epsilon^2} E\left[ \text{Var}\left\{\frac{\sqrt{n_r}\left|(\mathbb{P}_{n_r} - E )\left[\psi_{\theta_0}^{\text{eff}}(W,q,\theta,\widehat h^r) - \psi_{\theta_0}^{\text{eff}}(W,q,\theta,h)\right]\right|}{\|\psi_{\theta_0}^{\text{eff}}(W,q,\theta,\widehat h^r) - \psi_{\theta_0}^{\text{eff}}(W,q,\theta,h)\|}  \Big| \mathcal W_{-r}\right\}\right] \\
= & \epsilon^{-2},
\end{align*}
where the third row follows by the Markov's inequality. Therefore, we have 
$
\text{Res}_{1,r} = (\mathbb{P}_{n_r}-E) \left[\psi_{\theta_0}^{\text{eff}}(W,q,\theta,\widehat h^r) - \psi_{\theta_0}^{\text{eff}}(W,q,\theta,h)\right] = O_p(n_r^{-1/2} \|\psi_{\theta_0}^{\text{eff}}(W,q,\theta,\widehat h^r) - \psi_{\theta_0}^{\text{eff}}(W,q,\theta,h)\|) = o_p(n_r^{-1/2})$. Because we split the data as evenly as possible, $n_r/n = O_p(1)$ and therefore $\text{Res}_{1,r} = o_p(n^{-1/2})$. In addition, based on the mixed bias property, we obtain that 
\begin{align*}
|\text{Res}_{2,r}| & = |E[\psi_{\theta_0}^{\text{eff}}(W,q,\theta,\widehat h^r) - \psi_{\theta_0}^{\text{eff}}(W,q,\theta,h)]| \\
& = \left|\sum_{k=1}^KE[S_k(\widehat h_{a_k}^{r}-h_{a_k})(\widehat h_{b_k}^{r}-h_{b_k})]\right| \\
& \leq \sum_{k=1}^K\|S_k\| \times \|\widehat h_{a_k}^{r}-h_{a_k}\| \times \|\widehat h_{b_k}^{r}-h_{b_k}\| \\
& = \sum_{k=1}^K O_p(1) \times O_p(\xi_{a_k,n}) \times O_p(\xi_{b_k,n}) \\
& = O_p\left(\sum_{k=1}^K \xi_{a_k,n} \xi_{b_k,n}\right) = o_p(n^{-1/2}),
\end{align*}
uniformly for $\theta\in\Theta$, where the third row follows from the Cauchy-Schwarz inequality,  the fourth row follows from $E[S_k^2]<\infty$, and last equality holds by the rate of convergence condition in Theorem 3. We now have concluded \eqref{eq:thm3_1}. 
\end{proof}

\begin{proof}[Proof of Theorem 3]
We first prove the consistency such that $\widehat \theta^{de}$ converges in probability to $\theta_0$, based on the conditions listed in Theorem 3.  Define $\widehat G_n(\theta) = \Pn[\psi_{\theta_0}^{\text{eff}}(W,q,\theta,\widehat h)]$, $G_n(\theta) = \Pn[\psi_{\theta_0}^{\text{eff}}(W,q,\theta,h)]$, and  $G(\theta) = E[\psi_{\theta_0}^{\text{eff}}(W,q,\theta,h)]$. Our proof follows Corollary 3.2 in \cite{pakes1989simulation}. Specifically, we shall verify the following three conditions in Corollary 3.2 in \cite{pakes1989simulation}: PPC(i) $\|\widehat G_n(\widehat \theta^{de})\|<o_p(1)+\text{inf}_{\theta\in\Theta}\|\widehat G_n(\theta)\|$; PPC(ii) $\text{inf}_{|\theta-\theta_0|>\delta}\|\widehat G_n(\theta)\|>0$ for any $\delta>0$; and PPC(iii) $\text{sup}_{\theta\in\Theta}\frac{\|\widehat G_n(\theta)-G(\theta)\|}{1+\|\widehat G_n(\theta)\|+\|G(\theta)\|}=o_p(1)$. Notice that our condition (iv) implies that $\|\widehat G_n(\widehat \theta^{de})\|=o_p(n^{-1/2})$ and therefore PPC(i) holds. Moreover, condition (iii) in our Theorem 2 suggests that $G(\theta)$ is bounded away from $0$ for a small neighborhood of $\theta$ around $\theta_0$, which implies that PPC(ii) holds. In addition, we have
\begin{align*}
\|\widehat G_n(\theta)-G(\theta)\| & =  \|\widehat G_n(\theta)- G_n(\theta) + G_n(\theta) - G(\theta)\| \\
& \leq \|\widehat G_n(\theta)- G_n(\theta)\|  + \|G_n(\theta) - G(\theta)\| \\
& = o_p(n^{-1/2}) + o_p(1) = o_p(1)
\end{align*}
uniformly for $\theta\in\Theta$, where $\|\widehat G_n(\theta)- G_n(\theta)\| = o_p(n^{-1/2})$ due to Lemma \ref{lemma:s1} and $\|G_n(\theta) - G(\theta)\|=o_p(1)$ due to the P-Donsker in condition (ii) (or condition (ii$'$) in Remark 4). Thus, we have $\|\widehat G_n(\theta)-G(\theta)\|=o_p(1)$ and PPC(iii) holds. Then, $\widehat \theta^{de}$ is consistent.   

Finally, we prove the asymptotic normality such that $\sqrt{n}(\widehat \theta^{de}-\theta_0)$ converges to a zero-mean normal distribution with variance $E[\psi_{\theta_0}^{\text{eff}}(W,q,\theta_0,h)^2]<\infty$. We shall prove this by verifying the five regularity conditions in Theorem 3.3 of \cite{pakes1989simulation}, including PPT(i) $\widehat G_n(\widehat \theta^{de}) \leq o_p(n^{-1/2}) + \text{inf}_{\theta\in\Theta}\|\widehat G_n(\theta)\|$; PPT(ii) the derivative of $G(\theta)$ in terms of $\theta$ at $\theta_0$, denoted by $\Gamma$, is non-singular; PPT(iii) for all positive numbers $\delta_n=o_p(1)$, $\sup_{|\theta-\theta_0|<\delta_n} \frac{\|\widehat G_n(\theta)-G(\theta)+\widehat G_n(\theta_0)\|}{n^{-1/2}+\|\widehat G_n(\theta)\|+\|G(\theta)\|}=o_p(1)$; PPT(iv) $\sqrt{n} \widehat G_n(\theta_0)$ converges to a zero-mean normal distribution with finite variance; PPT(v) $\theta_0$ is an interior point of  $\Theta$. Our condition (iv) concludes PPT(i). Because $E[\phi(W,q,\theta,h)]\equiv 0$ for any $\theta$, we have
\begin{align}
\frac{d}{d \theta} E[\psi_{\theta_0}^{\text{eff}}(W,q,\theta,h)]\Big|_{\theta=\theta_0} & =  -\frac{d}{d \theta} E[B^{-1}g(W,q,\theta,\gamma)]\Big|_{\theta=\theta_0}-\frac{d}{d \theta} E[B^{-1}\phi(W,q,\theta,h)]\Big|_{\theta=\theta_0} \nonumber\\
& = -B^{-1} B + 0 = -1, \label{eq:thm3_3}
\end{align}
thus PPT(ii) holds with $\Gamma = -1$. In addition, we can show
\begin{align*}
& \sup_{|\theta-\theta_0|<\delta_n}\|\widehat G_n(\theta)-G(\theta)+\widehat G_n(\theta_0)\| \\
= &   \sup_{|\theta-\theta_0|<\delta_n}\|\widehat G_n(\theta)- G_n(\theta) + G_n(\theta) - G(\theta)+ G_n(\theta_0) - G_n(\theta_0) + \widehat G_n(\theta_0)\| +  \\
\leq &  \sup_{|\theta-\theta_0|<\delta_n}\|\widehat G_n(\theta)- G_n(\theta)\|  + \sup_{|\theta-\theta_0|<\delta_n}\|G_n(\theta) - G(\theta)+G_n(\theta_0)\| + \|\widehat G_n(\theta_0)- G_n(\theta_0)\| \\
= &  o_p(n^{-1/2}),
\end{align*}
where $\sup_{|\theta-\theta_0|<\delta_n}\|\widehat G_n(\theta)- G_n(\theta)\| = o_p(n^{-1/2})$ and $\|\widehat G_n(\theta_0)- G_n(\theta_0)\| = o_p(n^{-1/2})$ because of Lemma \ref{lemma:s1} and $\sup_{|\theta-\theta_0|<\delta_n}\|G_n(\theta) - G(\theta)+G_n(\theta_0)\|=o_p(n^{-1/2})$ because of the P-Donsker assumption in condition (ii) (or condition (ii$''$) in Remark 4). Therefore, $\sup_{|\theta-\theta_0|<\delta_n}\|\widehat G_n(\theta)-G(\theta)+\widehat G_n(\theta_0)\| = o_p(n^{-1/2})$, which suggests PPT(iii) holds. Next, PPT(iv) holds by Lemma \ref{lemma:s1} with the Central Limit Theorem that $\sqrt{n} \widehat G_n(\theta_0)$ converges to a normal distribution with mean $G(\theta_0)=E[\psi_{\theta_0}^{\text{eff}}(W,q,\theta_0,h)]=0$ and variance $E[\psi_{\theta_0}^{\text{eff}}(W,q,\theta_0,h)^2]<\infty$. Finally PPT(v) holds by our construction. Therefore, $\sqrt{n}(\widehat\theta^{de}-\theta_0)$ converges to a zero-mean normal distribution with variance $\Gamma^{-2}E[\psi_{\theta_0}^{\text{eff}}(W,q,\theta_0,h)^2]=E[\psi_{\theta_0}^{\text{eff}}(W,q,\theta_0,h)^2]$. This completes our proof. 
\end{proof}

\begin{proof}[Proof of Proposition \ref{thm:mr_machine_learning}]
Resume the notations in the proof of Theorem 3. We shall prove the consistency of $\widehat \theta^{de}$ by verifying conditions PPC(i)--PPC(iii). Proof of PPC(i) and PPC(ii) follows the same argument in the proof of Theorem 3. To verify PPC(iii), we first observe
\begin{align*}
\widehat G_n(\theta) - G_n(\theta) & = E[\psi_{\theta_0}^{\text{eff}}(W,q,\theta,\widehat h) - \psi_{\theta_0}^{\text{eff}}(W,q,\theta,h)] + o_p(1) \\
& = \sum_{k=1 }^K E\left[ S_{k}\{\widehat h_{a_k}-h_{a_k}\}\{\widehat h_{b_k}-h_{b_k}\}\right] + o_p(1) \\
& = o_p(1),
\end{align*}
uniformly for $\theta\in\Theta$, where the first equality follows from the Uniform Law of Large Numbers, the second equality follows from the mixed bias property (condition (iii)), the third equality follows from the assumption that the union model $\mathcal M_{\text{union}}=\cap_{k=1}^K\{\mathcal M_{a_k}\cup \mathcal M_{b_k}\}$ is correctly specified. Then, it follows that 
\begin{align*}
\sup_{\theta\in\Theta} \|\widehat G_n(\theta)-G(\theta)\| 
& \leq \sup_{\theta\in\Theta}\|\widehat G_n(\theta)- G_n(\theta)\|  + \sup_{\theta\in\Theta}\|G_n(\theta) - G(\theta)\| \\
& = o_p(1),
\end{align*}
where $\sup_{\theta\in\Theta}\|\widehat G_n(\theta)- G_n(\theta)\| = o_p(1)$ follows from the above discussion and $\sup_{\theta\in\Theta}\|G_n(\theta) - G(\theta)\|=o_p(1)$ follows from condition (ii). This further suggests that PPC(iii) also holds. We now complete the proof.
\end{proof}

\textit{Regularity conditions for Theorem 4.} Consider the scenario of using parametric working models for the nuisance functions, where we specify parametric models $h^{\text{par}}(w,\theta,\beta)$ for  $h \equiv h(w,F,\theta)$ based on
 finite-dimensional parameters $\beta$. 
 We assume that $\widehat \beta$ can be obtained based on the maximum likelihood estimation or the generalized estimating equations, and converges to $\beta^*$ with a parametric rate (i.e., $\widehat \beta = \beta^* + O_p(n^{-1/2})$).  Therefore, the final estimator $\widehat \theta^{de}$ is obtained by solving $\Pn[\psi_{\theta_0}^{\text{eff}}(W,q,\theta,h^{\text{par}}(W,\theta,\widehat \beta))]=0$.  Specific regularity conditions for Theorem 4 are presented below: (i) conditions in Theorem 2 hold with $B\neq 0$; (ii) the EIF of $\theta_0$ satisfies the mixed bias property with all $E[S_k^2]<\infty$; (iii) the class of functions $(\beta,\theta) \to \psi_{\theta_0}^{\text{eff}}(W,q,\theta,h^{\text{par}}(W,\theta,\beta))$ is P-Donsker  with $E[\psi_{\theta_0}^{\text{eff}}(W,q,\theta,h^{\text{par}}(W,\theta,\beta))^2]<\infty$ for $(\beta,\theta) \in \mathcal B \times \Theta$, where $\mathcal B$ is an open neighborhood of $\beta^*$; (iv) $E[\psi_{\theta_0}^{\text{eff}}(W,q,\theta_0,h^{\text{par}}(W,\theta,\beta))]$ is continuous and differentiable of $\theta$ and $\beta$ and the derivative $D_{\theta_0}:=\frac{d}{d\theta}E[\psi_{\theta_0}^{\text{eff}}(W,q,\theta_0,h^{\text{par}}(W,\theta,\beta^*))]|_{\theta=\theta_0}$ is non-singular; (v) $\widehat \beta$ is $\sqrt{n}$-consistent and asymptotically linear with the influence function $\psi_{\beta}(W)$ such that $\sqrt{n}(\widehat \beta - \beta^*) = \sqrt{n}\Pn[\psi_{\beta}(W)] + o_p(1)$ with $E[\psi_{\beta}(W)^2]<\infty$; (vi) $h_j^{\text{par}}(W,\theta,\beta^*) = h_j(W,F, \theta)$ if $\mathcal M_j$ is correctly specified; (vii) $\widehat\theta^{de}$ is an approximate solution of the debiased IEE such that  $\Pn[\psi_{\theta_0}^{\text{eff}}(W,q,\widehat\theta^{de},h^{\text{par}}(W,\widehat\theta^{de},\widehat \beta))]=o_p(n^{-1/2})$.

\begin{proof}[Proof of Theorem 4]
We first prove $\widehat \theta^{de}$ is $\sqrt{n}$-consistent if $\mathcal M_{\text{union}}$ is correctly specified. The mixed bias property suggests that, for all $\theta\in\Theta$,
\begin{align}
& E[\psi_{\theta_0}^{\text{eff}}(W,q,\theta,h^{\text{par}}(W,\theta,\beta^*))-\psi_{\theta_0}^{\text{eff}}(W,q,\theta,h)] \nonumber \\
= & \sum_{k=1}^KE[S_k\{h_{a_k}^{\text{par}}(W,\theta,\beta^*)-h_{a_k}(W,F,\theta)\}\{h_{b_k}^{\text{par}}(W,\theta,\beta^*)-h_{b_k}(W,F,\theta)\}] \nonumber \\
= & 0 \label{eq:thm4_1}
\end{align}
where last equality holds because of regularity condition (vi) and our assumption that the union model $\mathcal M_{\text{union}}:=\cap_{k=1}^K \{\mathcal M_{a_k}\cup \mathcal M_{b_k}\}$ is correctly specified. In particular, we have
\begin{equation}\label{eq:thm4_2}
E[\psi_{\theta_0}^{\text{eff}}(W,q,\theta_0,h^{\text{par}}(W,\theta_0,\beta^*))] = 0
\end{equation}
because $E[\psi_{\theta_0}^{\text{eff}}(W,q,\theta_0,h)]=0$. In addition, by the Uniform Law of Large Numbers and along with the P-Donsker assumption in condition (iii) and the $\sqrt{n}$-consistency of $\widehat\beta$ in condition (iv), we have that
\begin{align}
\Pn[\psi_{\theta_0}^{\text{eff}}(W,q,\theta,h^{\text{par}}(W,\theta,\widehat \beta))] & = E[\psi_{\theta_0}^{\text{eff}}(W,q,\theta,h^{\text{par}}(W,\theta,\beta^*))] + O_p(n^{-1/2}) \label{eq:thm4_3}
\end{align}
uniformly for $\theta \in \Theta$. Notice that $\frac{d}{d\theta}E[\psi_{\theta_0}^{\text{eff}}(W,q,\theta,h^{\text{par}}(W,\theta,\beta^*))]|_{\theta=\theta_0}$ is non-singular (condition (iv)), which implies that there exists a positive number $c$ such that
\begin{align*}
|\theta - \theta_0| & \leq c \times | E[\psi_{\theta_0}^{\text{eff}}(W,q,\theta,h^{\text{par}}(W,\theta,\beta^*)) - \psi_{\theta_0}^{\text{eff}}(W,q,\theta_0,h^{\text{par}}(W,\theta_0,\beta^*))]  | \\
& =  c \times | E[\psi_{\theta_0}^{\text{eff}}(W,q,\theta,h^{\text{par}}(W,\theta,\beta^*))]| \\
& = c \times | \Pn[\psi_{\theta_0}^{\text{eff}}(W,q,\theta,h^{\text{par}}(W,\theta,\widehat \beta))]| + O_p(n^{-1/2})
\end{align*}
in a small neighborhood of $\theta$ around $\theta_0$, where the second equality follows \eqref{eq:thm4_2} and the third equality follows \eqref{eq:thm4_3}. We then replace $\theta$ with $\widehat\theta^{de}$ in the previous result to obtain that
\begin{align*}
|\widehat\theta^{de} - \theta_0| \leq c \times | \Pn[\psi_{\theta_0}^{\text{eff}}(W,q,\widehat\theta^{de},h^{\text{par}}(W,\widehat\theta^{de},\widehat \beta))]| + O_p(n^{-1/2}) = O_p(n^{-1/2}),
\end{align*}
where the last equality holds because of condition (vii). This concludes that $\widehat\theta^{de}$ is $\sqrt{n}$-consistent.  

Next we prove the asymptotic normality of $\widehat \theta^{de}$. For brevity, we redefine $\psi_{\theta_0}^{\text{eff}}(W,q,\theta,h^{\text{par}}(W,\theta,\beta))$ as $G(W,\theta,\beta)$, and decompose
\begin{align*}
\Pn[G(W,\widehat\theta^{de},\widehat \beta)] = &  \Pn[G(W,\theta_0,\beta^*)] + (\Pn-E)[G(W,\widehat\theta^{de},\widehat \beta) - G(W,\theta_0,\beta^*)] \\
& +  E[G(W,\widehat\theta^{de},\widehat \beta) - G(W,\theta_0,\beta^*)],
\end{align*}
where the second term on the right-hand side is $o_p(n^{-1/2})$ due to the P-Donsker condition (iii) and the $\sqrt{n}$-consistency of $\widehat \beta$ and $\widehat \theta^{de}$. By Taylor expansion and noting  $E[G(W,\theta_0,\beta^*)]=0$, we can further decompose the third term on the right-hand side into
\begin{align*}
& E[G(W,\widehat\theta^{de},\widehat \beta) - G(W,\theta_0,\beta^*)] \\
= & \frac{d}{d\theta} E[G(W,\theta,\beta^*)]\Big|_{\theta=\theta_0} (\widehat\theta^{de}-\theta_0) + \frac{d}{d\beta} E[G(W,\theta_0,\beta)]\Big|_{\beta=\beta^*} (\widehat\beta-\beta^*) + o_p(n^{-1/2}) \\
= & D_{\theta_0} (\widehat\theta^{de}-\theta_0) + D_{\beta^*} (\widehat\beta-\beta^*) + o_p(n^{-1/2})
\end{align*}
where $D_{\theta_0}:=\frac{d}{d\theta} E[G(W,\theta,\beta^*)]|_{\theta=\theta_0}$ and $D_{\beta^*}:=\frac{d}{d\beta} E[G(W,\theta_0,\beta)]|_{\beta=\beta^*}$. Combining the above, we obtain
\begin{align*}
\Pn[G(W,\widehat\theta^{de},\widehat \beta)] &= \Pn[G(W,\theta_0,\beta^*)] + D_{\theta_0} (\widehat\theta^{de}-\theta_0) + D_{\beta^*} (\widehat\beta-\beta^*) + o_p(n^{-1/2})\\
& = \Pn[G(W,\theta_0,\beta^*)] + D_{\theta_0} (\widehat\theta^{de}-\theta_0) + \Pn[D_{\beta^*}\psi_{\beta}(W)] + o_p(n^{-1/2}),
\end{align*}
where the second row holds due to condition (v). Furthermore, because $\Pn[G(W,\widehat\theta^{de},\widehat \beta)]=o_p(n^{-1/2})$ (condition (vii)), we have that
$$
\sqrt{n}(\widehat\theta^{de}-\theta_0) = \sqrt{n}D_{\theta_0}^{-1}\Pn[G(W,\theta_0,\beta^*)-D_{\beta^*}\psi_{\beta}(W)] + o_p(1).
$$
This suggests that $\sqrt{n}(\widehat\theta^{de}-\theta_0)$ converges to a normal distribution with mean 0 and variance 
$$
V_{\theta_0} = D_{\theta_0}^{-2} E\left[\{G(W,\theta_0,\beta^*)-D_{\beta^*}\psi_{\beta}(W)\}^2\right].
$$
For completeness, we demonstrate that $V_{\theta_0}$ achieves the semiparametric efficiency lower bound $E[\psi_{\theta_0}^{\text{eff}}(W,q,\theta_0,h)^2]$ in the scenario that the parametric working models for all nuisance functions are correctly specified. In this scenario, $G(W,\theta,\beta^*) \equiv \psi_{\theta_0}^{\text{eff}}(W,q,\theta,h)$ and therefore $D_{\theta_0}:=\frac{d}{d\theta} E[G(W,\theta,\beta^*)]|_{\theta=\theta_0} = \frac{d}{d\theta} E[\psi_{\theta_0}^{\text{eff}}(W,q,\theta,h)]|_{\theta=\theta_0}=-1$ as shown in \eqref{eq:thm3_3}. Moreover, because $G(W,\theta,\beta^*)$ is the EIF, it is orthogonal to the influence function of the nuisance functions, which implies $D_{\beta^*}=0$. This concludes $V_{\theta_0}=E[\psi_{\theta_0}^{\text{eff}}(W,q,\theta,h)^2]$. 
\end{proof}

\section{Supporting information for Example 1}

\subsection{Derivation of the EIF based on Strategy II}


We calculate the EIF of $\theta_0$ based on Strategy II. To avoid the confusion between `$a$' in the threshold-transformed potential outcome $\bbI(Y_{a}\leq \theta)$ 
 and `$a$' in the data observation $w$ used for the identifying moment function $g(w,\tau,\theta,h)$ (the former is a fixed point and the latter is an observation of $W$), we shall rewrite $a$ in the threshold-transformed potential outcome as $a'$ such that we are interested in $\theta_0:=E[\bbI(Y_{a'}\leq \theta)]$.  We first derive the the adjustment term $\phi(w,\tau,\theta,h)$  following the point mass contamination approach \cite{hines2022demystifying}; in this approach, we permute the true observed data CDF $F$ toward the direction $\widetilde F$ of a point mass at single observation $\widetilde w$. Specifically, let $\bm{1}_{\widetilde w}(w)$ denote the Dirac measure at $\widetilde w$. Therefore, the joint density function of the observed data $W=\{Y,A,L\}$ under the parametric submodel $F_t$ is
$$
f_{t}(y,a,l) = t \bm{1}_{\widetilde y, \widetilde a, \widetilde l}(y,a,l) + (1-t) f(y,a,l),
$$
where $f(y,a,l)$ is the joint density function of $Y,A,L$. Based on the point mass contamination approach, the EIF of  $\bar g(F;\tau,\theta)$ is directly as
$$
\phi(\widetilde w,\tau,\theta,h) = \frac{d}{dt} \bar g(F_t;\tau,\theta)\Big|_{t=0},
$$
where 
\begin{align*}
\bar g(F_t;\tau,\theta) & = \int g(w,\tau,\theta,\gamma_t) dF(w) = \int \left\{\frac{\bbI(a=a') \bbI(y\leq\theta)}{f_t(a'|l)} -\tau\right\}f(y,a,l) dydadl \\
& = \int \left\{\frac{f(a'|l) F(\theta|a',l)}{f_t(a'|l)} \right\} f(l) dl -\tau = \int \left\{\frac{ F(\theta|a',l)f_t(l)}{f_t(a',l)} \right\} f(a',l) dl -\tau,
\end{align*}  
where $F(\theta|a,l)$ if the true probability of $Y\leq\theta$ conditional on $A=a$ and $L=l$, $f(a,l)$ and $f(a|l)$ are the true density functions of $\{A,L\}$ and $A|L$, and $f_t(l)$, $f_t(a|l)$, and $f_t(a,l)$ are the density functions of $L$, $A|L$, and $\{A,L\}$ under the parametric submodel. Then, by the chain rule, one can deduce that
\begin{align*}
\frac{d}{dt} \bar g(F_t;\tau,\theta)\Big|_{t=0} & = \int \frac{ F(\theta|a',l)f(a',l)}{f(a',l)}   \frac{d}{dt} f_t(l)\Big|_{t=0} dl - \int \frac{ F(\theta|a',l)f(l)}{f(a',l)} \frac{d}{dt} f_t(a',l)\Big|_{t=0}  dl \\
& = \int F(\theta|a',l) \{\bm{1}_{\widetilde l}(l)-f(l)\} dl - \int \frac{ F(\theta|a',l)}{f(a'|l)} \{\bm{1}_{\widetilde a, \widetilde l}(a',l)-f(a',l)\}  dl\\
& = F(\theta|a',\widetilde l) - E[F(\theta|a',L)] - \frac{\bbI(\widetilde a=a')F(\theta|a',\widetilde l)}{f(a'|\widetilde l)} + E[F(\theta|a',L)]\\
& = F(\theta|a',\widetilde l) - \frac{\bbI(\widetilde a=a')F(\theta|a',\widetilde l)}{f(a'|\widetilde l)}.
\end{align*}
This suggests that the adjustment term is
$$
\phi(W,\tau,\theta,h) = F_{Y|A,L}(\theta|a',L) - \frac{\bbI(A=a')F_{Y|A,L}(\theta|a',L)}{f_{A|L}(a'|L)},
$$
where the variance of $\phi(W,\tau,\theta,h)$ is bounded because of the positivity assumption. This further implies that the EIF of $\theta_0$ is
\begin{align*}
  & \psi_{\theta_0}^{\text{eff}}(W,q,\theta_0,h) \\
= &  - B^{-1} \Big\{g(W,q,\theta_0,\gamma) + \phi(W,q,\theta_0,h)\Big\} \\
= &  -\frac{1}{E[f_{Y|A,L}(\theta_0|a',L)]}\times\left[\frac{\bbI(A=a')}{f_{A|L}(a'|L)}\left\{\bbI(Y\leq\theta_0)-F_{Y|A,L}(\theta_0|a',L)\right\} + F_{Y|A,L}(\theta_0|a',L) - q\right],
\end{align*}
with $h=\{f_{A|L}(a'|L),F_{Y|A,L}(\theta_0|a',L)\}$.

\subsection{Derivation of the mixed bias property}

To derive the mixed bias property of $\psi_{\theta_0}^{\text{eff}}(W,q,\theta_0,h)$, we can show
\begin{align*}
&  E[-B\times \{\psi_{\theta_0}^{\text{eff}}(W,q,\theta,\widehat h)-\psi_{\theta_0}^{\text{eff}}(W,q,\theta,h)\}] \\
= & E\left[\frac{\bbI(A=a')}{\widehat f_{A|L}(a'|L)}\left\{\bbI(Y\leq\theta)-\widehat F_{Y|A,L}(\theta|a',L)\right\} + \widehat F_{Y|A,L}(\theta|a',L)\right]  \\
 & - E\left[\frac{\bbI(A=a')}{ f_{A|L}(a'|L)}\left\{\bbI(Y\leq\theta)-F_{Y|A,L}(\theta|a',L)\right\} +  F_{Y|A,L}(\theta|a',L)\right] \\
= & E\left[\frac{\bbI(A=a')}{\widehat f_{A|L}(a'|L)}\left\{\bbI(Y\leq\theta)-\widehat F_{Y|A,L}(\theta|a',L)\right\} + \widehat F_{Y|A,L}(\theta|a',L) - F_{Y|A,L}(\theta|a',L)\right] \\
= & E\left[\frac{f_{A|L}(a'|L)}{\widehat f_{A|L}(a'|L)}\left\{F_{Y|A,L}(\theta|a',L)-\widehat F_{Y|A,L}(\theta|a',L)\right\} + \widehat F_{Y|A,L}(\theta|a',L) - F_{Y|A,L}(\theta|a',L)\right] \\
= & E\left[ \frac{1}{\widehat f_{A|L}(a'|L)}\{\widehat f_{A|L}(a'|L)-f_{A|L}(a'|L)\}\{\widehat F_{Y|A,L}(\theta|a,L)-F_{Y|A,L}(\theta|a,L)\}\right] 
\end{align*}
Therefore, $E[\psi_{\theta_0}^{\text{eff}}(W,q,\theta, \widehat h)-\psi_{\theta_0}^{\text{eff}}(W,q,\theta, h)]$ equals to
$$
E\left[ S_{1}\{\widehat f_{A|L}(a'|L)-f_{A|L}(a'|L)\}\{\widehat F_{Y|A,L}(\theta|a',L)-F_{Y|A,L}(\theta|a',L)\}\right],
$$
with $S_1 = - \frac{1}{B\times \widehat f_{A|L}(a'|L)}$.

{

\subsection{Comparison to the estimator in \cite{firpo2007efficient}}\label{sec:sim_qte}

We compare the performance between the proposed method and the nonparametric efficient estimator by \cite{firpo2007efficient} for estimation of and inference about the QTE. Specifically, the estimating equation of $\theta_0:=Q_{Y_a}(q)$ in \cite{firpo2007efficient} coincides with the IEE in present work but they consider using a series logistic regression (SLR) to fit the propensity score $f_{A|L}(a|L)$. The SLR is a nonparametric approach for robust estimation of propensity scores \citep{hirano2003efficient}, and is built on a logistic power series of the baseline covariates $L$ to approximate the true propensity scores. Without the need to specify an outcome model, their estimator has been shown to be consistent, asymptotically normal, and semiparametrically efficient as long as the SLR can sufficiently capture the complexity of the true propensity score. In other words, the behavior of \cite{firpo2007efficient}'s approach critically depends on the performance of the SLR, and may be biased if the design matrix in the SLR is insufficient to approximate the true propensity score or can be unstable if the resulting propensity score estimate (i.e., $\widehat f_{A|L}(a|L)$) is near 0 for even a few data observations. In contrast, the proposed debiased IEE approach provides double protection against model misspecification, and therefore holds the promise to improve the SLR implementation proposed in \citet{firpo2007efficient}. 




\begin{table}[ht]
\centering
\caption{Numerical comparison between the nonparametric efficient estimator by \cite{firpo2007efficient} (denoted by `\cite{firpo2007efficient}') and the proposed machine learning-based debiased IEE estimator (denoted by `de-ml' ) for assessing $Q_{Y_0}(q)$ with $q\in\{0.25,0.5,0.75\}$ based on Example 1. We considered both correct (denoted by `T') or incorrect (denoted by `F') specifications of the propensity score models ($h_1$) and outcome model ($h_2$).\label{tab:sim_example1}}
\begin{tabular}{ccc rrrc}
  \hline
  \multicolumn{7}{c}{Panel A: $q=0.25$} \\
  \hline
    Method & $h_1$ & $h_2$ & Bias & RMSE & MAE & Coverage \\ 
  \hline
 Firpo (2007) & T &  & $-$0.011 & 1.043 & 0.814 & 0.951 \\ 
  & F & &  1.177 & 1.476 & 1.259 & 0.706 \\
 de-ml & T & T & 0.016 & 0.713 & 0.568 & 0.950 \\  
   & F & T &  $-$0.036 & 0.823 & 0.548 & 0.958 \\
   & T & F &  $-$0.042 & 0.992 & 0.781 & 0.936 \\ 
  & F & F &  0.633 & 1.055 & 0.849 & 0.820 \\ 
  \hline
  \hline
  \multicolumn{7}{c}{Panel B: $q=0.5$} \\
  \hline
    Method & $h_1$ & $h_2$ & Bias & RMSE & MAE & Coverage \\ 
  \hline
 Firpo (2007) & T &  & $-$0.056 & 0.758 & 0.607 & 0.961 \\ 
  & F & &  0.905 & 1.164 & 0.982 & 0.759 \\ 
 de-ml & T & T &  $-$0.020 & 0.625 & 0.506 & 0.947 \\ 
   & F & T &   0.025 & 0.624 & 0.490 & 0.960 \\ 
   & T & F &  $-$0.051 & 0.724 & 0.574 & 0.921 \\ 
  & F & F & 0.380 & 0.822 & 0.648 & 0.852 \\ 
  \hline
  \hline
  \multicolumn{7}{c}{Panel C: $q=0.75$} \\
  \hline
    Method & $h_1$ & $h_2$ & Bias & RMSE & MAE & Coverage \\ 
  \hline
 Firpo (2007) & T &  & 0.004 & 0.707 & 0.556 & 0.957 \\
  & F & & 0.558 & 0.918 & 0.743 & 0.875 \\ 
 de-ml & T & T &  0.026 & 0.623 & 0.496 & 0.954 \\ 
   & F & T &  0.013 & 0.661 & 0.528 & 0.943 \\ 
   & T & F &  $-$0.012 & 0.673 & 0.537 & 0.927 \\ 
  & F & F &  0.342 & 0.807 & 0.647 & 0.916 \\ 
  \hline
\end{tabular}
\end{table}

Below, we conduct a simulation study to evaluate the empirical performance of the debiased IEE estimator and the  estimator given by \cite{firpo2007efficient}. In this numerical study, we generate $n=1,000$ independent copies of the  observed data $W=\{L=\{L_1,L_2,L_3,L_4\},A,Y\}$ based on the following process. First, $L=\{L_1,L_2,L_3,L_4\}$ are generated based on four mutually independent standard normal distributions. Then, we simulate $A$ and $Y$ based on $A|L\sim \text{Bernoulli}(\text{expit}(-L_1+0.5L_2-0.25L_3-0.1L_4))$ and $Y|\{A,L\} \sim N(1.5A+10L_1+5L_2+5L_3+5L_4,\exp(2+A))$. We compare the bias, root mean square error (RMSE), mean absolute error (MAE), and the 95\% confidence interval coverage rate for estimating $Q_{Y_0}(q)$ with $q\in\{0.25, 0.5, 0.75\}$ between \cite{firpo2007efficient}'s estimator and the proposed debiased IEE with nuisance functions estimated through machine learners. For \cite{firpo2007efficient}'s approach, we fit the SLR with the order of the power series chosen based on a 10-fold cross-validation. For the proposed debiased IEE estimator, the propensity score $f_{A|L}(a|L)$ is estimated through Super Learner with generalized linear model, random forest, and extreme gradient boosting libraries. To estimate $F_{Y|A,L}(\theta|a,L)$, we consider an additive model 
$$Y=m_a(L)+v_a(L)\epsilon_a$$ 
within the strata of $A=a$, where $m_a(L)$ and $v_a(L)$ are the mean and variance structure and $\epsilon_a$ is a zero-mean error term. Then, $m_a(L)$ is estimated by regressing of $Y$ on $L$ among units with $A=a$ based on Super Learner with generalized linear model, random forest, and extreme gradient boosting libraries, $v_a(L)$ is similarly estimated by regressing of $(Y-\widehat m_a(L))^2$ on $L$ within units with $A=a$, and the CDF of $\epsilon_a$ (denoted by $F_{\epsilon_a}$) is estimated by kernel smoothing based on $(Y-\widehat m_a(L))/\sqrt{\widehat v_a(L)}$ among units within strata $A=a$. Given this additive model, an estimator of $F_{Y|A,L}(\theta|a,L)$ is given by $\widehat F_{\epsilon_a}((Y-\widehat m_a(L))/\sqrt{\widehat v_a(L)})$. We consider Wald-type confidence intervals. For \cite{firpo2007efficient}, we calculate the asymptotic variance based on their proposed sandwich variance estimator; for the proposed debiased IEE estimator, the asymptotic variance is calculated based on the  empirical variance of the EIF. 
To evaluate robustness of these two estimators, we consider both correct and incorrect specification of the nuisance functions. For correctly specified propensity score model, we input the true baseline covariates $L$ into the SLR or the Super Learner.  Otherwise, we input a set of transformed covariates $\widetilde L=[\widetilde L_1,\widetilde L_2,\widetilde L_3,\widetilde L_4]^T$ instead, where $\widetilde L_1 = \exp(0.5L_1)$, $\widetilde L_2 = {L_2}/{(1+L_1)}$, $\widetilde L_3 = (L_2L_3/25+0.6)^3$, and $\widetilde L_4=(L_2+L_4+20)^2$. Analogously, we input the true (or transformed) covariates into the Super Learner for $F_{Y|A,L}(\theta|a,L)$ when the outcome model is correctly (or incorrectly) specified.

Table \ref{tab:sim_example1} presents the performance of both estimators under correct and incorrect specification of the nuisance functions among 1,000 Monte Carlo experiments. As expected, \cite{firpo2007efficient} displays minimal bias with close-to-nominal confidence interval coverage when the SLR for the propensity score is correctly specified, but is biased with attenuated coverage otherwise. The debiased IEE approach provides minimal bias with nominal coverage when either the propensity score or the outcome model is correct, which empirically verifies the doubly robust property. Meanwhile, we found that the debiased IEE approach tends to be more efficient in finite samples, as its RMSE and MAE are consistently smaller than that of \cite{firpo2007efficient} even under misspecification. Notably, even if both models are misspecified, the debiased IEE approach still provides smaller RMSE and MAE with higher confidence interval coverage as compared to \cite{firpo2007efficient} with a misspecified SLR nuisance model.

}

\section{Supporting information for Example 2}

\subsection{Specification of the nuisance models and proof of the mixed bias property}

{In Example 2, Specification of the working models on the nuisance functions $h=\{\mu(L,\theta),\\ f_{A|L}(1|L),f_{A|M,L}(1|M,L),F_{Y|A,M,L}(\theta|1,M,L)\}$ can be flexible. Specifically, $f_{A|L}(1|L)$ and $f_{A|M,L}(1|M,L)$ can be obtained by regressions for binary outcomes (for example, logistic model or machine learners designed for a binary outcome). Similarly, $F_{Y|A,M,L}(\theta|1,M,L)$ can be estimated based on machine learning/parametric models designed for conditional CDFs. Estimation of $\mu(L,\theta)=E\left[F_{Y|A,M,L}(\theta|1,M,L)|A=0,L\right]$ is more involved as it includes a nested conditional expectation. If $M$ is low-dimensional, then 
$$
\widehat \mu (L,\theta) = \int_m\widehat F_{Y|A,M,L}(\theta|1,m,L) \widehat f_{M|A,L}(m|0,L) d m.
$$ An alternative strategy is by regression imputation \citep{zhou2022semiparametric,hsu2023doubly}, which is suitable when $M$ is multi-dimensional with many continuous components. We first calculate $\widehat \mu(\underline\theta_r,L)$ over a grid of $R$ values $\{\underline\theta_r\}_{r=1}^R$ in the outcome support $\mathcal Y$, where each $\widehat \mu(\underline\theta_r,L)$ is obtained by regressing $\widehat F_{Y|A,M,L}(\underline\theta_r|1,M,L)$ on $L$ among the units with $A=0$. Then $\mu(\theta,L)$ can be approximated by linearly interpolating the grid points $\{\underline\theta_r,\widehat \mu(\underline\theta_r,L)\}_{r=1}^R$ so that $$\widehat \mu(\theta,L) = \widehat \mu(\underline\theta_r,L) + \frac{\theta-\underline\theta_r}{\underline\theta_{r+1}-\underline\theta_r}\{\widehat \mu(\underline\theta_{r+1},L)-\widehat \mu(\underline\theta_{r},L)\},$$ 
for $\underline\theta_r<\theta<\underline\theta_{r+1}$.} 


We next show the mixed bias property of $\psi_{\theta_0}^{\text{eff}}(W,q,\theta_0,h)$. We first prove the following equality \begingroup\makeatletter\def\f@size{9}\check@mathfonts
\begin{equation}\label{eq:example3_1}
E\left[\frac{1-f_{A|M,L}(1|M,L)}{1-\widehat f_{A|L}(1|L)}\{F_{Y|A,M,L}(\theta|1,M,L)-\widehat \mu(L,\theta)\}\right] = E\left[\frac{1-f_{A|L}(1|L)}{1-\widehat f_{A|L}(1|L)}\{\mu(\theta,L)-\widehat\mu(\theta,L)\}\right].
\end{equation} \endgroup
Specifically, the left-hand side of \eqref{eq:example3_1} equals to \begingroup\makeatletter\def\f@size{9}\check@mathfonts
\begin{align*}
& E\left[E\left\{\frac{1-f_{A|M,L}(1|M,L)}{1-\widehat f_{A|L}(1|L)}\{F_{Y|A,M,L}(\theta|1,M,L)-\widehat \mu(L,\theta)\}\Big|L\right\}\right] \\
= & E\left[\frac{1}{1-\widehat f_{A|L}(1|L)}\int_mf_{A|M,L}(0|M,L)\{F_{Y|A,M,L}(\theta|1,M,L)-\widehat \mu(L,\theta)\}f_{M|L}(m|L)dm\right] \\
= & E\left[\frac{1}{1-\widehat f_{A|L}(1|L)}\int_m\frac{f_{M|A,L}(m|0,L)f_{A|L}(0|L)}{f_{M|L}(m|L)}\{F_{Y|A,M,L}(\theta|1,M,L)-\widehat \mu(L,\theta)\}f_{M|L}(m|L)dm\right] \\
= & E\left[\frac{f_{A|L}(0|L)}{1-\widehat f_{A|L}(1|L)}\int_m\{F_{Y|A,M,L}(\theta|1,M,L)-\widehat \mu(L,\theta)\}f_{M|A,L}(m|0,L)dm\right] \\
= & \text{ the right-hand side of \eqref{eq:example3_1}}.
\end{align*} \endgroup
Now, we can show that \begingroup\makeatletter\def\f@size{8.5}\check@mathfonts
\begin{align*}
& -B \times E[\psi_{\theta_0}^{\text{eff}}(W,q,\theta,\widehat h)-\psi_{\theta_0}^{\text{eff}}(W,q,\theta,h)] \\
= & E[\widehat \mu(L,\theta)-q] + E\left[\frac{A}{1-\widehat f_{A|L}(1|L)}\frac{1-\widehat f_{A|M,L}(1|M,L)}{\widehat f_{A|M,L}(1|M,L)}\left\{\bbI(Y\leq\theta) - \widehat F_{Y|A,M,L}(\theta|1,M,L)\right\}\right] \\
& + E\left[\frac{1-A}{1-\widehat f_{A|L}(1|L)}\left\{\widehat F_{Y|A,M,L}(\theta|1,M,L)-\widehat \mu(L,\theta)\right\}\right] - E[\mu(L,\theta)-q] \\
= &  E\left[\frac{f_{A|M,L}(1|M,L)}{1-\widehat f_{A|L}(1|L)}\frac{1-\widehat f_{A|M,L}(1|M,L)}{\widehat f_{A|M,L}(1|M,L)}\left\{F_{Y|A,M,L}(\theta|1,M,L) - \widehat F_{Y|A,M,L}(\theta|1,M,L)\right\}\right] \\
& + E\left[\frac{1-f_{A|M,L}(1|M,L)}{1-\widehat f_{A|L}(1|L)}\left\{\widehat F_{Y|A,M,L}(\theta|1,M,L)-\widehat \mu(L,\theta)\right\}\right] \\
& +E[\widehat \mu(L,\theta)-\mu(L,\theta)] \\
= & E\left[\frac{1-\widehat f_{A|M,L}(1|M,L)}{\{1-\widehat f_{A|L}(1|L)\}\widehat f_{A|M,L}(1|M,L)}\{ \widehat f_{A|M,L}(1|M,L)- f_{A|M,L}(1|M,L)\}\left\{\widehat F_{Y|A,M,L}(\theta|1,M,L) -  F_{Y|A,M,L}(\theta|1,M,L)\right\}\right] \\
& + E\left[\frac{1}{1-\widehat f_{A|L}(1|L)}\{ \widehat f_{A|M,L}(1|M,L)- f_{A|M,L}(1|M,L)\}\left\{\widehat F_{Y|A,M,L}(\theta|1,M,L) -  F_{Y|A,M,L}(\theta|1,M,L)\right\}\right] \\
& + E\left[\frac{1-f_{A|M,L}(1|M,L)}{1-\widehat f_{A|L}(1|L)}\{F_{Y|A,M,L}(\theta|1,M,L)-\widehat \mu(L,\theta)\}\right] +E[\widehat \mu(L,\theta)-\mu(L,\theta)]  \\
& \quad \text{(by equation \eqref{eq:example3_1}, we have below)} \\
= & E\left[\frac{1-\widehat f_{A|M,L}(1|M,L)}{\{1-\widehat f_{A|L}(1|L)\}\widehat f_{A|M,L}(1|M,L)}\{ \widehat f_{A|M,L}(1|M,L)- f_{A|M,L}(1|M,L)\}\left\{\widehat F_{Y|A,M,L}(\theta|1,M,L) -  F_{Y|A,M,L}(\theta|1,M,L)\right\}\right] \\
& + E\left[\frac{1}{1-\widehat f_{A|L}(1|L)}\{ \widehat f_{A|M,L}(1|M,L)- f_{A|M,L}(1|M,L)\}\left\{\widehat F_{Y|A,M,L}(\theta|1,M,L) -  F_{Y|A,M,L}(\theta|1,M,L)\right\}\right] \\
& + E\left[\frac{1-f_{A|L}(1|L)}{1-\widehat f_{A|L}(1|L)}\{\mu(L,\theta)-\widehat \mu(L,\theta)\}\right] +E[\widehat \mu(L,\theta)-\mu(L,\theta)] \\
= & E\left[-\frac{1}{1-\widehat f_{A|L}(1|L)}\{\widehat f_{A|L}(1|L)-f_{A|L}(1|L)\}\{\widehat \mu(L,\theta)-\mu(L,\theta)\}\right]  \\
& + E\left[\frac{1}{\{1-\widehat f_{A|L}(1|L)\}\widehat f_{A|M,L}(1|M,L)}\{ \widehat f_{A|M,L}(1|M,L)- f_{A|M,L}(1|M,L)\}\left\{\widehat F_{Y|A,M,L}(\theta|1,M,L) -  F_{Y|A,M,L}(\theta|1,M,L)\right\}\right]
\end{align*}\endgroup
This suggests that the EIF of $\theta_0$ satifies the mixed bias property with $K=2$, $[\{a_1,b_1\},\{a_2,b_2\}]=[\{1,2\},\{3,4\}]$, and $S_1 = \frac{1}{B\times\{1-\widehat f_{A|L}(1|L)\}}$ and $S_2 = -\frac{1}{B\times\{1-\widehat f_{A|L}(1|L)\}\widehat f_{A|M,L}(1|M,L)}$. Here, both $E[S_1^2]<\infty$ and $E[S_2^2]<\infty$ if both $\widehat f_{A|M,L}(1|M,L)$ and $\widehat f_{A|L}(1|L)$ are bounded away from 1 and 0, respectively.


{

\subsection{Comparison to the inverse CDF estimator in \cite{hsu2023doubly}}

In the context of quantile mediation analysis, \cite{hsu2023doubly} proposed a double machine learning estimator of $Q_{Y_{1M_0}}(q)$. In their approach, one first estimates the entire CDF $F_{Y_{1M_0}}$ based on a set of discrete grid points and then inverts this grid-based CDF to obtain the desired quantile estimate of $\theta_0:=Q_{Y_{1M_0}}(q)$. That is, they first propose an estimator of $\tau_0(\theta):=F_{Y_{1M_0}}(\theta)$ with a fixed threshold $\theta\in \mathcal Y$ based on its  efficient influence function  (i.e., $\psi_{\tau_0(\theta)}^{\text{eff}}$ in Section 4.1 of the main manuscript), where the nuisance functions are estimated through machine learners (with a primary focus on the Post-Lasso approach; see  \citet{belloni2017program}). Specifically, for a fixed $\theta$ they consider estimating $\tau_0(\theta)$ by solving
\begin{equation}\label{eq:hsu_ee}
\mathbb{P}_n[g(W,\tau,\theta,\widehat \gamma)+\phi(W,\tau,\theta,\widehat h)] = 0
\end{equation}
with respect to $\tau$, where $g(W,\tau,\theta, \gamma)$ and $\phi(W,\tau,\theta,h)$ are defined in Section 4.1 of the main manuscript. To assess the quantile, they then choose $R$ grid points in the support of outcome, say $\underline{\theta}_1<\underline{\theta}_2<\cdots < \underline{\theta}_R$, and obtain $\widehat F_{Y_{1M_0}}(\underline{\theta}_r)$ for each $r$ based on \eqref{eq:hsu_ee}. Next, they use the data pairs, $\{\underline{\theta}_r,\widehat F_{Y_{1M_0}}(\underline{\theta}_r)\}_{r=1}^R$, to construct an approximate estimator for $F_{Y_{1M_0}}$, and then empirically invert this CDF estimate to obtain $\theta_0:=Q_{Y_{1M_0}}(q)$. On the other hand, our debiased IEE approach estimates $\theta_0$ directly by solving 
$$
\mathbb{P}_n[g(W,q,\theta,\widehat \gamma)+\phi(W,q,\theta,\widehat h)] = 0
$$
with respect to $\theta$. In other words, the proposed debiased IEE estimator leverages the inverse estimating equation of \eqref{eq:hsu_ee} to solve for $\theta_0$. \cite{hsu2023doubly} show that their estimator is consistent, asymptotically normal, and semiparametrically efficient under certain regularity conditions for the nuisance function estimators (these conditions are met by the Post-Lasso approach). Their approach is asymptotically equivalent to our debiased IEE approach when nuisance functions are consistently estimated through machine learners.


Nevertheless, there are several differences between the proposed debiased IEE approach and \cite{hsu2023doubly}. First, \cite{hsu2023doubly} leverages the entire CDF as an intermediary in order to study the quantile estimand, even though the CDF itself is not the target estimand. The proposed approach, in contrast, obtains the quantile estimate directly by solving the inverse estimating equation without the need to assess the entire CDF. Second, \cite{hsu2023doubly} additionally requires specifying a set of discrete grid points to approximates the CDF $F_{Y_{1M_0}}$, and one has to decide on the number and the locations of these grid points.  Because this is a discrete approximation to the continuous CDF, there could be approximation errors when the number of grid points (i.e., $R$) is relatively small or the set of points are sparse around $Q_{Y_{1M_0}}(q)$. Moreover, in \cite{hsu2023doubly}, we could not find a clear guideline on the choice of $R$ and specification of the locations of $\{\underline \theta_r\}_{r=1}^R$. In contrast, the debiased IEE approach does not require any specification of grid points and avoids such decisions that may affect the empirical performance of the final estimator.  

\subsection{Supporting information for the simulation study in Section 4.2 of main manuscript}

In Section 4.2 of the paper, we conduct numerical studies to investigate the finite-sample performance of the proposed debiased IEE approach and  \cite{hsu2023doubly}. One thousand copies of the observed data $W=\{L=\{L_1,L_2,L_3,L_4\},A,M=\{M_1,M_2\},Y\}$ are generated based on the following process. First, baseline covariates $L=\{L_1,L_2,L_3,L_4\}$ are simulated based on four mutually independent standard normal distributions. Then, we simulate the treatment $A$ based on a Bernoulli distribution with $f_{A|L}(1|L)=\text{expit}(-1L_1+0.5L_2-0.25L_3-0.1L_4)$. Next, given $A$ and $L$, the mediators $M_1$ and $M_2$ are drawn based on a bivariate normal distribution 
$$
\left(\begin{matrix}
M_1 \\ M_2
\end{matrix}\right) \sim N\left(\left[\begin{matrix}
0.5A+2L_1+L_2+L_3+L_4 \\ A-L_1-L_2-L_3-L_4
\end{matrix}\right],\left[\begin{matrix}
1 & 0.2 \\ 0.2 & 1
\end{matrix}\right]\right)
$$
Finally, we generate the outcome $Y$ based on $Y|\{A,M,L\}\sim N(2 + 1.5A + M_1 + M_2+10L_1+5L_2+5L_3+5L_4,\exp(2+A))$. We focus on assessing $Q_{1M_0}(q)$ among three estimators, including \texttt{hsu-pl($R$)}, \texttt{de-pl($R$)}, and \texttt{de-ml}. Below, we provide procedures on calculating these estimators. 


As a common ground for comparison, the same set of nuisance functions are required for all estimators, including  $f_{A|L}(1|L)$, $f_{A|M,L}(1|M,L)$, $F_{Y|A,M,L}(\theta|1,M,L)$, and $\mu(L,\theta)=E[F_{Y|A,M,L}(\theta|1,M,L)|A=0,L]$. For \texttt{hsu-pl($R$)}, we follow the procedures in \cite{hsu2023doubly} and use the Post-Lasso approach to estimate all of the nuisance functions. Specifically, we estimate $f_{A|L}(1|L)$ and $f_{A|M,L}(1|M,L)$ via the Post-Lasso with a logistic link function, with the Lasso penalty parameter chosen based on suggestions from \cite{belloni2017program}. For each threshold $\underline\theta_r$, we estimate $F_{Y|A,M,L}(\underline\theta_r|1,M,L)$ by regressing $\mathbb{I}(Y\leq \underline\theta_r)$ on $M$ and $L$ among treated units based on the Post-Lasso with a Probit link function, and estimate $\mu(L,\underline\theta_r)$ by regressing $\widehat F_{Y|A,M,L}(\underline\theta_r|1,M,L)$ on $L$ among control units based on the Post-Lasso approach with a Probit link function. To make the Lasso-based nuisance estimation more flexible, we input a third degree polynomial of the covariates $L$, which creates a total of 34 control variables including the original covariates as well as their higher-order and interaction terms. We consider four choices of $R\in\{4,10,40,100\}$ to investigate the robustness of \texttt{hsu-pl($R$)} against different number of grid points. Since we cannot find guidelines on specification of $\{\underline\theta_r\}_{r=1}^R$ in \cite{hsu2023doubly}, we follow the suggestions in \cite{belloni2017program} to specify the grid of thresholds, which similarly considered the inverse CDF approach for studying quantiles. According to \cite{belloni2017program}, the locations of the thresholds are chosen based on quantiles of the empirical distribution of the observed outcome, where $\underline \theta_1$ and $\underline \theta_R$ are fixed at 5\% and 95\% quantile of the observed outcome and the other intermediate thresholds are evenly located in-between so that $\underline \theta_r$ locates at the $(0.05+0.9\frac{r-1}{R-1})\times 100\%$ quantile of the observed outcome. 

To ensure comparability, we recycle the nuisance function estimates from \texttt{hsu-pl($R$)} to construct \texttt{de-pl($R$)}. In the \texttt{de-pl($R$)}, we need to estimate $F_{Y|A,M,L}(\theta|1,M,L)$ and $\mu(L,\theta)$ for all $\theta\in\mathcal Y$ (see Remark 2), which are approximated based on the data pairs $\{\underline\theta_r,\widehat F_{Y|A,M,L}(\underline\theta_r|1,M,L)\}_{r=1}^R$ and $\{\underline\theta_r,\widehat \mu(L,\underline\theta_r)\}_{r=1}^R$, where $\theta$ not on the grid $\{\underline\theta_r\}_{r=1}^R$ are linearly interpolated. Note that \texttt{de-pl($R$)} and \texttt{hsu-pl($R$)} leverage the same nuisance function estimates, and the only difference lies in that \texttt{hsu-pl($R$)} estimates $\theta_0$ by inverting a discrete CDF estimate whereas \texttt{de-pl($R$)} directly estimates $\theta_0$ by solving the debiased IEE. 

To demonstrate potential improvement of using other modern machine learners, we consider an alternative debiased IEE estimator, denoted by \texttt{de-ml}, with nuisance functions  estimated via Super Learner that taps into the generalized linear regression, random forest, and extreme gradient boosting libraries. Specifically, we estimate $f_{A|L}(1|L)$ and $f_{A|M,L}(1|M,L)$ by regressing $A$ on $L$ and $\{M,L\}$ via Super Learner, respectively. To estimate $F_{Y|A,M,L}(\theta|1,M,L)$, we consider an additive model of $Y|\{A=1,M,L\}$ (in a similar fashion to the additive model of $F_{Y|A,L}(\theta|a,L)$ in Section \ref{sec:sim_qte} of the Supplementary Material), where its conditional mean and variance are estimated through Supper Learner and the density of the error term is assessed through kernel smoothing. Observing $\mu(L,\theta) = \int_m F_{Y|A,M,L}(\theta|1,M,L)  f_{M|A,L}(m|A=0,L) dm$, we estimate $\mu(L,\theta)$ through $\int_{m_2} \int_{m_1} \widehat F_{Y|A,M,L}(\theta|1,M,L) \widehat f_{M_1|A,L}(m_1|A=0,L) \widehat f_{M_2|A,M_1,L}(m_2|A=0,M_1,L) dm$, where the integral is evaluated based on the Monte Carlo integration. To obtain $\widehat f_{M_1|A,L}(m_1|A=0,L)$ and $\widehat f_{M_2|A,M_1,L}(m_2|A=0,M_1,L) dm$, we again consider two additive models for $M_1|\{A=0,L\}$ and $M_2|\{A=0,M_1,L\}$ respectively, with their conditional mean and variance estimated through Super Learner and their error terms estimated via kernel smoothing. One major advantage of \texttt{de-ml} over \texttt{de-pl($R$)} is that \texttt{de-ml} estimates the $\mu(L,\theta)$ and $F_{Y|A,M,L}(\theta|1,M,L)$ directly without any discretization. 

We compare the performance of \texttt{hsu-pl($R$)}, \texttt{de-pl($R$)}, and \texttt{de-ml} on estimation of $Q_{Y_{1M_0}}(q)$ with $q\in\{0.1, 0.25, 0.5, 0.75,0.9\}$. The simulation results of $Q_{Y_{1M_0}}(0.9)$ are shown and discussed in Section 4.2 of the main manuscript. The corresponding simulation results for other quantile levels are provided in Figure \ref{fig:QME-10}--\ref{fig:QME-75}. Overall, the simulations results are qualitatively similar to that of $Q_{Y_{1M_0}}(0.9)$. One exception is that when assessing median estimands under $q=0.5$ (Figure \ref{fig:QME-50}), the \texttt{hsu-pl($R$)} has adequate performance even with small number of thresholds.


}
\section{Example 3: Survivor quantile causal effect}\label{sec:sqce}

\subsection{Theoretical development}\label{sec:sqce_theory}


We further exemplify our general results under the \emph{truncation-by-death} set up, where 
a non-mortality outcome is subject to truncation by a terminal event. Suppose that we observe a set of baseline covariates $L$, a treatment $A\in\{0,1\}$, a survival status $M$ with $M=1$ for survival and $M=0$ for death, and an outcome $Y$; $Y=\star$ is considered not well defined for units with $M=0$. Denote $Y_a$ and $M_a$ as the potential outcome and survival status had the unit received treatment $a$. Under the principal stratification framework \citep{frangakis2002principal}, the study population is divided into four principal strata based on the joint potential values of the survival status $V=\{M_1,M_0\}\in\{0,1\}^{\otimes 2}$ with $\{1,1\}$ the always-survivors stratum for units who would always survive until measurement of $Y$. 
Since the pair of final potential outcomes are well-defined only for always-survivors, the survivor average causal effect (SACE), defined as $E[Y_1-Y_0|V=\{1,1\}]$, is a causal estimand of wide interest. 
Under principal ignorability and monotonicity, \citet{jiang2022multiply} has developed the EIF and related estimators for SACE. 

To study the causal effect for the entire distribution of the non-mortality outcome among the always-survivors, we define the survivor quantile causal effect (SQCE), $\text{SQCE}(q)=Q_{Y_1|V}(q|\{1,1\})-Q_{Y_0|V}(q|\{1,1\})$. To focus ideas, we only consider estimation of $\theta_0:=Q_{Y_0|V}(q|\{1,1\})$, as the development for $Q_{Y_1|V}(q|\{1,1\})$ is analogous and deferred in Section \ref{sec:sqce_parallel_results}. Under several assumptions, including (I) consistency and positivity ($Y_a=Y$ and $M_a=M$ if $A=a$, and $f_{A|L}(a|l)>0$ and $f_{M|A,L}(1|0,L)>0$ for any $a$ and $l$), (II) ignorability ($\{Y_1,Y_0,M_1,M_0\}\perp A|L$), (III) monotonicity ($M_1\geq M_0$), and (IV) principal ignorability ($Y_a\perp V|L$ for $a\in\{0,1\}$), we can identify the mean of the threshold-transformed potential outcome  $\tau_0(\theta):=E[\bbI(Y_0\leq\theta)]$ as $\frac{E[f_{M|A,L}(1|0,L)F_{Y|A,M,L}(\theta|1,1,L)]}{E[f_{M|A,L}(1|0,L)]}$. In other words, the identifying moment function for $\tau_0(\theta)$ is
$$
g(W,\tau,\theta,\gamma) = f_{M|A,L}(1|0,L)\{F_{Y|A,M,L}(\theta|0,1,L) - \tau\},
$$
where $\gamma = \{f_{M|A,L}(1|0,L),F_{Y|A,M,L}(\theta|0,1,L)\}$. If we further assume uniqueness of the $q$-quantile of $Y_1|V=\{1,1\}$ (Assumption 2), Theorem 1 suggests that $\theta_0$ is also identified with $E[g(W,q,\theta_0,\gamma)]=0$, where a plug-in IEE estimator $\widehat\theta^{pi}$ can be obtained by solving $\Pn[\widehat f_{M|A,L}(1|0,L)\{\widehat F_{Y|A,M,L}(\theta|0,1,L) - q\}]=0$ in terms of $\theta$. 

The debiased IEE can be employed to improve performance of $\widehat\theta^{pi}$, where one should derive the EIF for $\theta_0$ at first. Of note, the EIF for $\tau_0(\theta)$ has been derived in \cite{jiang2022multiply} with the normalizing denominator $C(\theta)=-E[f_{M|A,L}(1|0,L)]$, the adjustment term
\begin{align*}
\phi(W,\tau,\theta,h) = & \frac{(1-A)M}{1-f_{A|L}(1|L)}\{\bbI(Y\leq\theta)-F_{Y|A,M,L}(\theta|0,1,L)\} \\
& + \frac{(1-A)\{F_{Y|A,M,L}(\theta|0,1,L)-\tau\}}{1-f_{A|L}(1|L)} \{M-f_{M|A,L}(1|0,L)\},
\end{align*}
and nuisance functions $h=\{f_{M|A,L}(1|0,L),F_{Y|A,M,L}(\theta|0,1,L),f_{A|L}(1|L)\}$. For ease of notation, we abbreviate the three nuisance functions as $h_1$ to $h_3$. According to Theorem 2 with Strategy I, $\psi_{\theta_0}^{\text{eff}}(W,q,\theta_0,h)$ has the following explicit expression
\begin{equation}\label{eq:example2_eif}
-\frac{1}{B} \left[h_1(h_2-q)+\frac{(1-A)M}{1-h_3}\{\bbI(Y\leq\theta_0)-h_2\}+\frac{(1-A)\{h_2-q\}}{1-h_3} \{M-h_1\}\right],
\end{equation}
where $B := \frac{d}{d\theta}E[g(W,q,\theta,\gamma)]|_{\theta=\theta_0}=E[f_{M|A,L}(1|0,L)f_{Y|A,M,L}(\theta_0|0,1,L)]$. Therefore, the debiased IEE can be constructed by setting the empirical average of \eqref{eq:example2_eif} to 0 and replacing each $h_j$ with its estimate $\widehat h_j$. The debiased IEE estimator $\widehat \theta^{de}$ can be obtained accordingly. 

{We take a moment to discuss estimation of the nuisance functions $h$. Because both $M$ and $A$ are binary variables, $\widehat f_{M|A,L}(1|0,L)$ and $\widehat f_{A|L}(1|L)$ can be obtained by binary regression (e.g., logistic regression) or machine learners for binary classification problems. To estimate $F_{Y|A,M,L}(\theta|0,1,L)$, one can fit a flexible parametric model to represent the entire CDF $F_{Y|A,M,L}(y|0,1,L)$.  
Alternatively, as inspired by \cite{kennedy2017non}, we can assume an additive model for $F_{Y|A,M,L}(y|0,1,L)$ such that 
$$Y=m(L)+v(L)\epsilon$$ 
within strata $\{A=0,M=1\}$, where $m(L)$ and $v(L)$ are the mean and variance structure and $\epsilon$ is a zero-mean error term. Then, $m(L)$ can be estimated by regressing of $Y$ on $L$ among units within strata $\{A=0,M=1\}$ either through machine learners or generalized linear models, $v(L)$ can be similarly estimated by regressing of $(Y-\widehat m(L))^2$ on $L$, and the density of $\epsilon$ (denoted by $f_{\epsilon}$) can be estimated by using kernel smoothing based on $(Y-\widehat m(L))/\widehat v(L)$ among units within strata $\{A=0,M=1\}$. It follows that $\widehat F_{Y|A,M,L}(\theta|0,1,L)=\int_{-\infty}^\theta \widehat f_{Y|A,M,L}(y|0,1,L) dy$, where   $\widehat f_{Y|A,M,L}(y|0,1,L)=\widehat f_\epsilon\left((y-\widehat m(L))/\widehat v(L)\right)$ is the density estimator of $f_{Y|A,M,L}(y|0,1,L)$.}

We next show the mixed bias property of $\psi_{\theta_0}^{\text{eff}}(W,q,\theta,h)$. Specifically, because $E[\phi(W,q,\theta,h)]=0$, we have $E[\psi_{\theta_0}^{\text{eff}}(W,q,\theta,h)]=-B \times E[g(W,q,\theta,h)]$, and thus
\begin{align*}
& -B \times E[\psi_{\theta_0}^{\text{eff}}(W,q,\theta,\widehat h)-\psi_{\theta_0}^{\text{eff}}(W,q,\theta, h)] \\
= & E\left[\widehat f_{M|A,L}(1|0,L)\left\{\widehat F_{Y|A,M,L}(\theta|0,1,L)-q\right\}\right] \\
& + E\left[\frac{(1-A)M}{1-\widehat f_{A|L}(1|L)}\left\{\bbI(Y\leq\theta)-\widehat F_{Y|A,M,L}(\theta|0,1,L)\right\}\right] \\
 & + E\left[\frac{(1-A)\{\widehat F_{Y|A,M,L}(\theta|0,1,L)-q\}}{1-\widehat f_{A|L}(1|L)}\left\{M-\widehat f_{M|A,L}(1|0,L)\right\}\right] \\
 & - E[f_{M|A,L}(1|0,L)\left\{F_{Y|A,M,L}(\theta|0,1,L)-q\right\}] \\
 = & E\left[\widehat f_{M|A,L}(1|0,L)\left\{\widehat F_{Y|A,M,L}(\theta|0,1,L)-q\right\}\right] \\
  & + E\left[\frac{(1-f_{A|L}(1|L))f_{M|A,L}(1|0,L)}{1-\widehat f_{A|L}(1|L)}\left\{F_{Y|A,M,L}(\theta|0,1,L)-\widehat F_{Y|A,M,L}(\theta|0,1,L)\right\}\right] \\
 & + E\left[\frac{\{1-f_{A|L}(1|L)\}\{\widehat F_{Y|A,M,L}(\theta|0,1,L)-q\}}{1-\widehat f_{A|L}(1|L)}\left\{f_{M|A,L}(1|0,L)-\widehat f_{M|A,L}(1|0,L)\right\}\right] \\
 & - E[f_{M|A,L}(1|0,L)\left\{F_{Y|A,M,L}(\theta|0,1,L)-q\right\}] \\
 = &  E\left[-\frac{\widehat F_{Y|A,M,L}(\theta|0,1,L)-q}{1-\widehat f_{A|L}(1|L)}\left\{\widehat f_{A|L}(1|L)-f_{A|L}(1|L)\right\}\left\{\widehat f_{M|A,L}(1|0,L)-f_{M|A,L}(1|0,L)\right\}\right]\\
  & + E\left[-\frac{f_{M|A,L}(1|0,L)}{1-\widehat f_{A|L}(1|L)}\left\{\widehat f_{A|L}(1|L)-f_{A|L}(1|L)\right\}\left\{\widehat F_{Y|A,M,L}(\theta|0,1,L)-F_{Y|A,M,L}(\theta|0,1,L)\right\}\right] 
\end{align*}
This suggests that $\psi_{\theta_0}^{\text{eff}}(W,q,\theta,h)$ satisfies the mixed bias property with $K=2$, $[\{a_1,b_1\},\{a_2,b_2\}]=[\{1,2\},\{1,3\}]$, and $S_1 = \frac{\widehat F_{Y|A,M,L}(\theta|0,1,L)-q}{B\{1-\widehat f_{A|L}(1|L)\}}$ and $S_2 = \frac{f_{M|A,L}(1|0,L)}{B\{1-\widehat f_{A|L}(1|L)\}}$, where $E[S_1^2]<\infty$ and $E[S_2^2]<\infty$ if $\widehat f_{A|L}(1|L)$ is bounded away from 1. 
Therefore, if one considers using parametric working models to obtain $\widehat h$, then $\widehat\theta^{de}$ is consistent and asymptotically normal if the union model $\{\mathcal M_1 \cup \mathcal M_2\} \cap \{\mathcal M_1 \cup \mathcal M_3\}$ is correctly specified (Theorem 4). That is, $\widehat\theta^{de}$ is a doubly robust estimator under either $\mathcal M_1$ or  $\mathcal M_2 \cap \mathcal M_3$. Otherwise, if $\widehat h$ is obtained by data-adaptive machine learners with the rate of convergence satisfying $\xi_{1,n}\xi_{2,n}+\xi_{1,n}\xi_{3,n}=o(n^{-{1}/{2}})$, $\widehat\theta^{de}$ is consistent and asymptotically normal with its asymptotic variance attaining the semiparametric efficiency lower bound (Theorem 3). {When machine learners are used, one can estimate $\text{Var}(\widehat\theta^{de})$ based on the empirical variance of the EIF \eqref{eq:example2_eif}, where $B$ can be estimated by $\widehat B=\mathbb{P}_n[\widehat f_{M|A,L}(1|0,L)\widehat f_{Y|A,M,L}(\theta_0|0,1,L)]$. Here, $\widehat f_{M|A,L}(1|0,L)$ is part of $\widehat h$ and $\widehat f_{Y|A,M,L}(\widehat\theta^{de}|0,1,L) = \widehat f_\epsilon((\widehat \theta^{de}-\widehat m(L))/\widehat v(L))$ if the additive model $Y=m(L)+v(L)\epsilon$ is used for $F_{Y|A,M,L}(y|0,1,L)$.}

\subsection{Parallel results on estimation of  $Q_{Y_1|V}(q|\{1,1\})$}\label{sec:sqce_parallel_results}

Next, we briefly describe the parallel results for $\theta_0:=Q_{Y_1|V}(q|\{1,1\})$. In this case, the identifying moment function for $\tau_0(\theta):=E[\bbI(Y_1\leq\theta)|V=\{1,1\}]$ is
$$
g(W,\tau,\theta,\gamma) = f_{M|A,L}(1|0,L)\{F_{Y|A,M,L}(\theta|1,1,L) - \tau\},
$$
with $\gamma = \{f_{M|A,L}(1|0,L),F_{Y|A,M,L}(\theta|1,1,L)\}$. The EIF of $\tau_0(\theta)$ is drived in \cite{jiang2022multiply} with the form $\psi_{\tau_0(\theta)}^{\text{eff}}(W,\tau_0(\theta),\theta,h)=-C^{-1}\left\{g(W,\tau_0(\theta),\theta,\gamma)+\phi(W,\tau_0(\theta),\theta,h)\right\}$ with $C=-E[f_{M|A,L}(1|0,L)]$, the adjustment term
\begin{align*}
\phi(W,\tau,\theta,h) = & f_{M|A,L}(1|0,L)\frac{AM}{f_{M|A,L}(1|1,L)f_{A|L}(1|L)}\{\bbI(Y\leq\theta)-F_{Y|A,M,L}(\theta|1,1,L)\} \\
& + \frac{(1-A)\{F_{Y|A,M,L}(\theta|1,1,L)-\tau\}}{1-f_{A|L}(1|L)} \{M-f_{M|A,L}(1|0,L)\},
\end{align*}
and nuisance functions $h=\{f_{M|A,L}(1|0,L),F_{Y|A,M,L}(\theta|1,1,L),f_{M|A,L}(1|1,L),f_{A|L}(1|L)\}$. Therefore, $\psi_{\theta_0}^{\text{eff}}(W,q,\theta_0,h)$ can be obtained accordingly based on Theorem 2 and is given by 
\begin{align*}
& -\frac{1}{B}\Big[f_{M|A,L}(1|0,L)\{F_{Y|A,M,L}(\theta_0|1,1,L) - q\} \\
 & + f_{M|A,L}(1|0,L)\frac{AM}{f_{M|A,L}(1|1,L)f_{A|L}(1|L)}\{\bbI(Y\leq\theta_0)-F_{Y|A,M,L}(\theta|1,1,L)\} \\
& + \frac{(1-A)\{F_{Y|A,M,L}(\theta_0|1,1,L)-q\}}{1-f_{A|L}(1|L)} \{M-f_{M|A,L}(1|0,L)\}\Big],
\end{align*}
where $B:=\frac{d}{d\theta}E[g(W,q,\theta,\gamma)]|_{\theta=\theta_0}=E[f_{M|A,L}(1|0,L)f_{Y|A,M,L}(\theta_0|1,1,L)]$. Furthermore, we can show \begingroup\makeatletter\def\f@size{9.5}\check@mathfonts 
\begin{align*}
& -B \times E[\psi_{\theta_0}^{\text{eff}}(W,q,\theta,\widehat h)-\psi_{\theta_0}^{\text{eff}}(W,q,\theta,h)] \\
= & E\left[\widehat f_{M|A,L}(1|0,L)\left\{\widehat F_{Y|A,M,L}(\theta|1,1,L)-q\right\}\right] \\
& + E\left[\widehat f_{M|A,L}(1|0,L)\frac{AM}{\widehat f_{M|A,L}(1|1,L)\widehat f_{A|L}(1|L)}\{\bbI(Y\leq\theta)-\widehat F_{Y|A,M,L}(\theta|1,1,L)\}\right] \\
 & + E\left[\frac{(1-A)\{\widehat F_{Y|A,M,L}(\theta|1,1,L)-q\}}{1-\widehat f_{A|L}(1|L)}\left\{M-\widehat f_{M|A,L}(1|0,L)\right\}\right] \\
 & - E[f_{M|A,L}(1|0,L)\left\{F_{Y|A,M,L}(\theta|1,1,L)-q\right\}] \\
= & E\left[\widehat f_{M|A,L}(1|0,L)\left\{\widehat F_{Y|A,M,L}(\theta|1,1,L)-q\right\}\right] \\
& + E\left[\widehat f_{M|A,L}(1|0,L)\frac{f_{M|A,L}(1|1,L) f_{A|L}(1|L)}{\widehat f_{M|A,L}(1|1,L)\widehat f_{A|L}(1|L)}\{F_{Y|A,M,L}(\theta|1,1,L)-\widehat F_{Y|A,M,L}(\theta|1,1,L)\}\right] \\
 & + E\left[\frac{\{1-f_{A|L}(1|L)\}\{\widehat F_{Y|A,M,L}(\theta|1,1,L)-q\}}{1-\widehat f_{A|L}(1|L)}\left\{f_{M|A,L}(1|0,L)-\widehat f_{M|A,L}(1|0,L)\right\}\right] \\
 & - E[f_{M|A,L}(1|0,L)\left\{F_{Y|A,M,L}(\theta|1,1,L)-q\right\}] \\
 = &  + E\left[-\left\{\widehat f_{M|A,L}(1|0,L)-  f_{M|A,L}(1|0,L)\right\}\left\{\widehat F_{Y|A,M,L}(\theta|1,1,L)-F_{Y|A,M,L}(\theta|1,1,L)\right\}\right]\\
 & + E\left[-\frac{\widehat F_{Y|A,M,L}(\theta|1,1,L)-q}{1-\widehat f_{A|L}(1|L)}\left\{\widehat f_{M|A,L}(1|0,L)-f_{M|A,L}(1|0,L)\right\}\left\{\widehat f_{A|L}(1|L)- f_{A|L}(1|L)\right\}\right] \\
 & + E\left[\frac{\widehat f_{M|A,L}(1|0,L)}{\widehat f_{M|A,L}(1|1,L)}\left\{\widehat f_{M|A,L}(1|1,L)-  f_{M|A,L}(1|1,L)\right\}\left\{\widehat F_{Y|A,M,L}(\theta|1,1,L)-F_{Y|A,M,L}(\theta|1,1,L)\right\}\right]\\
  & + E\left[\frac{\widehat f_{M|A,L}(1|0,L)f_{M|A,L}(1|1,L)}{\widehat f_{A|L}(1|L)\widehat f_{M|A,L}(1|1,L)}\left\{\widehat F_{Y|A,M,L}(\theta|1,1,L)-F_{Y|A,M,L}(\theta|1,1,L)\right\}\left\{\widehat f_{A|L}(1|L)- f_{A|L}(1|L)\right\}\right] 
\end{align*}\endgroup
If we let $h_1$--$h_4$ to denote the four nuisance functions $f_{M|A,L}(1|0,L)$, $F_{Y|A,M,L}(\theta|1,1,L)$, $f_{M|A,L}(1|1,L)$, and $f_{A|L}(1|L)$, respectively, then it is clear that $\psi_{\theta_0}^{\text{eff}}(W,q,\theta,h)$ satisfies the mixed bias property with $K=4$, $[\{a_1,b_1\},\{a_2,b_2\},\{a_3,b_3\},\{a_4,b_4\}] = [\{1,2\},\{1,4\},\{2,3\},\{2,4\}]$, and $S_1=B^{-1}$, $S_2=\frac{\widehat F_{Y|A,M,L}(\theta|1,1,L)-q}{B\{1-\widehat f_{A|L}(1|L)\}}$, $S_3 = -\frac{\widehat f_{M|A,L}(1|0,L)}{B\times \widehat f_{M|A,L}(1|1,L)}$, and $S_4 = - \frac{\widehat f_{M|A,L}(1|0,L)f_{M|A,L}(1|1,L)}{B\times\widehat f_{A|L}(1|L)\widehat f_{M|A,L}(1|1,L)}$. Notice that all $E[S_k^2]<\infty$ if the estimated nuisance $\widehat f_{M|A,L}(1|1,L)$ and $\widehat f_{A|L}(1|L)$ are bounded away from 0. Therefore, if $\widehat h$ is obtained by  machine learners with the rate of convergence satisfying $\xi_{1,n}\xi_{2,n}+\xi_{1,n}\xi_{4,n}+\xi_{2,n}\xi_{3,n}+\xi_{2,n}\xi_{4,n}=o(n^{-\frac{1}{2}})$, $\widehat\theta^{de}$ is CAN with its variance attaining the semiparametric efficiency lower bound. Otherwise, if we consider using parametric working models to obtain $\widehat h$, then $\widehat\theta^{de}$ is CAN if the union model $\{\mathcal M_1 \cup \mathcal M_2\} \cap \{\mathcal M_1 \cup \mathcal M_4\} \cap \{\mathcal M_2 \cup \mathcal M_3\} \cap \{\mathcal M_2 \cup \mathcal M_4\}$ is correctly specified. In other words, $\widehat\theta^{de}$ is CAN if either $\mathcal M_1 \cap \mathcal M_2$, $\mathcal M_1 \cap \mathcal M_3 \cap \mathcal M_4$, or $\mathcal M_2 \cap  \mathcal M_4$ is correctly specified.

\begin{figure}[ht]
\begin{center}
\includegraphics[width=0.99\textwidth]{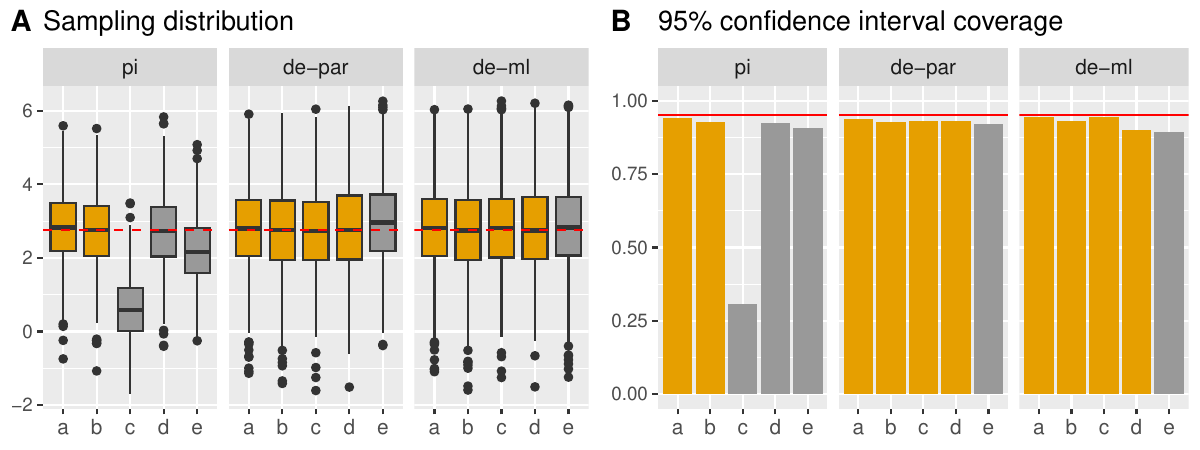}
\end{center}
\caption{Sampling distributions (Panel A) and 95\% confidence interval coverage probability (Panel B) among estimators of $Q_{Y_0|V}(0.5|\{1,1\})$, in scenarios (a) all nuisance models correctly specified; (b) $\widehat f_{A|L}(1|L)$ incorrect; (c) $\widehat f_{M|A,L}(1|0,L)$ incorrect;  (d) $\widehat F_{Y|A,M,L}(\theta|0,1,L)$ incorrect; (e) all nuisance models incorrect.
For each scenario, the box/bar filled with orange color indicates that the estimator is consistent based on theory. The red dotted line in Panel A is the true value of $Q_{Y_0}(0.5)$.}
\label{fig:Q0_50}
\end{figure}

\begin{figure}[ht]
\begin{center}
\includegraphics[width=0.99\textwidth]{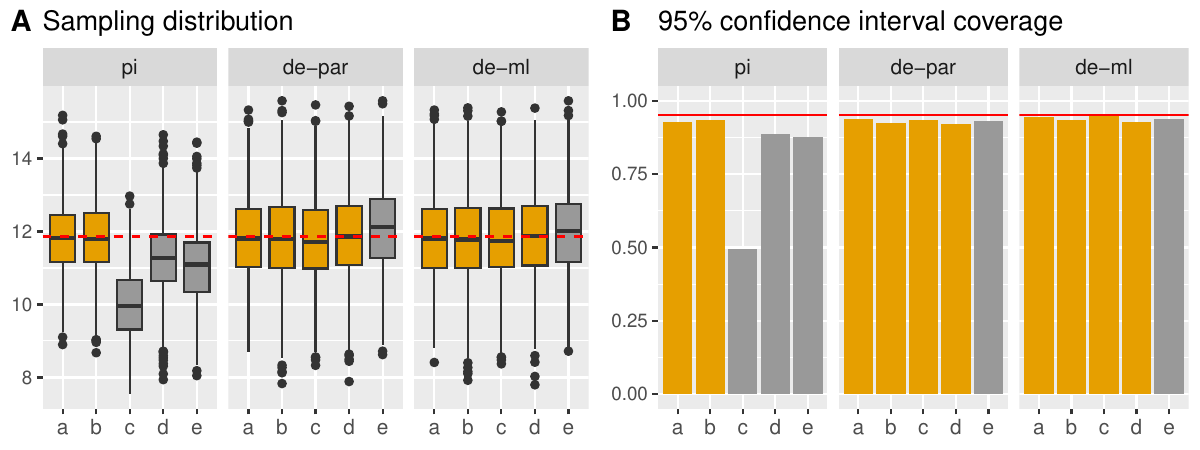}
\end{center}
\caption{Sampling distributions (Panel A) and 95\% confidence interval coverage probability (Panel B) among estimators of $Q_{Y_0|V}(0.75|\{1,1\})$, in scenarios (a) all nuisance models correctly specified; (b) $\widehat f_{A|L}(1|L)$ incorrect; (c) $\widehat f_{M|A,L}(1|0,L)$ incorrect;  (d) $\widehat F_{Y|A,M,L}(\theta|0,1,L)$ incorrect; (e) all nuisance models incorrect.
For each scenario, the box/bar filled with orange color indicates that the estimator is consistent based on theory. The red dotted line in Panel A is the true value of $Q_{Y_0}(0.75)$.}
\label{fig:Q0_75}
\end{figure}

{

\subsection{A simulation study on $Q_{Y_0|V}(q|\{1,1\})$}\label{sec:sim}

We investigate the empirical performance of the proposed estimators in Section \ref{sec:sqce_theory}. A thousand Monte Carlo experiments with $n = 1,000$ units are generated based on the following process. First, four baseline covariates $L=[L_1,L_2,L_3,L_4]^T$ are generated based on four mutually independent standard normal distributions. Then, we generate the treatment $A$ by $f_{A|L}(1|L)=\text{expit}(-L_1+0.5L_2-0.25L_3-0.1L_4)$, the observed survival status $M$ by $f_{M|A,L}(1|A,L)=\text{expit}(-1+2A+L_1-0.8L_2+0.6L_3-L_4)$, where $\text{expit}(x)=(1+\exp(-x))^{-1}$. Finally, for survivors ($M=1$), we generate the outcome $Y$ based on $Y|\{A,M=1,L\}\sim N(1+1.5A+10L_1+5L_2+5L_3+5L_4,\exp(2+A))$, where $Y=\star$ among the non-survivors ($M=0$). We consider assessing $Q_{Y_0|V}(q|\{1,1\})$ with $q\in\{0.25,0.5,0.75\}$; additional simulation studies for $Q_{Y_1|V}(q|\{1,1\})$ and $\text{SQCE}(q)$ are provided in Section \ref{sec:qme_additional_sim}. 

We compare the following three approaches: (i) the plug-in IEE estimator $\widehat Q_{Y_0}^{pi}(q)$ that uses parametric models for $\{f_{M|A,L}(1|0,L),F_{Y|A,M,L}(\theta|0,1,L)\}$, (ii) a debiased IEE estimator $\widehat Q_{Y_0}^{de\text{-}par}(q)$ with parametric models for $h$, (iii) a debiased IEE estimator $\widehat Q_{Y_0}^{de\text{-}ml}(q)$ with machine learners for $h$. When parametric models are used, we fit a logistic regression of $A$ on $L$ to obtain $\widehat f_{A|L}(1|L)$ and a logistic regression of $A$ on $L$ among individuals with $A=0$ to obtain $\widehat f_{M|A,L}(1|0,L)$; to obtain $\widehat F_{Y|A,M,L}(\theta|0,1,L)$, we consider a heteroskedastic Gaussian model of $Y$ on $\{A,L\}$ among the survivors; the conditional mean is obtained by fitting a linear regression of $Y$ on $\{A,L\}$ and the conditional variance is obtained by fitting $\{Y-\widehat E[Y|A=0,M=1,L]\}^2$ on $\{A,L\}$. We consider five-fold cross-fitting when machine learners are used to estimate the nuisance. Specifically, $\widehat f_{A|L}(1|L)$ and $\widehat f_{M|A,L}(1|0,L)$ are obtained by Super Learner \citep{phillips2023practical} with random forest, extreme gradient boosting, and generalize linear model libraries. To fit $F_{Y|A,M,L}(\theta|0,1,L)$, we consider the additive model in Section \ref{sec:sqce_theory}, where the mean and variance structures ($m(L)$ and $v(L)$) are estimated through Super Learner with random forest, extreme gradient boosting, and generalized linear model libraries. We evaluate each method under 5 scenarios: (a) all models are correctly specified; (b) $\widehat f_{A|L}(1|L)$ is incorrectly specified; (c) $\widehat f_{M|A,L}(1|0,L)$ is incorrectly specified; (d) $\widehat F_{Y|A,M,L}(\theta|0,1,L)$ is incorrectly specified; (e) all models are incorrectly specified. For correctly specified models, we input the true baseline covariates $L$ into the parametric models or the Super Learner. Otherwise, we input a set of transformed covariates $\widetilde L=[\widetilde L_1,\widetilde L_2,\widetilde L_3,\widetilde L_4]^T$ instead, where $\widetilde L_1 = \exp(0.5L_1)$, $\widetilde L_2 = {L_2}/{(1+L_1)}$, $\widetilde L_3 = (L_2L_3/25+0.6)^3$, and $\widetilde L_4=(L_2+L_4+20)^2$. 

\begin{figure}
\begin{center}
\includegraphics[width=0.99\textwidth]{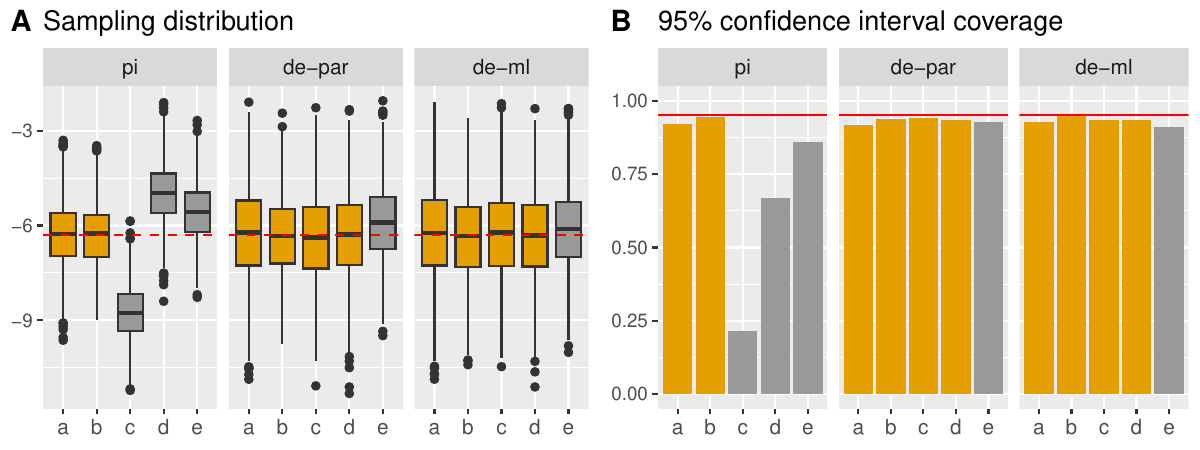}
\end{center}
\caption{Sampling distribution (Panel A) and 95\% confidence interval coverage (Panel B) among estimators of $Q_{Y_0|L}(0.25|\{1,1\})$, in scenarios (a) all  models correct; (b) $\widehat f_{A|L}(1|L)$ incorrect; (c) $\widehat f_{M|A,L}(1|0,L)$ incorrect; (d) $\widehat F_{Y|A,M,L}(\theta|0,1,L)$ incorrect; (e) all  models incorrect. For each scenario, the box/bar filled with orange color indicates that the estimator is consistent based on theory. The red dotted line in Panel A is the true value of the estimand.}
\label{fig:Q0_25}
\end{figure}

Figure \ref{fig:Q0_25}(A) displays the sampling distribution of $\widehat Q_{Y_0|V}^{pi}(0.25|\{1,1\})$, $\widehat Q_{Y_0|V}^{de\text{-}par}(0.25|\{1,1\})$, and $\widehat Q_{Y_0|V}^{de\text{-}ml}(0.25|\{1,1\})$ among 1,000 Monte Carlo samples. Based on Section \ref{sec:sqce_theory}, we know that $\widehat Q_{Y_0|V}^{pi}(0.25|\{1,1\})$ is consistent in scenarios (a)--(b), and $\widehat Q_{Y_0|V}^{de\text{-}par}(0.25|\{1,1\})$ and $\widehat Q_{Y_0|V}^{de\text{-}ml}(0.25|\{1,1\})$ are consistent in scenarios (a)--(d). All estimators perform as anticipated---they concentrate around the true value when the required nuisance functions are correctly specified but diverge from the true value otherwise. In particular, $\widehat Q_{Y_0|V}^{de\text{-}par}(0.25|\{1,1\})$ and $\widehat Q_{Y_0|V}^{de\text{-}ml}(0.25|\{1,1\})$ provides additional robustness against model misspecification in scenarios (c) and (d), comparing to the plug-in estimator $\widehat Q_{Y_0|V}^{pi}(0.25|\{1,1\})$. Figure \ref{fig:Q0_25}(B) displays the coverage probability based on a 95\% Wald-type confidence interval among each estimator, where the asymptotic variance of $\widehat Q_{Y_0|V}^{pi}(0.25|\{1,1\})$ and $\widehat Q_{Y_0|V}^{de\text{-}par}(0.25|\{1,1\})$ are calculated based on nonparametric bootstrap and the asymptotic variance of $\widehat Q_{Y_0|V}^{de\text{-}ml}(0.25|\{1,1\})$ are calculated based on the empirical variance of the EIF. 
We observe that all methods present close-to-nominal coverage in scenarios that they are expected to be consistent. We also have investigated the performance of $Q_{Y_0|V}(0.5|\{1,1\})$ and $Q_{Y_0|V}(0.75|\{1,1\})$ among each estimator (see Figures \ref{fig:Q0_50} and \ref{fig:Q0_75}) and the results are qualitatively similar.

\subsection{Additional Simulation Studies on $Q_{Y_1|V}(q|\{1,1\})$ and $\text{SQCE}(q)$}\label{sec:qme_additional_sim}

We conduct additional simulation studies to assess the performance of our estimators on $Q_{Y_1|V}(q|\{1,1\})$ and $\text{SQCE}(q)=Q_{Y_1|V}(q|\{1,1\})-Q_{Y_0|V}(q|\{1,1\})$, with $q\in\{0.25,0.5,0.75\}$. Simulation results for $Q_{Y_1|V}(q|\{1,1\})$ can be found in Section \ref{sec:sim}. 

Similar to the assessment of $Q_{Y_0|V}(q|\{1,1\})$, we consider three estimators of $Q_{Y_1|V}(q|\{1,1\})$ in Section \ref{sec:sqce_parallel_results}: (i) a plug-in IEE estimator $\widehat Q_{Y_1|V}^{pi}(q|\{1,1\})$, (ii) a debiased IEE $\widehat Q_{Y_1|V}^{de\text{-}par}(q|\{1,1\})$ with parametric models for nuisance $h$, and (iii) a debiased IEE estimator $\widehat Q_{Y_1|V}^{de\text{-}ml}(q|\{1,1\})$ with machine learners for nuisances $h$. As detailed in Section \ref{sec:sqce_theory}, the nuisance functions $h$ for $Q_{Y_0|V}(q|\{1,1\})$ include $f_{A|L}(1|L)$, $f_{M|A,L}(1|0,L)$, $f_{M|A,L}(1|1,L)$, $F_{Y|A,M,L}(\theta|1,1,L)$. As $f_{M|A,L}(1|0,L)$ and $f_{A|L}(1|L)$ are also used in the estimation of $Q_{Y_0|V}(q|\{1,1\})$, specification of the working models of these two nuisance functions are same to these discussion in Section \ref{sec:sqce_theory}. Specifications of $\{f_{M|A,L}(1|1,L),F_{Y|A,M,L}(\theta|1,1,L)\}$ follow those of $\{f_{M|A,L}(1|0,L),F_{Y|A,M,L}(\theta|0,1,L)\}$ in Section \ref{sec:sim}, but here we target on the treated units ($A=1$) rather than the control units ($A=0$). We also consider 5 scenarios regarding correct and incorrect specification of the working models: (a) all models are correctly specified; (b) $\widehat f_{A|L}(1|L)$ is incorrectly specified; (c) $\widehat f_{M|A,L}(1|0,L)$ and $f_{M|A,L}(1|0,L)$ are incorrectly specified; (d) $\widehat F_{Y|A,M,L}(\theta|1,1,L)$ is incorrectly specified; (e) all models are incorrectly specified. For correctly (or incorrectly) specified models, we input the true covariates $L$ (or the transformed covariates $\widetilde L$) into the parametric model or Super Learner. By theory, $\widehat Q_{Y_1|V}^{de\text{-}par}(q|\{1,1\})$ and $\widehat Q_{Y_1|V}^{de\text{-}ml}(q|\{1,1\})$ are consistent in Scenarios (a)--(d), whereas $\widehat Q_{Y_1|V}^{pi}(q|\{1,1\})$ is only consistent in Scenarios (a)--(b).  The simulation results, including the sampling distribution and 95\% confidence interval coverage rate, are displayed in Supplementary Material Figure \ref{fig:Q1_25}--\ref{fig:Q1_75} for $Q_{Y_1|V}(0.25)$, $Q_{Y_1|V}(0.5|\{1,1\})$, and $Q_{Y_1|V}(0.75|\{1,1\})$, respectively. Overall, the simulation results are qualitatively similar to our results on $Q_{Y_0|V}(q|\{1,1\})$, where  $\widehat Q_{Y_1|V}^{de\text{-}par}(q|\{1,1\})$ and $\widehat Q_{Y_1|V}^{de\text{-}ml}(q|\{1,1\})$ provide minimal bias with nominal coverage in Scenario (a)--(d) when they are expected to be consistent.

For completeness, we also assess the performance of $\widehat{\text{SQCE}}^{pi}(q)=\widehat Q_{Y_1|V}^{pi}(q|\{1,1\})-\widehat Q_{Y_0|V}^{pi}(q|\{1,1\})$, $\widehat{\text{SQCE}}^{de\text{-}par}(q|\{1,1\})=\widehat Q_{Y_1|V}^{de\text{-}par}(q|\{1,1\})-\widehat Q_{Y_0|V}^{de\text{-}par}(q|\{1,1\})$, and $\widehat{\text{SQCE}}^{de\text{-}ml}(q)=\widehat Q_{Y_1|V}^{de\text{-}ml}(q|\{1,1\})-\widehat Q_{Y_0|V}^{de\text{-}ml}(q)$. Note that there are a total of 5 nuisance functions used in $Q_{Y_0|V}(q|\{1,1\})$ and $Q_{Y_1|V}(q|\{1,1\})$, including $f_{A|L}(1|L)$, $f_{M|A,L}(1|0,L)$, $f_{M|A,L}(1|1,L)$, $F_{Y|A,M,L}(\theta|1,1,L)$, $F_{Y|A,M,L}(\theta|0,1,L)$. We consider the following five scenarios regarding correct and incorrect specification of nuisance functions: (a) all models are correctly specified; (b) $\widehat f_{A|L}(1|L)$ is incorrectly specified; (c) $\widehat f_{M|A,L}(1|0,L)$ and $f_{M|A,L}(1|0,L)$ are incorrectly specified; (d) $\widehat F_{Y|A,M,L}(\theta|1,1,L)$ and $\widehat F_{Y|A,M,L}(\theta|0,1,L)$ incorrectly specified; (e) all models are incorrectly specified. By theory,  $\widehat{\text{SQCE}}^{de\text{-}par}(q)$ and $\widehat{\text{SQCE}}^{de\text{-}ml}(q)$ are still consistent in Scenarios (a)--(d) but $\widehat{\text{SQCE}}^{pi}(q)$ is only consistent in Scenarios (a)--(b). We present the simulation results of $\text{SQCE}(0.25)$, $\text{SQCE}(0.5)$, and $\text{SQCE}(0.75)$ in  Supplementary Material Figures \ref{fig:Qd_25}--\ref{fig:Qd_75}. Overall, these results align with our expectations. In scenarios (a) through (d), both $\widehat{\text{SQCE}}^{de\text{-}par}(q)$ and $\widehat{\text{SQCE}}^{de\text{-}ml}(q)$ exhibit minimal bias with close-to-nominal coverage.

\subsection{Application to the National Jobs Corps study}

The National Jobs Corps study is a vocationally focused training program for disadvantaged youths aged between 16 to 24 in United States \citep{schochet2001national}, where the data set is publicly available at the R package \texttt{causalweight} with 9,240 participants. As a randomized experiment, participants were randomized to either participate the training program immediately ($A=1$) or not participate it until three years later ($A=0$). In this analysis, we aim to study the effectiveness of treatment on participants' weekly earnings during the fourth year after assignment ($Y$). However, the outcome is truncated by `death' (the employment status $M$), where $Y$ is only well defined among the participants who are employed during the fourth year after assignment ($M=1$) but is not well defined otherwise (also see \citealp{zhang2009likelihood}). We therefore assess the survivor quantile causal effect among these who are always-employed with $M(1)=M(0)=1$, $\text{SQCE}(q)=Q_{Y_1|V}(q|\{1,1\})-Q_{Y_0|V}(q|\{1,1\})$, for $q$ ranged from 0.05 to 0.95 with a 0.01 increment. We adjust for all 28 baseline covariates ($L$) available in the \texttt{causalweight} package, including gender, age, race, education, marital status, general health condition, smoking status, weekly earnings at assignment, household size, parents' years of education, welfare receipt during childhood, and others.

\begin{figure}
\begin{center}
\includegraphics[width=0.99\textwidth]{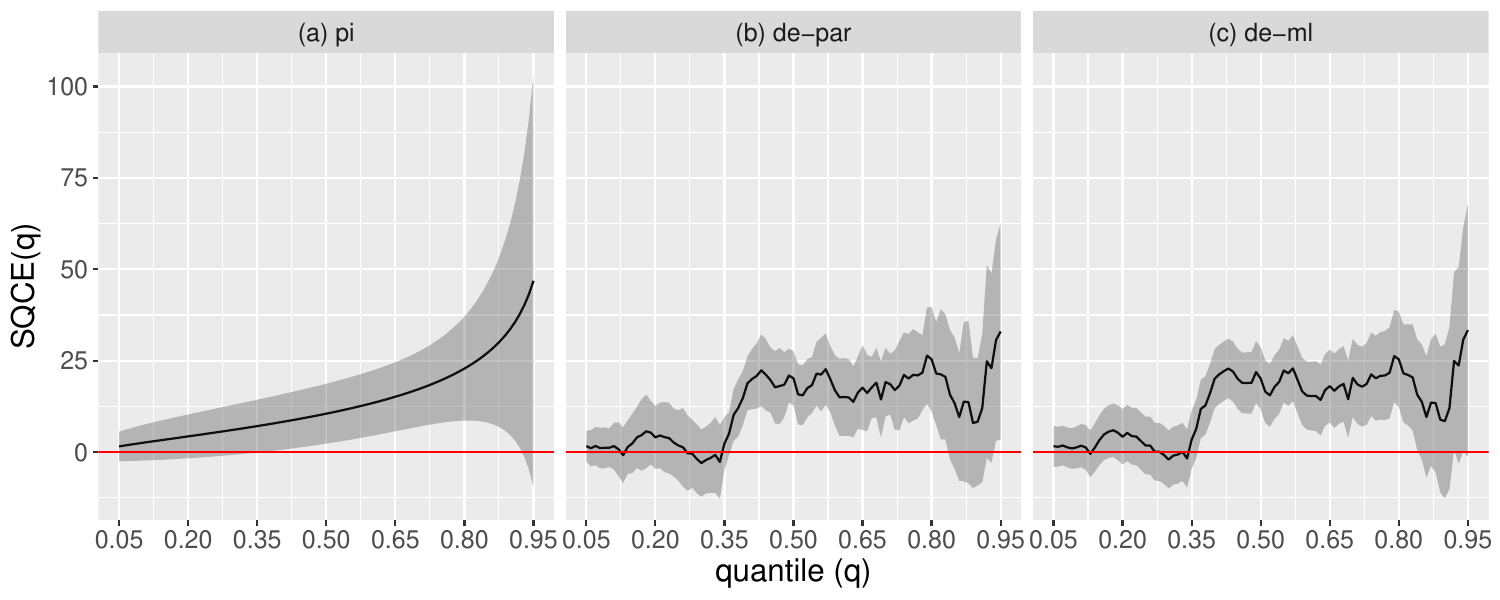}
\end{center}
\caption{Analysis of the survivor quantile causal effect (SQCE) with its 95\% confidence interval for the National Jobs Corps study, based on $\widehat{\text{SQCE}}^{pi}(q)$ (Panel a), $\widehat{\text{SQCE}}^{de\text{-}par}(q)$ (Panel b), and $\widehat{\text{SQCE}}^{de\text{-}ml}(q)$ (Panel c)}
\label{fig:SQCE-JC}
\end{figure}

We follow the parametric working models and machine learners in Sections \ref{sec:sim}--\ref{sec:qme_additional_sim} to estimate the nuisance functions. One exception is our specification of 
the two outcome probabilities $F_{Y|A,M,L}(\theta|0,1,L)$ and $F_{Y|A,M,L}(\theta|1,1,L)$. Because $Y$ is always non-negative and also right-skewed, we consider after a log-transformation, $\log(Y)|\{A=1,M=1,L\}$ and $\log(Y)|\{A=0,M=1,L\}$ follow the heteroskedastic Gaussian model in Section \ref{sec:sim}.  Similarly, we also assume the logarithm of outcome follows the additive model when machine learners are used. We consider three estimators $\widehat{\text{SQCE}}^{pi}(q)$, $\widehat{\text{SQCE}}^{de\text{-}par}(q)$, and $\widehat{\text{SQCE}}^{de\text{-}ml}(q)$; the estimated results are given in Figure \ref{fig:SQCE-JC}. All estimators indicate a positive quantile effect among survivors, demonstrating that the treatment effectively enhances participants' weekly wages. Both $\widehat{\text{SQCE}}^{de\text{-}par}(q)$ and $\widehat{\text{SQCE}}^{de\text{-}ml}(q)$ yield similar point estimates across different quantiles. As $q$ increases, both estimators suggests that the treatment effect first strengthens around $q\in[0.3,0.5]$ and then plateaus afterwards. In contrast, $\widehat{\text{SQCE}}^{pi}(q)$ indicates that the treatment effect progressively strengthens as $q$ increases.


}

\section{Example 4: Quantile treatment effect with a time-varying treatment}

Consider the estimation of treatment effect in a longitudinal observational study with $T$ time periods ($T\geq 2$), where in each time period we observe a treatment with certain covariates. Specifically, for each $t=1,\dots,T$, let $A_t\in\{0,1\}$ be the treatment administrated during period $t$ and $L_t$ be a set of time-varying covariates measured just at the beginning of period $t$. The outcome $Y$ is measured at the end of the $T$-th period. To summarize, we observe $W=\{L_1,A_1,L_2,A_2,\dots,L_T,A_T,Y\}$. Following the standard notation in longitudinal causal inference \citep{bang2005doubly}, denote $\bar A_{t} =\{A_1,\cdots,A_t\}$ and $\bar L_{t} =\{L_1,\cdots,L_t\}$ as the observed treatment and covariates history up until period $t$. Similarly, let $\bar a_T \in \{0,1\}^{\otimes T}$ be a possible treatment regimen of the time-varying treatment $\bar A_T$. For each possible $\bar a_T$, let $Y_{\bar a_T}$ be the potential outcome that this unit would have observed if this unit would follow $\bar a_T$. Existing literature mainly considers estimating the mean of $Y_{\bar a_T}$, based on which a mean treatment effect $E[Y_{\bar a_T}-Y_{\bar a_T'}]$ can be analyzed, with only a few exceptions considering estimating the quantile of $Y_{\bar a_T}$, $\theta_0:=Q_{Y_{\bar a_T}}(q)$. For example, under a marginal structural quantile model, \cite{hogan2004marginal} and \cite{cheng2024doubly} develop an inverse probability estimator and a doubly robust estimator for $\theta_0$, respectively. Here, in contrast to previous work using a marginal structural model to restrict the form of the $q$-quantile of $Y_{\bar a_T}$, we directly estimate $\theta_0$ without any parametric restrictions.

In order to use the IEEs assessing $\theta_0:=Q_{Y_{\bar a_T}}(q)$, one need first find an identifying moment function for the expectation of the threshold-transformed potential outcome $\tau_0(\theta):=E[\bbI(Y_{\bar a_T}\leq\theta)]$. According to \cite{bang2005doubly}, under assumptions (I) consistency ($Y_{\bar a_T}=Y$ if $\bar A_T=\bar a_T$), (II) sequential ignorability ($Y_{\bar a_T}\perp A_t|\bar L_{t},\bar A_{t-1}=\bar a_{t-1}$ for any $t$ and $\bar a_{t-1}$), and (III) positivity (if $f_{\bar A_{t-1},\bar L_t}(\bar a_{t-1},\bar l_t)>0$, then $f_{A_t|\bar A_{t-1},\bar L_t}(a_t|\bar a_{t-1},\bar l_t)>0$ for any $t$, $\bar a_t$, and $\bar l_t$), $\tau_0(\theta)$ can be identified by the following inverse probability of treatment weighted threshold outcome,
$$
\tau_0(\theta) = \frac{\bbI(\bar A_T=\bar a_T)}{\bar \pi_{T}(\bar a_T,\bar L_T)}\mathbb{I}(Y\leq \theta),
$$
where $\bar \pi_{t}(\bar a_t,\bar L_t) = \prod_{r=1}^t f_{A_r|\bar A_{r-1},\bar L_{r}}(a_r|\bar a_{r-1},\bar L_{r})$ is the cumulative probability for receiving the treatment $\bar a_t$ during the first $t$ periods. Thus, an identifying moment function for $\tau_0(\theta)$ is
$$
g(W,\tau,\theta,\gamma) = \frac{\bbI(\bar A_T=\bar a_T)}{\bar \pi_{T}(\bar a_T,\bar L_T)}\mathbb{I}(Y\leq \theta) -\tau
$$
with $\gamma=\{f_{A_1|L_{1}}(a_1|L_1),f_{A_2|A_1,\bar L_{2}}(a_2|a_1,\bar L_2),\dots, f_{A_T|\bar A_{T-1},\bar L_{T}}(a_T|\bar a_{T-1},\bar L_T) \}$ a sequence of longitudinal treatment probabilities. Notice that the $T$ nuisance functions in $\gamma$ are essentially a sequence of regressions with a binary response variable, which can be obtained by logistic regressions or machine learners designed for binary classifications. Therefore, if one additionally assumes that the $q$-quantile of the distribution of $Y_{\bar a_T}$ is unique, Theorem 2 suggests that $\theta_0$ is also identified and the IEE can be constructed by
$$
\Pn\left[\frac{\bbI(\bar A_T=\bar a_T)}{\widehat{\bar \pi}_{T}(\bar a_T,\bar L_T)}\mathbb{I}(Y\leq \theta) -q\right] =  0,
$$
where $ \widehat{\bar \pi}_{t}(\bar a_t,\bar L_t)=\prod_{r=1}^t \widehat f_{A_r|\bar A_{r-1},\bar L_{r}}(a_r|\bar a_{r-1},\bar L_{r})$. Solving the IEE in terms of $\theta$ leads to a convenient estimator of $\theta_0$. 

To improve the performance of direct using the identifying moment function, one can derive the EIF of $\theta_0$. There are a productive line of research investigating nonparametric efficient estimation of the mean of potential outcome in longitudinal settings (see, for example,  \cite{tran2019double,bodory2022evaluating,chernozhukov2022automatic}), and the EIF of $\tau_0(\theta)$ is given by $\psi_{\tau_0(\theta)}^{\text{eff}}(W,\tau_0(\theta),\theta,h)=-\frac{1}{C(\theta)}\left\{g(W,\tau_0(\theta),\theta,\gamma) + \phi(W,\tau_0(\theta),\theta,h)\right\}$ with $C(\theta)=-1$ and the adjustment term
\begin{align*}
\phi(W,\tau,\theta,h) = &  \sum_{t=1}^{K-1} \frac{\bbI(\bar A_t=\bar a_t)}{\bar \pi_{t}(\bar a_t,\bar L_t)} \left\{\mu_{t+1}(\bar a_{t+1},\bar L_{t+1},\theta)-\mu_{t}(\bar a_{t},\bar L_{t},\theta)\right\} \\
&  + \mu_{1}(a_1,L_1,\theta)- \frac{\bbI(\bar A_T=\bar a_T)}{\bar \pi_{T}(\bar a_T,\bar L_T)} \mu_{T}(\bar a_{T},\bar L_{T},\theta).
\end{align*}
Therefore, $h$ contains $2T$ nuisance functions, where $\{h_1,\dots,h_T\}$  are $\gamma$ in the identifying moment function and $\{h_{T+1},\dots,h_{2T}\}$ are $\mu_1(a_1,L_1,\theta)$, $\mu_2(\bar a_2,\bar L_2,\theta)$, $\cdots$, $\mu_T(\bar a_T,\bar L_T,\theta)$. Here, $\mu_{t}(\bar a_{t},\bar L_t,\theta)$, $t=1,\dots,T$, are characterized by the following nested outcome expectations:
\begin{equation}\label{eq:example5_1}
\left\{\begin{aligned}
&\mu_{T}(\bar a_{T},\bar L_T,\theta):=F_{Y|\bar A_T,\bar L_T}(\theta|\bar a_T,\bar L_T) \\
&\mu_{t}(\bar a_{t},\bar L_t,\theta):= E[\mu_{t+1}(\bar a_{t+1},\bar L_{t+1},\theta)|\bar A_t=\bar a_t,\bar L_{t}], \quad \text{for $t=1,\dots,T-1$}.
\end{aligned}\right.
\end{equation}
Based on Theorem 2, it is immediate that the EIF of $\theta_0$ has the following explicit form:
\begin{align*}
\psi_{\theta_0}^{\text{eff}}(W,q,\theta_0,h) = &  -\frac{1}{B}\Big\{\frac{\bbI(\bar A_T=\bar a_T)}{\bar \pi_{T}(\bar a_T,\bar L_T)} \left\{\bbI(Y\leq\theta_0) - \mu_{T}(\bar a_{T},\bar L_{T},\theta_0)\right\} \\
& + \sum_{t=1}^{K-1} \frac{\bbI(\bar A_t=\bar a_t)}{\bar \pi_{t}(\bar a_t,\bar L_t)} \left\{\mu_{t+1}(\bar a_{t+1},\bar L_{t+1},\theta_0)-\mu_{t}(\bar a_{t},\bar L_{t},\theta_0)\right\} \\
&  + \mu_{1}(a_1,L_1,\theta_0)- q\Big\},
\end{align*}
where $B=\frac{d}{d\theta}E\left[\frac{\bbI(\bar A_T=\bar a_T)}{\widehat{\bar \pi}_{T}(\bar a_T,\bar L_T)}\mathbb{I}(Y\leq \theta) -q\right]|_{\theta=\theta_0} = \frac{d}{d\theta}E\left[\frac{\bbI(\bar A_T=\bar a_T)}{\widehat{\bar \pi}_{T}(\bar a_T,\bar L_T)}F_{Y|\bar A_T,\bar L_T}(\theta|\bar a_T,\bar L_T) -q\right]|_{\theta=\theta_0}=E\left[\frac{\bbI(\bar A_T=\bar a_T)}{\widehat{\bar \pi}_{T}(\bar a_T,\bar L_T)}f_{Y|\bar A_T,\bar L_T}(\theta_0|\bar a_T,\bar L_T)\right]$. Thus, the debiased estimator $\widehat \theta^{de}$ is obtained by solving $\Pn[\psi_{\theta_0}^{\text{eff}}(W,q,\theta,\widehat h)]=0$ in terms of $\theta$. Here, $\widehat h$ is certain estimate of $h$, where estimation of the first $T$ components (i.e., $\widehat \gamma$) is introduced before, and below we discuss several strategies for estimating nested outcome expectations \eqref{eq:example5_1}. First, we can fit a sequence of parametric or non-parametric models for the conditional densities $f_{Y|\bar A_T,\bar L_T}$ and $f_{L_t|\bar A_{t-1},\bar L_{t-1}}$, $t=2,\dots,T$, then we have $\widehat \mu_{T}(\bar a_{T},\bar L_T,\theta)=\widehat F_{Y|\bar A_T,\bar L_T}(\theta|\bar a_T,\bar L_T)$ and  $$\widehat \mu_{t}(\bar a_{t},\bar L_t,\theta)=\int_{l_{t+1}}\widehat \mu_{t+1}(\bar a_{t+1},\bar L_{t+1},\theta) \widehat f_{L_{t+1}|\bar A_t,\bar L_t}(l_{t+1}|\bar a_t,\bar L_t)dl_t, \quad t=T-1,\dots,1.$$ However, this may be computationally infeasible when $L_t$ is multidimensional with many continuous components. A possible remedy is using the iterative conditional regression (ICR) approach in \cite{cheng2024doubly}, where we specify a sequence of parametric working models for the distributions  $Y_{\bar a_T}|\{\bar A_t=\bar a_t,\bar L_t\}$, $t=T,\dots,1$, and then solve an iterative estimating equation to obtain  $\widehat \mu_t(\bar a_t,\bar L_t,\theta)$, $t=1,\dots,T$, directly. Notice that ICR is a purely parametric method and may not easily incorporate nonparametric methods or machine learners. Alternatively, one can use the ``regression imputation" approach as we discussed in Example 2, where we first select a grid of points $\underline\Theta:=\{\underline\theta_r\}_{r=1}^R$ in the support of $Y$. Then, for each $\underline\theta_r \in \underline\Theta$, we obtain $\widehat\mu_T(\bar a_T,\bar L_T,\underline\theta_r)=\widehat F_{Y|\bar A_T,\bar L_T}(\theta|\bar a_T,\bar L_T)$ based on machine learning/parametric methods for conditional CDFs and then, iteratively, obtain $\widehat\mu_t(\bar a_t,\bar L_t,\underline\theta_r)$, $t=T-1,\dots,1$, by regressing $\widehat\mu_{t+1}(\bar a_{t+1},\bar L_{t+1},\underline\theta_r)$ on $\bar L_{t}$ based on units with $\bar A_t=\bar a_t$. Finally, $\mu_t(\bar a_t,\bar L_t,\theta)$ can be approximated by linearly interpolating the grids of pairs $\{\underline\theta_r,\widehat \mu_t(\bar a_t,\bar L_t,\underline\theta_r)\}_{r=1}^R$ so that $\widehat \mu_t(\bar a_t,\bar L_t,\theta) = \widehat \mu_t(\bar a_t,\bar L_t,\underline\theta_r) + \frac{\theta-\underline\theta_r}{\underline\theta_{r+1}-\underline\theta_r}\{\widehat  \mu_t(\bar a_t,\bar L_t,\underline\theta_{r+1})-\widehat  \mu_t(\bar a_t,\bar L_t,\underline\theta_r)\}$, where $\underline\theta_r<\theta<\underline\theta_{r+1}$.

We next analyze the asymptotic properties of $\widehat\theta^{de}$. After some algebra, we can show that
\begingroup\makeatletter\def\f@size{9.5}\check@mathfonts 
\begin{align*}
& E[\psi_{\theta_0}^{\text{eff}}(W,q,\theta,\widehat h)-\psi_{\theta_0}^{\text{eff}}(W,q,\theta,h)] \\
= & \sum_{t=1}^T E\left[S_t\left\{\widehat f_{A_t|\bar A_{t-1},\bar L_t}(a_t|\bar a_{t-1},\bar L_t)-f_{A_t|\bar A_{t-1},\bar L_t}(a_t|\bar a_{t-1},\bar L_t)\right\}\left\{\widehat \mu_t(a_t,L_t,\theta)- \mu_t(a_t,L_t,\theta)\right\}\right] 
\end{align*}\endgroup
with $S_1=-\frac{1}{B\widehat f_{A_1|L_1}(a_1|L_1)}$ and $S_t=-\frac{\bar \pi_{t-1}(\bar a_{t-1},\bar L_{t-1})}{B\widehat{\bar \pi}_{t-1}(\bar a_{t-1},\bar L_{t-1})\widehat f_{A_t|\bar A_{t-1},\bar L_{t-1}}(a_t|\bar a_{t-1},\bar L_{t-1})}$ if $t>1$. Notice that all $E[S_t^2]<\infty$, $t=1,\dots,T$, if all $\widehat f_{A_t|\bar A_{t-1},\bar L_{t-1}}(a_t|\bar a_{t-1},\bar L_{t-1})$ are bounded away from 0. Therefore, $\psi_{\theta_0}^{\text{eff}}(W,q,\theta,h)$ satisfies the mixed bias property (10), with $K=T$ and $\{a_k,b_k\}=\{k,T+k\}$, $k=1,\dots, K$. This indicates when all nuisance functions $h$ are consistently estimated based on machine learners with a cross-fitting procedure, $\widehat \theta^{de}$ is CAN and efficient if the rate of convergence of $\widehat h$ satisfies $\sum_{t}^T \xi_{t,n}\xi_{T+t,n}=o_p(n^{-\frac{1}{2}})$. Otherwise, if parametric working models are used to fit $h$, then $\widehat \theta^{de}$ is CAN under the union model $\cap_{t=1}^T \{\mathcal M_t \cup \mathcal M_{T+t}\}$. In other words,  $\widehat \theta^{de}$ is CAN if, for all $t\in\{1,\dots,T\}$, either $\mathcal M_t$ or $\mathcal M_{T+t}$ is correctly specified, which includes a total of $2^T$ opportunities to achieve consistency. However, the last $T$ components in $h$ (i.e., the nested outcome expectations in \eqref{eq:example5_1}) are not variationally independent, where correct specification of $\mathcal M_{T+t}$ depends on correct specification of all $\mathcal M_{T+t+1},\dots,\mathcal M_{2T}$. It follows that $\widehat \theta^{de}$ degenerates to a $(T+1)$-tuple robust estimator such that $\widehat \theta^{de}$ is CAN if, for any $t=0,\dots, T$, the first $t$ treatment probabilities (i.e., $\mathcal M_1,\dots, \mathcal M_t$) and the last $T-t$ outcome expectations (i.e., $\mathcal M_{T+t+1},\dots,\mathcal M_{2T}$) are correctly specified.


\begin{figure}[ht]
\begin{center}
\includegraphics[width=0.99\textwidth]{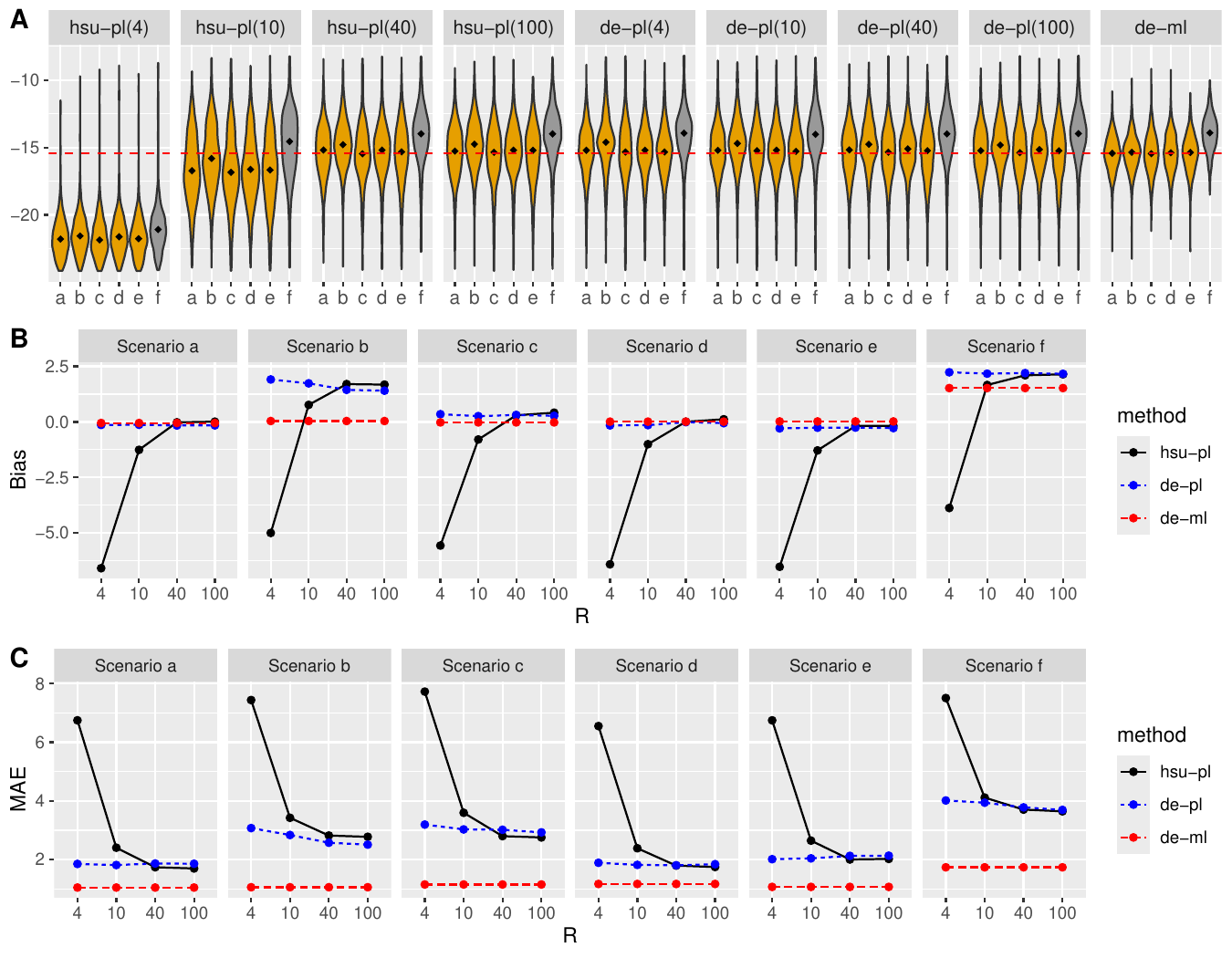}
\end{center}
\caption{Simulation results of $Q_{Y_{1M_0}}(0.1)$ in Example 2. Panel A: Violin plots of the point estimates of $Q_{Y_1M_0}(0.1)$ under 6 scenarios (Scenarios a--f) regarding correct and incorrect specification of nuisance models. The red dotted line is the true value of $Q_{Y_1M_0}(0.1)$. For each scenario, the violin plot filled with orange color indicate the estimator is consistent based on theory. Panels B and C: Bias and MAE for point estimates of $Q_{1M_0}(0.1)$ with different number of thresholds (i.e., $R$) used to construct \texttt{hsu-pl($R$)} and \texttt{de-pl($R$)}. The bias and MAE of \texttt{de-ml} (red horizontal line) are added as reference.}
\label{fig:QME-10}
\end{figure}

\begin{figure}[ht]
\begin{center}
\includegraphics[width=0.99\textwidth]{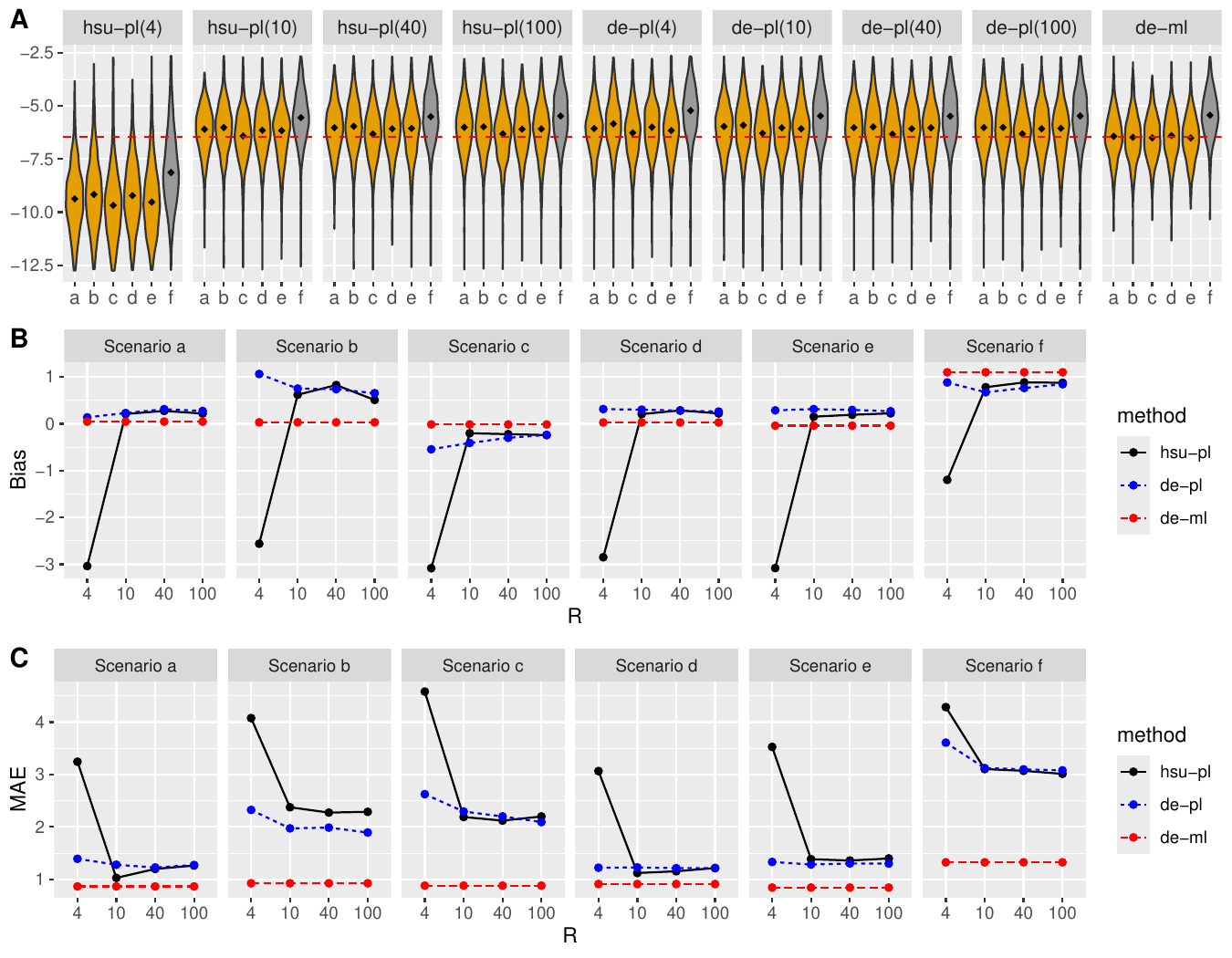}
\end{center}
\caption{Simulation results of $Q_{Y_{1M_0}}(0.25)$ in Example 2. Panel A: Violin plots of the point estimates of $Q_{Y_1M_0}(0.25)$ under 6 scenarios (Scenarios a--f) regarding correct and incorrect specification of nuisance models. The red dotted line is the true value of $Q_{Y_1M_0}(0.25)$. For each scenario, the violin plot filled with orange color indicate the estimator is consistent based on theory. Panels B and C: Bias and MAE for point estimates of $Q_{1M_0}(0.25)$ with different number of thresholds (i.e., $R$) used to construct \texttt{hsu-pl($R$)} and \texttt{de-pl($R$)}. The bias and MAE of \texttt{de-ml} (red horizontal line) are added as reference.}
\label{fig:QME-25}
\end{figure}

\begin{figure}[ht]
\begin{center}
\includegraphics[width=0.99\textwidth]{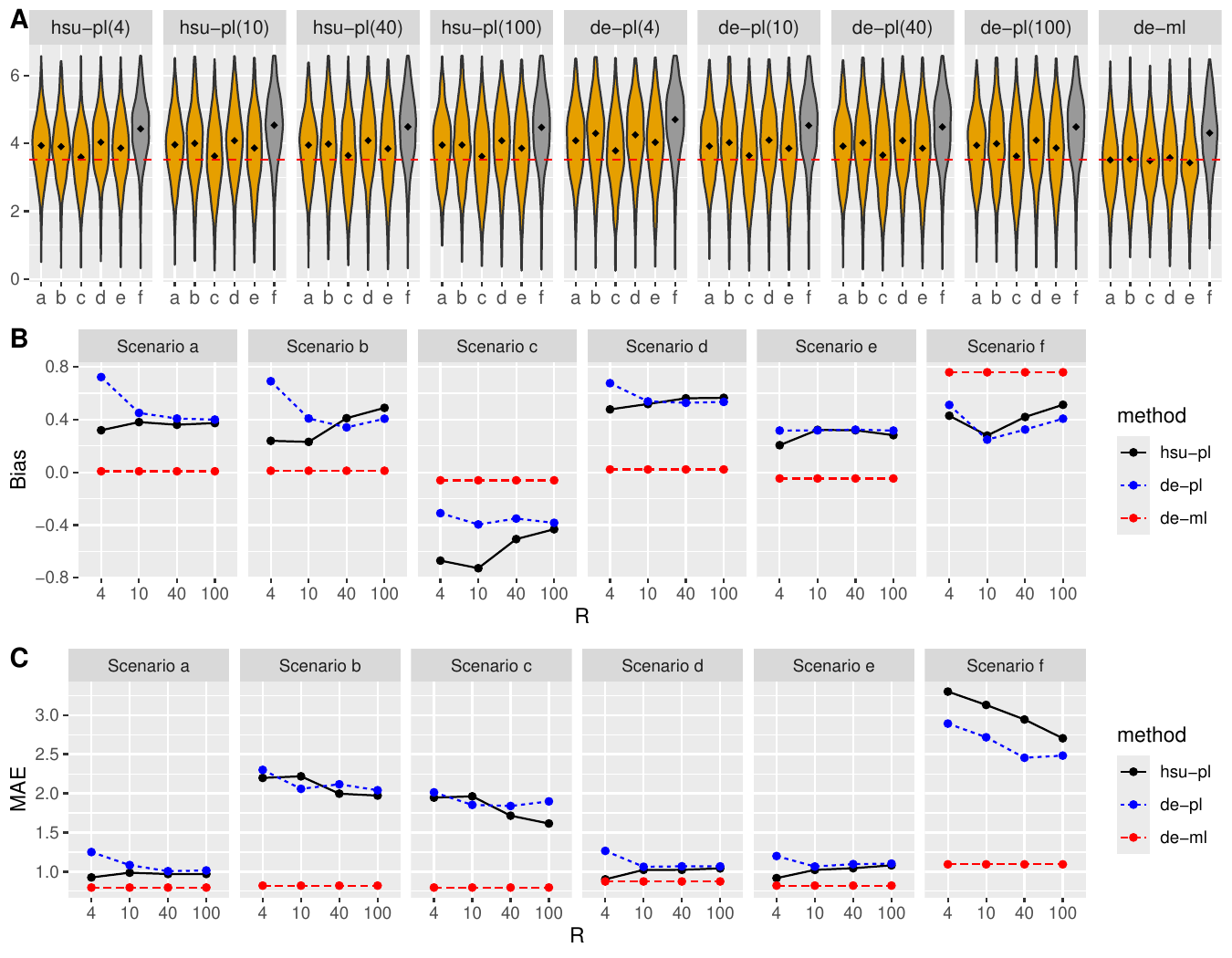}
\end{center}
\caption{Simulation results of $Q_{Y_{1M_0}}(0.5)$ in Example 2. Panel A: Violin plots of the point estimates of $Q_{Y_1M_0}(0.5)$ under 6 scenarios (Scenarios a--f) regarding correct and incorrect specification of nuisance models. The red dotted line is the true value of $Q_{Y_1M_0}(0.5)$. For each scenario, the violin plot filled with orange color indicate the estimator is consistent based on theory. Panels B and C: Bias and MAE for point estimates of $Q_{1M_0}(0.5)$ with different number of thresholds (i.e., $R$) used to construct \texttt{hsu-pl($R$)} and \texttt{de-pl($R$)}. The bias and MAE of \texttt{de-ml} (red horizontal line) are added as reference.}
\label{fig:QME-50}
\end{figure}

\begin{figure}[ht]
\begin{center}
\includegraphics[width=0.99\textwidth]{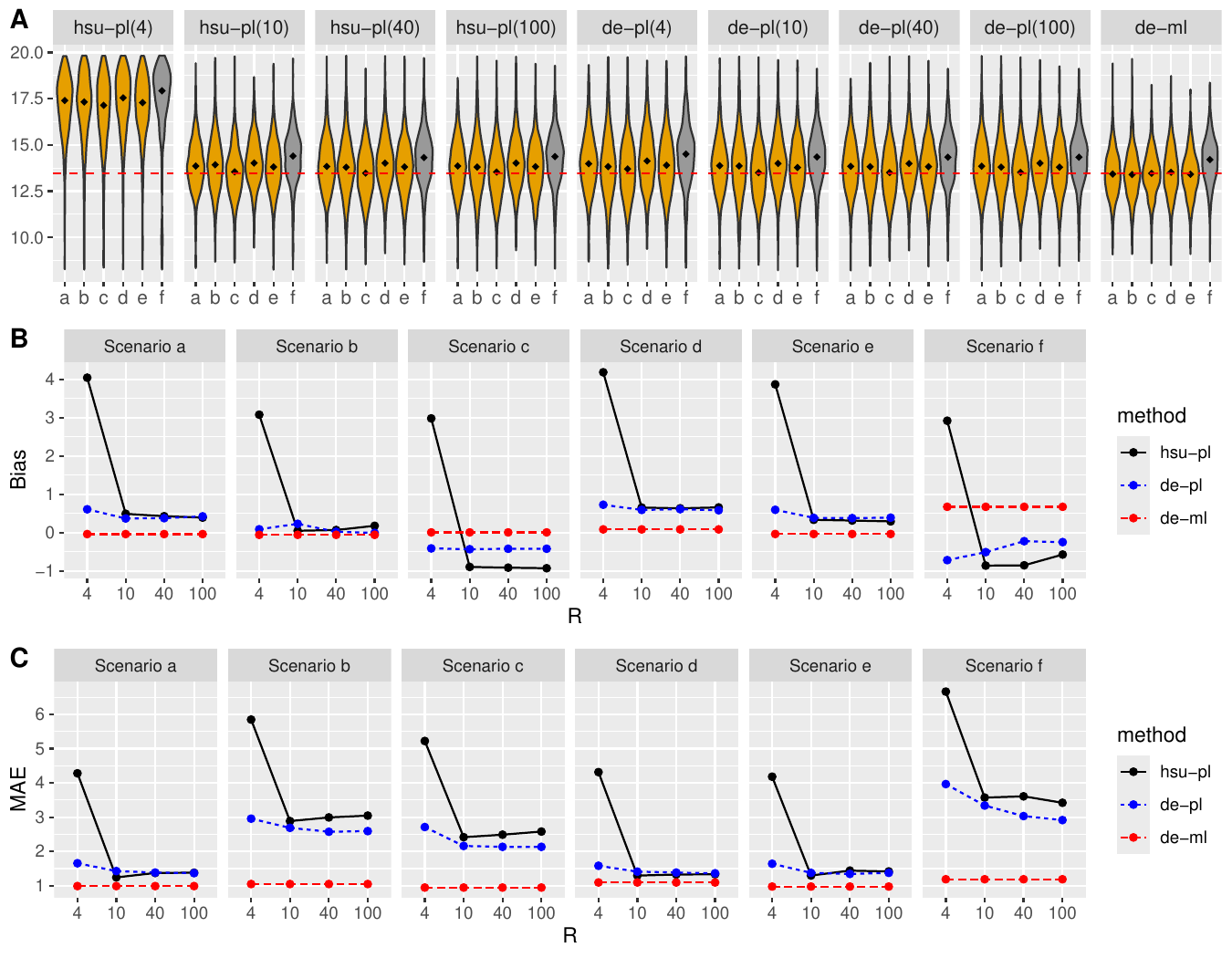}
\end{center}
\caption{Simulation results of $Q_{Y_{1M_0}}(0.75)$ in Example 2. Panel A: Violin plots of the point estimates of $Q_{Y_1M_0}(0.75)$ under 6 scenarios (Scenarios a--f) regarding correct and incorrect specification of nuisance models. The red dotted line is the true value of $Q_{Y_1M_0}(0.75)$. For each scenario, the violin plot filled with orange color indicate the estimator is consistent based on theory. Panels B and C: Bias and MAE for point estimates of $Q_{1M_0}(0.75)$ with different number of thresholds (i.e., $R$) used to construct \texttt{hsu-pl($R$)} and \texttt{de-pl($R$)}. The bias and MAE of \texttt{de-ml} (red horizontal line) are added as reference.}
\label{fig:QME-75}
\end{figure}

\begin{figure}[ht]
\begin{center}
\includegraphics[width=0.99\textwidth]{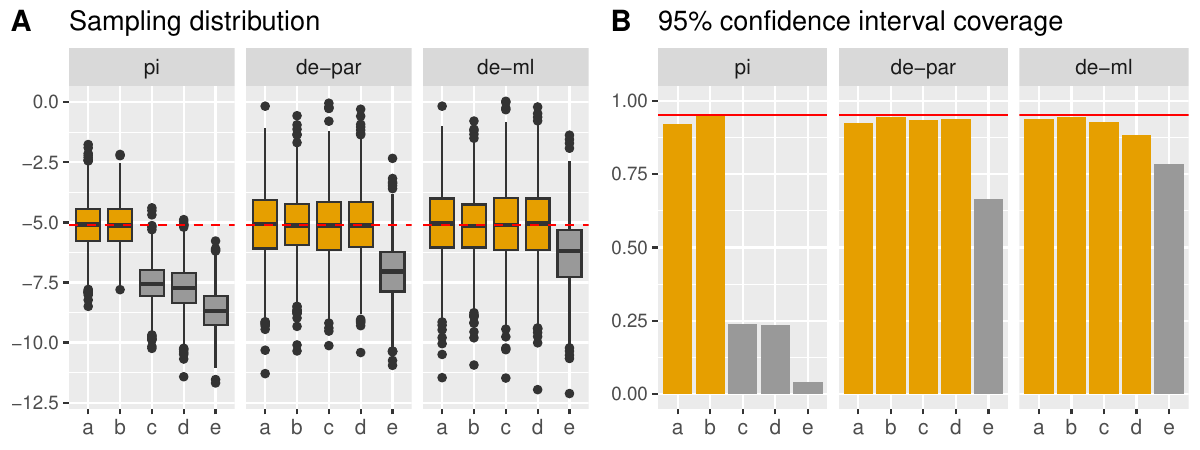}
\end{center}
\caption{Sampling distributions (Panel A) and 95\% confidence interval coverage probability (Panel B) among estimators of $Q_{Y_1|V}(0.25|\{1,1\})$, in scenarios (a) all nuisance models correctly specified; (b) $\widehat f_{A|L}(1|L)$ incorrect; (c) $\widehat f_{M|A,L}(1|0,L)$ and $\widehat f_{M|A,L}(1|1,L)$ incorrect; (d) $\widehat F_{Y|A,M,L}(\theta|1,1,L)$ incorrect; (e) all nuisance models incorrect.
For each scenario, the box/bar filled with orange color indicates that the estimator is consistent based on theory. The red dotted line in Panel A is the true value of $Q_{Y_1}(0.25)$.}
\label{fig:Q1_25}
\end{figure}

\begin{figure}[ht]
\begin{center}
\includegraphics[width=0.99\textwidth]{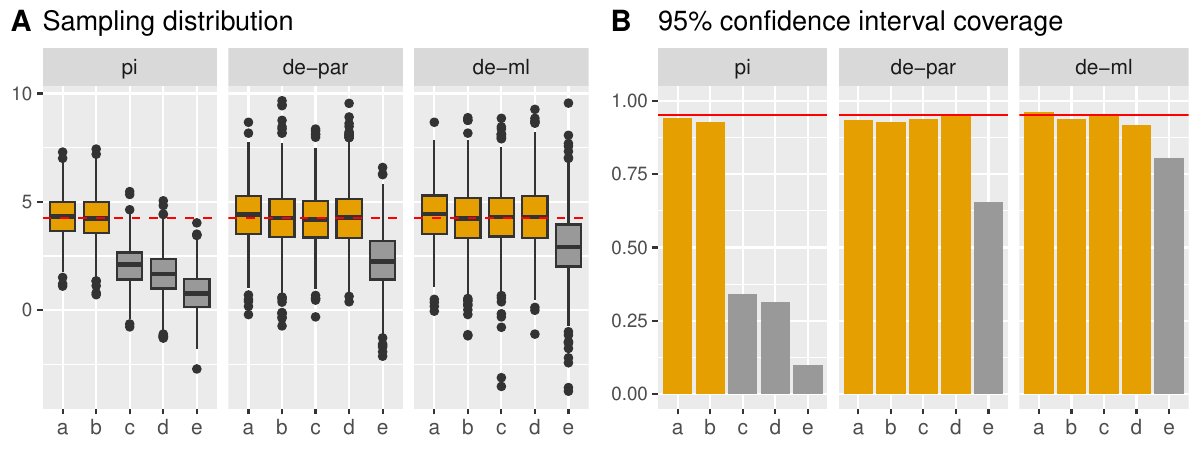}
\end{center}
\caption{Sampling distributions (Panel A) and 95\% confidence interval coverage probability (Panel B) among estimators of $Q_{Y_1|V}(0.5|\{1,1\})$, in scenarios (a) all nuisance models correctly specified; (b) $\widehat f_{A|L}(1|L)$ incorrect; (c) $\widehat f_{M|A,L}(1|0,L)$ and $\widehat f_{M|A,L}(1|1,L)$ incorrect; (d) $\widehat F_{Y|A,M,L}(\theta|1,1,L)$ incorrect; (e) all nuisance models incorrect.
For each scenario, the box/bar filled with orange color indicates that the estimator is consistent based on theory. The red dotted line in Panel A is the true value of $Q_{Y_1}(0.5)$.}
\label{fig:Q1_50}
\end{figure}

\begin{figure}[ht]
\begin{center}
\includegraphics[width=0.99\textwidth]{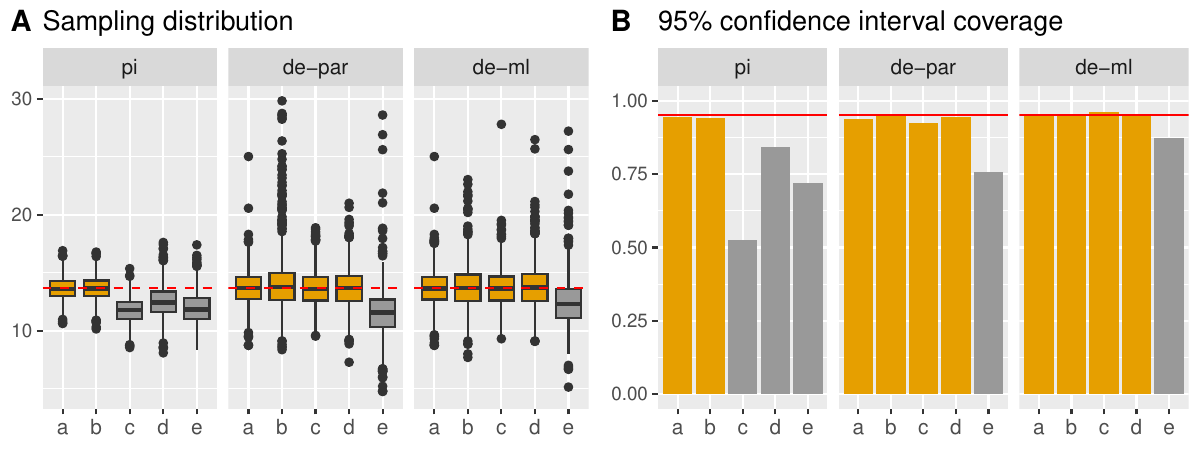}
\end{center}
\caption{Sampling distributions (Panel A) and 95\% confidence interval coverage probability (Panel B) among estimators of $Q_{Y_1|V}(0.75|\{1,1\})$, in scenarios (a) all nuisance models correctly specified; (b) $\widehat f_{A|L}(1|L)$ incorrect; (c) $\widehat f_{M|A,L}(1|0,L)$ and $\widehat f_{M|A,L}(1|1,L)$ incorrect; (d) $\widehat F_{Y|A,M,L}(\theta|1,1,L)$ incorrect; (e) all nuisance models incorrect.
For each scenario, the box/bar filled with orange color indicates that the estimator is consistent based on theory. The red dotted line in Panel A is the true value of $Q_{Y_1}(0.75)$.}
\label{fig:Q1_75}
\end{figure}

\begin{figure}[ht]
\begin{center}
\includegraphics[width=0.99\textwidth]{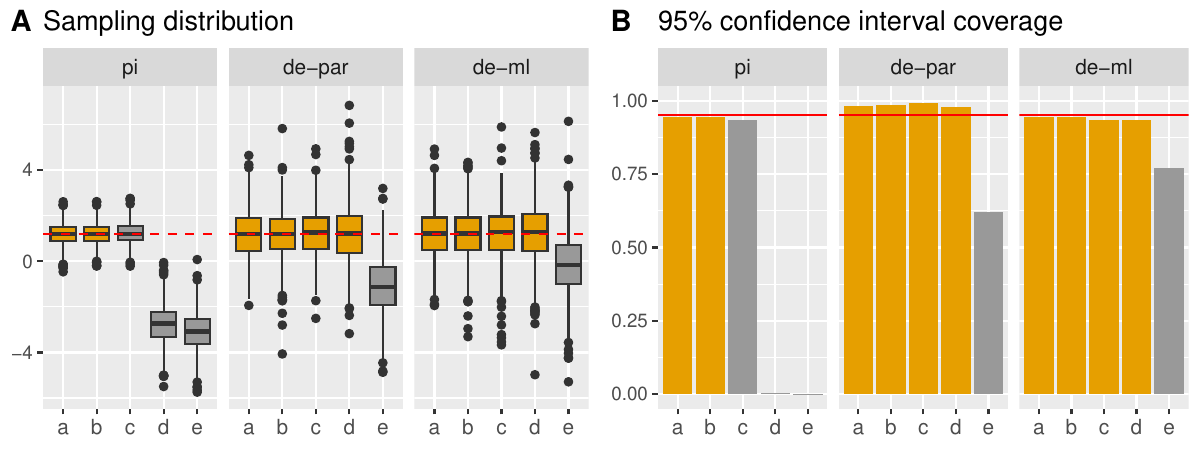}
\end{center}
\caption{Sampling distributions (Panel A) and 95\% confidence interval coverage probability (Panel B) among estimators of $\text{SQCE}(0.25)$, in scenarios (a) all nuisance models correctly specified; (b) $\widehat f_{A|L}(1|L)$ incorrect; (c) $\widehat f_{M|A,L}(1|0,L)$ and $\widehat f_{M|A,L}(1|1,L)$ incorrect; (d) $\widehat F_{Y|A,M,L}(\theta|1,1,L)$ and $\widehat F_{Y|A,M,L}(\theta|0,1,L)$ incorrect; (e) all nuisance models incorrect.
For each scenario, the box/bar filled with orange color indicates that the estimator is consistent based on theory. The red dotted line in Panel A is the true value of $\text{SQCE}(0.25)$.}
\label{fig:Qd_25}
\end{figure}

\begin{figure}[ht]
\begin{center}
\includegraphics[width=0.99\textwidth]{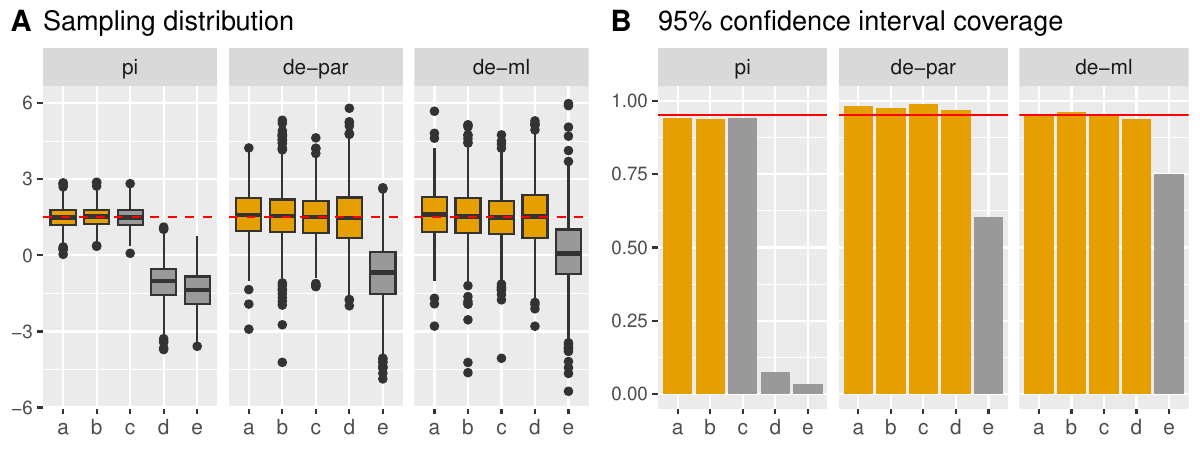}
\end{center}
\caption{Sampling distributions (Panel A) and 95\% confidence interval coverage probability (Panel B) among estimators of $\text{SQCE}(0.5)$, in scenarios (a) all nuisance models correctly specified; (b) $\widehat f_{A|L}(1|L)$ incorrect; (c) $\widehat f_{M|A,L}(1|0,L)$ and $\widehat f_{M|A,L}(1|1,L)$ incorrect; (d) $\widehat F_{Y|A,M,L}(\theta|1,1,L)$ and $\widehat F_{Y|A,M,L}(\theta|0,1,L)$ incorrect; (e) all nuisance models incorrect.
For each scenario, the box/bar filled with orange color indicates that the estimator is consistent based on theory. The red dotted line in Panel A is the true value of $\text{SQCE}(0.5)$.}
\label{fig:Qd_50}
\end{figure}

\begin{figure}[ht]
\begin{center}
\includegraphics[width=0.99\textwidth]{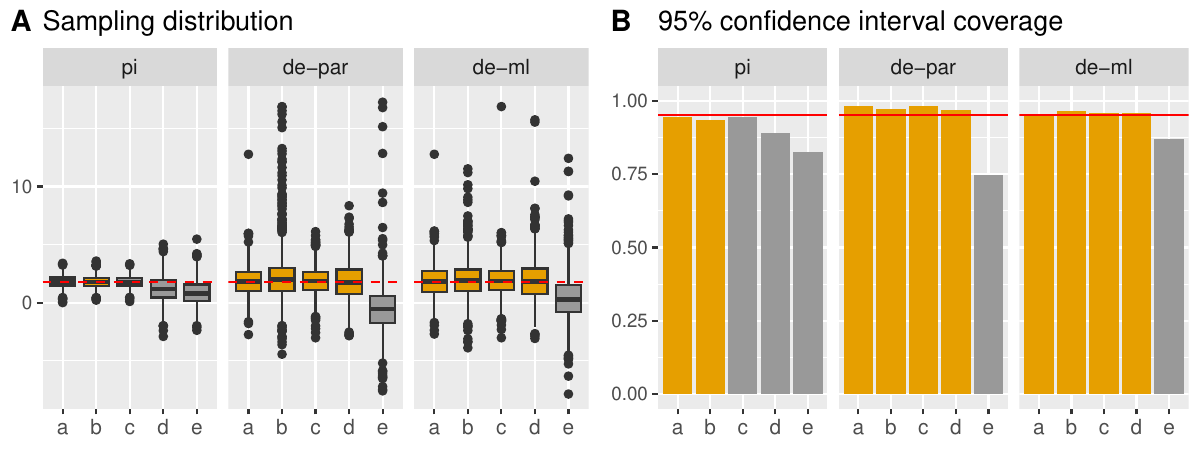}
\end{center}
\caption{Sampling distributions (Panel A) and 95\% confidence interval coverage probability (Panel B) among estimators of $\text{SQCE}(0.75)$, in scenarios (a) all nuisance models correctly specified; (b) $\widehat f_{A|L}(1|L)$ incorrect; (c) $\widehat f_{M|A,L}(1|0,L)$ and $\widehat f_{M|A,L}(1|1,L)$ incorrect; (d) $\widehat F_{Y|A,M,L}(\theta|1,1,L)$ and $\widehat F_{Y|A,M,L}(\theta|0,1,L)$ incorrect; (e) all nuisance models incorrect.
For each scenario, the box/bar filled with orange color indicates that the estimator is consistent based on theory. The red dotted line in Panel A is the true value of $\text{SQCE}(0.75)$.}
\label{fig:Qd_75}
\end{figure}

\end{document}